\documentclass[12pt]{article}
\usepackage[T1]{fontenc}
\usepackage{lmodern}
\usepackage[margin=25mm]{geometry}
\usepackage{amsmath,amssymb,booktabs,graphicx,microtype}
\usepackage[font=small]{caption}
\usepackage[numbers,sort&compress]{natbib}
\usepackage[hidelinks]{hyperref}
\usepackage{xurl}
\title{DeepTensorMoment: adaptive local representations for fast universal interatomic potentials}
\author{Xin Chen\textsuperscript{1}, Lifang Wang\textsuperscript{1}, Xingyu Gao\textsuperscript{1}, Bangkai Zhu\textsuperscript{1},\\
Yucheng Ouyang\textsuperscript{2}, Ying Liu\textsuperscript{2}, Haifeng Liu\textsuperscript{1}, Haifeng Song\textsuperscript{1,}\thanks{Correspondence: \texttt{song\_haifeng@iapcm.ac.cn}}\\[4pt]
\small \textsuperscript{1}National Key Laboratory of Computational Physics,\\
\small Institute of Applied Physics and Computational Mathematics, Beijing, China\\
\small \textsuperscript{2}State Key Lab of Processors, Institute of Computing Technology,\\
\small Chinese Academy of Sciences, Beijing, China}
\date{}
\begin{document}
\maketitle
\begin{abstract}
Which local computations justify their cost in fast universal interatomic potentials, and when do more accurate forces improve physical predictions?
We introduce DeepTensorMoment (DTM), a family of strictly local equivariant potentials with adjustable angular resolution, atom-centred tensor-moment refinement and adaptive neighbour weighting.
Across alloy and OMat24 benchmarks, finer angular resolution improves force accuracy, and refinement brings further gains while reducing the benefit of explicit cubic readout.
At equal parameter count, weighting neighbours with refined rather than initial moments lowers force errors at similar throughput.
On OMat24, DTM and DPA4C share the measured high-throughput frontier.
DTM models span mean absolute force errors of 73.0--159.6\,meV/\AA{} on the official validation set and deliver 0.35--105.5 million atom-steps/s for a 16-element alloy on one RTX~5090 GPU.
Across 13 DTM models, lower validation force errors generally accompany more accurate predictions of formation energy, bulk modulus and thermal conductivity.
\end{abstract}

Materials design requires understanding how chemical composition and atomic structure determine stability and physical properties. 
High-throughput screening and inverse design assess many candidate materials through structural relaxation and property calculations~\cite{chen2022m3gnet,mattergen2025,prome2025}. 
Detailed studies of promising candidates also require large simulation cells and long trajectories to capture defect motion, phase transformations and transport. Direct electronic-structure calculations are often prohibitively costly at these scales. 
Machine-learned interatomic potentials (MLIPs) reduce this cost by learning atomic interactions from first-principles data~\cite{chen2020tensoralloy,zhang2018deeppotse}. 
For these applications, a useful potential must combine accurate predictions across diverse compositions with fast force evaluation.

Early MLIPs established accurate descriptions of individual chemical systems~\cite{behler2007,chen2018atomic,zhang2018deeppotse,schutt2017schnet,shapeev2016mtp}. 
Larger training collections and more expressive representations have since enabled universal MLIPs to describe many chemical systems with a single model.
Examples include DPA families~\cite{dpa12024,dpa22024,dpa32026}, MACE and Allegro foundation models~\cite{macefoundation2025,kavanagh2026foundation}, and OMat24-trained EquiformerV2 models~\cite{liao2024equiformerv2,omat24}. 
Compact local representations offer a route to fast evaluation across diverse chemistries. 
UNEP-v1 describes elemental metals and their alloys, and NEP89 extends the neuroevolution-potential approach to inorganic and organic materials across 89 elements~\cite{song2024unep,liang2026nep89}. 
DPA4C achieves competitive accuracy at high throughput by combining compressed inference with a compact representation of local correlations~\cite{li2026dpa4c}.

Which additions to these compact representations improve fitting accuracy enough to justify their computational cost, and when do those gains carry over to physical predictions? 
A strictly local model can retain finer detail about neighbour distances and directions~\cite{bartok2013soap,drautz2019ace}, form richer correlations from the collected information through internal processing or readout~\cite{batatia2025design}, or use the surrounding environment to reweight individual neighbours~\cite{dpa12024}. 
The value of each operation depends on what the representation already retains. 
Its cost also depends on where it is applied. 
Updating each neighbour separately can be more expensive than updating their combined features, even when both operations use the same number of trainable parameters.
A connected model family lets us vary these operations under common training and evaluation conditions, measure their cost and test whether gains in force accuracy carry over to physical properties.

Here we introduce DeepTensorMoment (DTM), a family of strictly local equivariant potentials built on TensorAlloy and TensorMD~\cite{chen2020tensoralloy,chen2021machinelearning,ouyang2025tensormd,ouyang2026smallfootprint}. 
DTM represents each atomic neighbourhood by tensor moments with adjustable angular resolution. 
It can refine these moments at the central atom and, at selected stages, use the updated moments to reweight individual neighbours within the same cutoff neighbourhood.
On the 16-element UNEP-v1 dataset (HEA16) and on OMat24, finer angular resolution and a first refinement block improve force accuracy.
Explicit cubic readout reduces force error more in the base models than after the first refinement block.
At fixed parameter count, using updated moments to weight neighbours improves force accuracy at nearly the same throughput as using the initial moments. 
Across 18 models, the mean absolute error (MAE) in forces ranges from 73.0 to 159.6\,meV/\AA{} on the official OMat24 validation set.
DTM shares the high-throughput frontier with DPA4C.
On a single RTX~5090 GPU, throughput ranges from 0.35 to 105.5 million atom-steps/s for a 16-element body-centred cubic alloy.
Within DTM, lower validation force MAE closely tracks improvements in formation energies and bulk moduli. 
Thermal-conductivity errors also tend to decrease as force accuracy improves, although rare phonon instabilities can change the model ranking by mean conductivity error.

\section*{Results}
\subsection*{Three representation choices within DTM}

\begin{figure}[p]\centering\makebox[\linewidth][c]{\includegraphics[width=183mm]{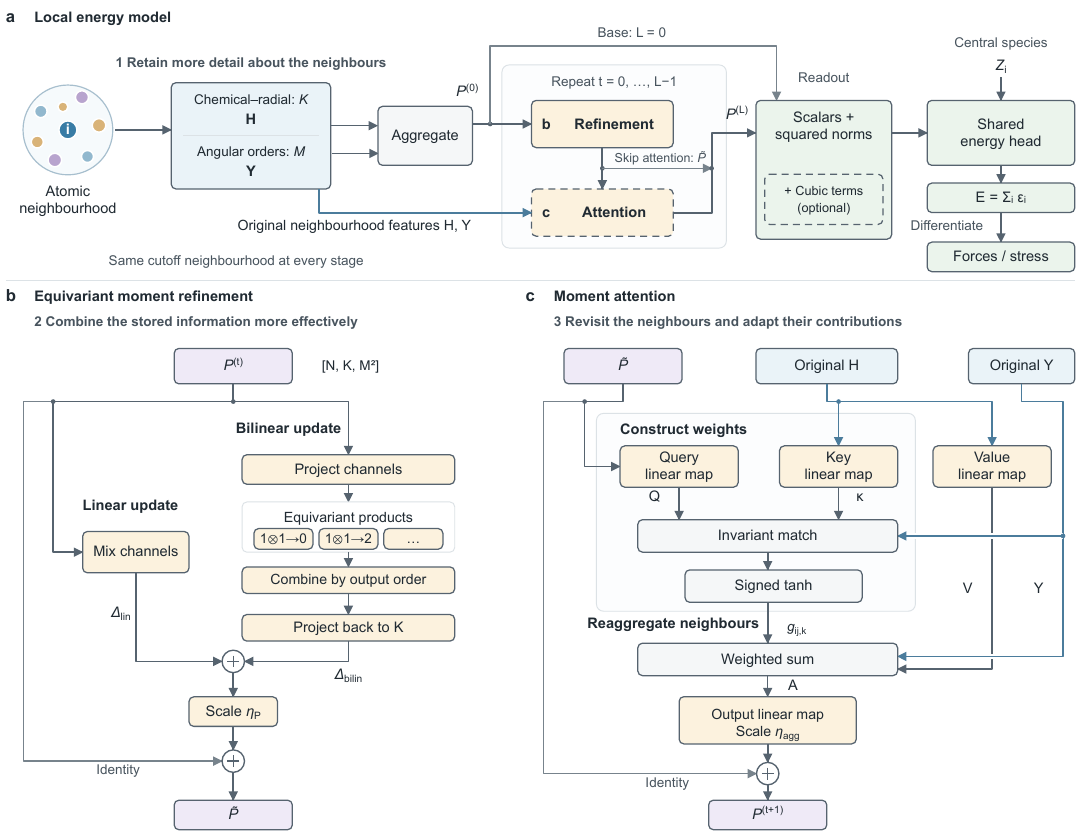}}
\caption{\textbf{Three representation choices within DTM.}
\textbf{a}, Radial width $K$ and angular resolution $M$ set the detail retained in the initial tensor moments. The base model ($L=0$) reads them out directly. Refined models apply $L$ stages before the scalar-and-norm readout, optionally augmented by cubic features. A shared network uses these invariants and the central species to predict atomic energies. Differentiating the total energy gives forces and stress. 
\textbf{b}, A refinement block mixes channels and couples angular orders through tensor products, then adds the scaled updates to the central moments. Their shape is $[N,K,M^2]$ for $N$ atoms and $M=\ell_{\max}+1$ angular orders. 
\textbf{c}, Moment attention forms queries from refined moments and keys and values from the original radial features. The queries, keys and neighbour directions give signed weights for a new sum over the same neighbourhood.}\label{fig:architecture}\end{figure}

DTM sums contributions from neighbours within a fixed cutoff ($r_{\mathrm c}=6\,\text{\AA}$ by default) to form a set of \emph{tensor moments} at each central atom. 
It can refine these moments over successive stages and, at selected stages, use the updated moments to reweight the original neighbours. 
A shared network predicts each atomic energy from invariant features of the final moments and the identity of the central atom (Fig.~\ref{fig:architecture}a).

\emph{First, retain more detail about the neighbours.} 
For a central atom $i$ and neighbour $j$ with atomic numbers $Z_i$ and $Z_j$, a learned chemical--radial weight $h_{ij,k}=h_k(Z_i,Z_j,r_{ij})$ encodes their identities and separation in channel $k=1,\dots,K$.
A real irreducible Cartesian harmonic $\mathbf{Y}_\ell(\widehat{\mathbf{r}}_{ij})\in\mathbb{R}^{2\ell+1}$ describes the direction from $i$ to $j$ at angular order $\ell$. Summing their products over the cutoff neighbourhood $\mathcal{N}_i$ gives the initial tensor moments,
\begin{equation}\label{eq:init_moments}
\mathbf{P}_{ik\ell}^{(0)} = \sum_{j \in \mathcal{N}_i} h_{ij,k} \, \mathbf{Y}_\ell(\widehat{\mathbf{r}}_{ij}).
\end{equation}

The model retains $M=\ell_{\max}+1$ angular orders, from $\ell=0$ to $\ell_{\max}$.
The initial state $\mathbf{P}_i^{(0)}$ contains $KM^2$ components per atom.
Increasing $K$ gives the model more channels to distinguish chemical species and distances. Increasing $M$ retains finer directional patterns.

\emph{Second, combine the stored information more effectively.} 
Once the model has summed the neighbour contributions, it can form richer correlations among the resulting moments. 
At each stage $t=0,\dots,L-1$, an ordinary \emph{equivariant moment-refinement block} (P-block, Fig.~\ref{fig:architecture}b) updates the central moments using linear and bilinear terms,
\begin{equation}\label{eq:pblock}
\widetilde{\mathbf{P}}_i^{(t)} = \mathbf{P}_i^{(t)} + \eta_{\mathrm{P}}^{(t)} \Big[ \mathbf{\Delta}_{\mathrm{lin}}\!\big(\mathbf{P}_i^{(t)}\big) + \mathbf{\Delta}_{\mathrm{bilin}}\!\big(\mathbf{P}_i^{(t)}, \mathbf{P}_i^{(t)}\big) \Big].
\end{equation}

The linear term mixes channels within each angular order, the bilinear term couples angular orders through low-rank equivariant tensor products, and $\eta_{\mathrm P}^{(t)}$ sets the learned scale of the update. 
Allegro also uses equivariant tensor products, applying them to features associated with each centre--neighbour pair~\cite{musaelian2023allegro}, while DTM applies them to the moments already summed at the central atom, so an ordinary block requires no further access to individual neighbours. 
The model can also form correlations at readout. An optional \emph{cubic readout} (C1) contracts triples of final moments to give additional invariant features $\mathbf{G}_i^{\mathrm{cub}}$~\cite{shapeev2016mtp}.

\emph{Third, revisit the neighbours and adapt their contributions.} 
At selected stages $t\in\mathcal{J}$, \emph{moment attention} (MA, Fig.~\ref{fig:architecture}c) uses the refined central moments $\widetilde{\mathbf{P}}_i^{(t)}$ to determine how strongly each neighbour contributes to the next update. 
It forms equivariant queries $\mathbf{Q}_{ik\ell}^{(t)}$ from the refined moments and scalar keys $\kappa_{ij,k\ell}^{(t)}$ and values $v_{ij,k}^{(t)}$ from the original chemical--radial features $\mathbf{h}_{ij}$. 
Matching each query to the neighbour direction gives a signed gate $g_{ij,k}^{(t)}\in(-1,1)$ and a new weighted sum over neighbours,
\begin{equation}\label{eq:ma}
\begin{aligned}
g_{ij,k}^{(t)} &= \tanh\!\left( \sum_{\ell=0}^{M-1} c_\ell \, \kappa_{ij,k\ell}^{(t)} \big\langle \mathbf{Q}_{ik\ell}^{(t)}, \mathbf{Y}_\ell(\widehat{\mathbf{r}}_{ij}) \big\rangle \right), \\
\mathbf{P}_{ik\ell}^{(t+1)} &= \widetilde{\mathbf{P}}_{ik\ell}^{(t)} + \eta_{\mathrm{agg}}^{(t)} \left[\mathcal{W}_{\mathrm{O},\ell}^{(t)}\!\left( \sum_{j \in \mathcal{N}_i} (\mathbf{g}_{ij}^{(t)} \odot \mathbf{v}_{ij}^{(t)}) \otimes \mathbf{Y}_\ell(\widehat{\mathbf{r}}_{ij}) \right)\right]_k.
\end{aligned}
\end{equation}

Here $c_\ell=[(2\ell+1)M]^{-1/2}$, and $\mathcal{W}_{\mathrm O,\ell}^{(t)}$ mixes channels. 
$\mathbf{P}_i^{(t+1)}=\widetilde{\mathbf{P}}_i^{(t)}$ for stages without attention. 
In the Cartesian tensor model HotPP, successive stages pass updated features between neighbouring atoms~\cite{wang2024hotpp}.
DTM updates the query at the central atom and revisits the original neighbour features, preserving the cutoff neighbourhood of each atomic energy.
The signed weights allow neighbour contributions to reinforce or cancel, and summation preserves permutation invariance~\cite{zaheer2017deepsets,lee2019settransformer}. 
At fixed depth, widths and angular resolution, the work per atom scales linearly with neighbour count.

After $L$ stages ($L=0$ for an unrefined base model), the standard readout (C0) passes the scalar moments, squared norms of the higher-order moments and central-element features $\boldsymbol{\chi}(Z_i)$ to the shared energy network $f_{\mathrm E}$.
C1 includes the bracketed cubic features as additional inputs,
\begin{equation}\label{eq:readout}
\begin{aligned}
\mathbf{G}_i &= \Big(P_{ik0}^{(L)},\,
 \big(\|\mathbf{P}_{ik\ell}^{(L)}\|^2\big)_{\ell=1}^{M-1}\Big)_{k=1}^{K}
 \mathbin{\Vert} \big[\mathbf{G}_i^{\mathrm{cub}}\big]_{\mathrm{C1}}
 \mathbin{\Vert} \boldsymbol{\chi}(Z_i), \\
\varepsilon_i &= E_{Z_i}^{(0)} + f_{\mathrm E}(\mathbf G_i),
 \qquad E_{\mathrm{tot}} = \sum_{i=1}^N \varepsilon_i .
\end{aligned}
\end{equation}

Here $\Vert$ between feature blocks denotes concatenation, and $E_{Z_i}^{(0)}$ is a trainable elemental reference energy. 
Analytical derivatives of the total energy give conservative forces and virial stresses.

\subsection*{Accuracy gains and evaluation cost}

Starting from an unrefined base DTM model, which additions improve force accuracy, and what do they cost in simulation speed? 
We use matched comparisons among 25 architectures trained on the HEA16 dataset to distinguish the gains from retaining finer angular detail, refining the central moments and reweighting individual neighbours.
For each comparison, we relate the change in validation force error to the change in molecular-dynamics throughput and check whether the accuracy gains carry over to the HEA16 testing subset and three external alloy datasets (S1--S3).
We report throughput for an equiatomic body-centred cubic alloy of the HEA16 elements (BCC16) on the RTX~5090 because all four material--GPU combinations show similar trends (Supplementary Table~19).

\begin{figure}[p]\centering\makebox[\linewidth][c]{\begin{minipage}{183mm}\centering\includegraphics[width=\linewidth]{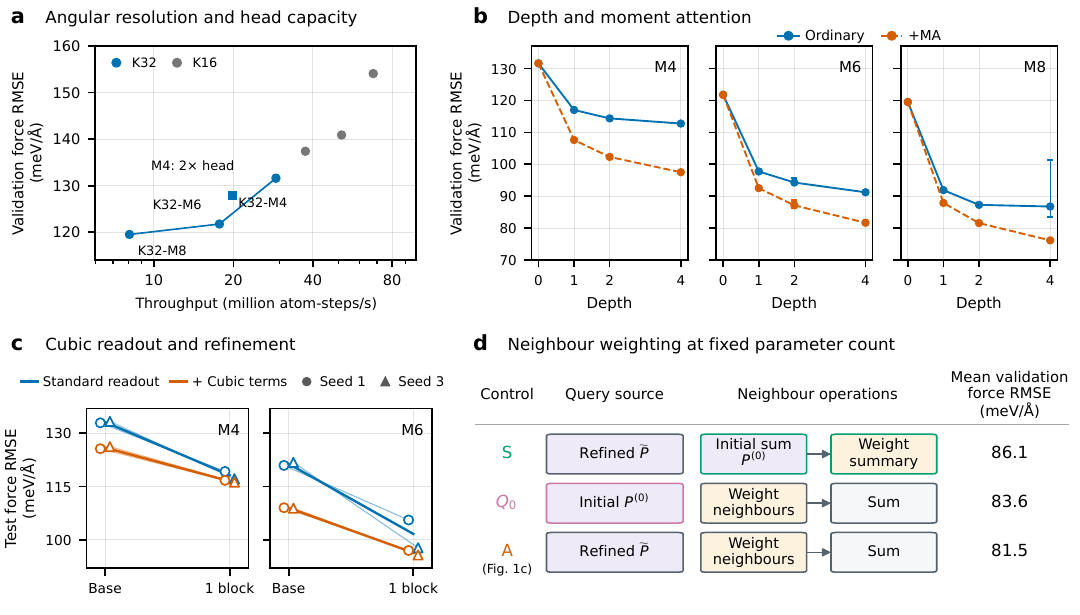}
\captionsetup{font=footnotesize}\caption{\textbf{Accuracy gains from local computation in HEA16-trained models.} 
\textbf{a}, Radial, angular and energy-network-width controls versus RTX~5090 throughput on BCC16.
\textbf{b}, Refinement depth with and without moment attention (MA). 
Panels \textbf{a,b} show six-run means and ranges of validation force RMSE.
At $M=8$, one of the six runs with four ordinary refinement blocks showed a late increase in validation force RMSE, widening the range at depth 4 (Supplementary Fig.~3b,c).
\textbf{c}, Test force RMSE with standard (C0) or cubic (C1) readout before and after one block.
Circles and triangles denote seeds 1 and 3, respectively, and thick lines connect the two-seed means.
\textbf{d}, Neighbour weighting and mean validation force RMSE for S, $Q_0$ and A (419,993 trainable parameters each).
$P^{(0)}$ and $\widetilde P$ denote initial and refined moments, respectively.
Coloured outlines mark changes in query source or pooling relative to standard moment attention (A, Fig.~\ref{fig:architecture}c).
Errors are averaged over seeds 1 and 3 (individual values in Supplementary Table~2).}\label{fig:hea-refinement}
\end{minipage}}\end{figure}

In the base model, retaining finer angular detail improves force accuracy more than widening the energy network (Fig.~\ref{fig:hea-refinement}a). 
Starting from $M=4$, increasing $M$ to 6 lowers mean validation force root-mean-square error (RMSE) by 7.5\%.
Doubling the energy-network width at $M=4$ lowers it by 2.8\%. 
From a common baseline of 29.0 million atom-steps/s, throughput falls to 17.7 million with $M=6$ and 19.8 million with the wider network.
Increasing $M$ from 6 to 8 gives a smaller accuracy gain.
Angular resolution determines which directional patterns the model retains. 
A wider energy network can learn a more flexible function of its inputs, but for a fixed encoder it assigns the same energy to environments with identical inputs~\cite{pozdnyakov2020incompleteness}.

Adding a first ordinary refinement block or explicit cubic readout (C1) to the base model improves force accuracy, with the block giving the larger gain (Fig.~\ref{fig:hea-refinement}b,c).
On the three external alloy datasets combined (S1--S3), the block alone lowers force RMSE by 11.0\% at $M=4$ and 20.0\% at $M=6$.
C1 alone lowers it by 3.95\% and 10.13\%, respectively.
Once the block is present, adding C1 lowers force RMSE by a further 0.83\% at $M=4$ and 1.93\% at $M=6$, while reducing throughput by 37--39\% (Supplementary Tables~12 and~19).
For a non-scalar output of the ungated block, write the refined moment as $\widetilde{\mathbf P}=U+V$, where $U=\mathcal L(\mathbf P)$ includes the incoming moment and its linear update, and $V=\mathcal B(\mathbf P,\mathbf P)$ is the bilinear update.
The standard readout gives $\|\widetilde{\mathbf P}\|^2=\|U\|^2+2\langle U,V\rangle+\|V\|^2$.
The cross term is cubic in the incoming moments, which helps explain why explicit cubic readout adds less after refinement (Supplementary Eq.~(S31)).

At matched angular resolution and refinement depth, adding attention lowers validation force RMSE in all 54 paired comparisons (Fig.~\ref{fig:hea-refinement}b).
It also lowers force MAE for every matched pair on both the testing subset and the external alloy structures. 
The only two exceptions to the force-RMSE improvement occur on the testing subset and trace to the same highly compressed Pt--V structure (Supplementary Fig.~2 and Supplementary Section~2.4).
After excluding HEA16 test structures with unusually short interatomic distances, we fitted one scale factor per structure whose reference forces met the threshold in Methods.
This factor accounts for an overall tendency to predict forces that are too large or too small.
Across 18 matched comparisons, 89--96\% of the decrease in squared force error from moment attention remained after accounting for this scale.
The improvement therefore came mainly from better predictions of force directions and of the relative magnitudes of forces on different atoms (Supplementary Section~2.6).
The gain extends to the four-block models, showing that reweighting individual neighbours remains useful even after several stages of processing the pooled moments.

Starting from one ordinary refinement block, the model can refine the central moments a second time or revisit individual neighbours through moment attention.
At $M=8$ on HEA16, adding a second block (M8-P2) gives slightly lower mean validation force RMSE and higher throughput than adding attention (DTM-HEA16-L, M8-P1+MA), with comparable force errors on the external alloy structures (Supplementary Fig.~1a and Supplementary Table~1).
Further processing at the central atom is therefore the more economical next step for DTM.
Repeated neighbour weighting reaches lower errors at greater depth: the four-block attention model gives the lowest force RMSE on both the testing subset and the combined external alloy datasets.

On OMat24, finer angular resolution again improves force accuracy in the base models, and refinement reduces the additional gain from cubic readout.
At $M=4$, with three hidden layers of width 128 in the energy network, cubic readout lowers validation force MAE by 3.8\% in the base model and by 0.44\% after one ordinary block (Supplementary Tables~16 and~17).
Cubic readout remains useful at several operating points: four of the five cubic-readout models lie on the measured force-MAE--throughput frontier for BCC16 on the RTX~5090 (Fig.~\ref{fig:frontiers}b and Supplementary Tables~16 and~20).
Among the OMat24-trained DTM models, the four-block attention model gives the lowest official validation force MAE.

\subsection*{Force gains at fixed parameter count}

Adding moment attention also increases model size, so does the way it weights neighbours improve force accuracy when the number of trainable parameters is held fixed?
We compare three HEA16 models at $K=32$ and $M=6$, each with four refinement blocks and 419,993 trainable parameters (Fig.~\ref{fig:hea-refinement}d).
Standard moment attention ($\mathrm{A}$) uses refined central moments to weight individual neighbours before summation.
The pooled-summary control ($\mathrm{S}$) uses refined moments to weight a summary of the neighbour contributions, while the initial-moment control ($Q_0$) uses the initial moments to weight individual neighbours.

Both $\mathrm{S}$ and $\mathrm{A}$ use the refined central moments to set their weights, but they apply those weights at different stages of the neighbour sum.
Because $\mathrm{S}$ operates on the pooled summary, it can scale the contributions to a given moment channel together and complete the update at the central atom.
By revisiting the neighbours before summation, $\mathrm{A}$ can assign different weights to two neighbours contributing to the same channel according to their features and directions (Fig.~\ref{fig:architecture}c).
On HEA16, $\mathrm{A}$ lowers validation force RMSE by 5.37\% relative to $\mathrm{S}$ (Fig.~\ref{fig:hea-refinement}d), while $\mathrm{S}$ runs 1.94--2.50 times as fast (Supplementary Table~19).

Both $\mathrm{A}$ and $Q_0$ revisit individual neighbours, so their comparison tests what the model gains by updating the moments that set the weights.
In $\mathrm{A}$, angular correlations formed during refinement can change how strongly each neighbour contributes to the next sum, whereas the query in $Q_0$ has access only to the initial moments.
$\mathrm{A}$ lowers validation force RMSE by 2.53\% relative to $Q_0$, and their measured throughputs differ by at most 4.2\% (Fig.~\ref{fig:hea-refinement}d and Supplementary Table~19).
$\mathrm{A}$ also gives the lowest force MAE and RMSE of the three models on the HEA16 testing subset and on the three external alloy datasets combined (Supplementary Tables~3 and~4).

The same ordering holds on the official OMat24 validation set, where force MAE decreases from 82.9\,meV/\AA{} for $\mathrm{S}$ to 80.1 for $Q_0$ and 78.1 for $\mathrm{A}$ (Supplementary Table~18).
These controls show why parameter count alone cannot guide model design.
Models with the same parameter count can differ substantially in throughput, and even at similar throughput the choice of query affects force accuracy.
When throughput is the priority, these results favour further processing of the moments already stored at the central atom.
When the model weights neighbours individually, using refined central moments to set the weights improves force accuracy at similar measured throughput.

\subsection*{Competitive accuracy and speed}

\begin{figure}[p]\centering\includegraphics[width=\linewidth,height=0.86\textheight,keepaspectratio]{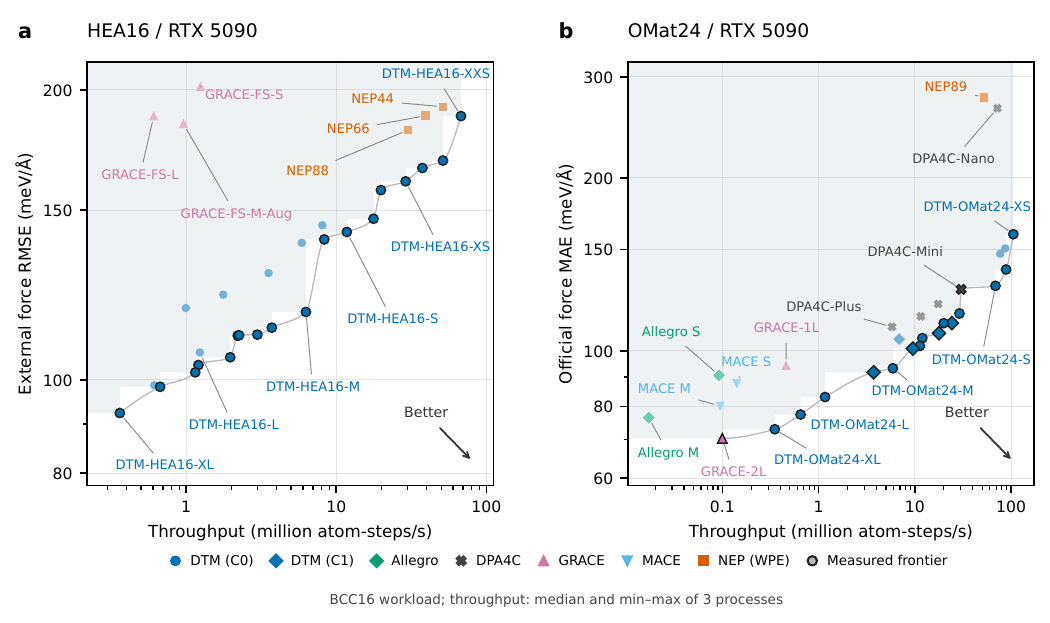}
\caption{\textbf{Accuracy and throughput across model families.} 
\textbf{a}, Force RMSE across external alloy sets S1--S3 for 25 HEA16-trained DTM models and external models.
Supplementary Table~1 gives force errors for all 25 DTM models and identifies the labelled architectures.
Supplementary Table~10 compares six of these models with the external models.
\textbf{b}, Force MAE over all Cartesian components on the official OMat24 validation set (1,025,361 structures) for 30 models with matched throughput measurements, including 5 with cubic readout.
DTM circles and diamonds denote standard and cubic readouts.
RTX~5090/BCC16 throughputs show the median and range of three independent processes.
Dark-edged markers identify nondominated measurements, grey curves guide the eye, and shading marks the dominated region.
NEP timings use GPUMD-WPE~\cite{bu2026wpe}.
Supplementary Table~16 gives the OMat24 force errors and Supplementary Tables~19 and~20 give the HEA16 and OMat24 throughputs, respectively.}
\label{fig:frontiers}\end{figure}

Are these accuracy--throughput choices competitive against other model families? 
On HEA16, the NEP44/66/88 series illustrates the strategy of enriching the initial representation~\cite{wu2026torchnep}. 
These models contract angular moments directly into rotationally invariant many-body descriptors and increase the number of radial and angular descriptor channels across the series. 
On the external multicomponent alloy sets S1--S3~\cite{song2024unep}, the unrefined DTM-HEA16-XS gives 11.5\% lower force RMSE than NEP88 (160.7 versus 181.7\,meV/\AA{}) at nearly equal RTX~5090/BCC16 throughput (29.0 versus 30.1 million atom-steps/s, Fig.~\ref{fig:frontiers}a).
With one refinement block and moment attention, DTM-HEA16-L also gives lower formation-energy MAE and RMSE than every evaluated NEP and GRACE-FS~\cite{shuang2026comparison} model on each of the five external datasets (Supplementary Table~9).

On the official OMat24 validation set (1,025,361 structures), DTM and DPA4C share the high-throughput portion of the measured force-MAE frontier (Fig.~\ref{fig:frontiers}b).
The five evaluated DPA4C models aggregate neighbours once and form higher-order correlations through invariant contractions at readout~\cite{li2026dpa4c}.
DTM-OMat24-M applies one learned equivariant refinement block before its standard readout.
On the same BCC16 cell, DTM-OMat24-M and DPA4C-Plus deliver 5.91 and 5.79 million atom-steps/s, respectively.
At this speed, the one-block DTM-OMat24-M gives 15.3\% lower force MAE (93.2 versus 110.0\,meV/\AA{}).
At higher throughput, the unrefined DTM-OMat24-S runs 2.28 times as fast as DPA4C-Mini with a 1.4\% larger force MAE.
DTM-OMat24-XS is faster than NEP89~\cite{liang2026nep89} with GPUMD-WPE~\cite{gpumd2025,bu2026wpe} on BCC16, while NEP89 is faster on elemental Cu (Supplementary Tables~16 and~20).

At each moment-attention stage, correlations learned at the central atom guide the next neighbour sum.
At $M=8$, the one-stage model gives an official force MAE of 83.1\,meV/\AA{} and the four-stage model gives 73.0\,meV/\AA{}.
The one-stage model runs 3.3 times as fast.
On BCC16, the four-stage DTM-OMat24-XL delivers 0.35 million atom-steps/s, compared with 0.10 for GRACE-2L~\cite{grace2026} at 70.4\,meV/\AA{}.
We take the base DTM-OMat24-S, the one-block DTM-OMat24-M and the two- and four-stage attention models DTM-OMat24-L and DTM-OMat24-XL into the physical benchmarks to test which predictions benefit from the additional computation.

\subsection*{Physical predictions guide model choice}

\begin{figure}[p]\centering\includegraphics[width=\linewidth,height=0.86\textheight,keepaspectratio]{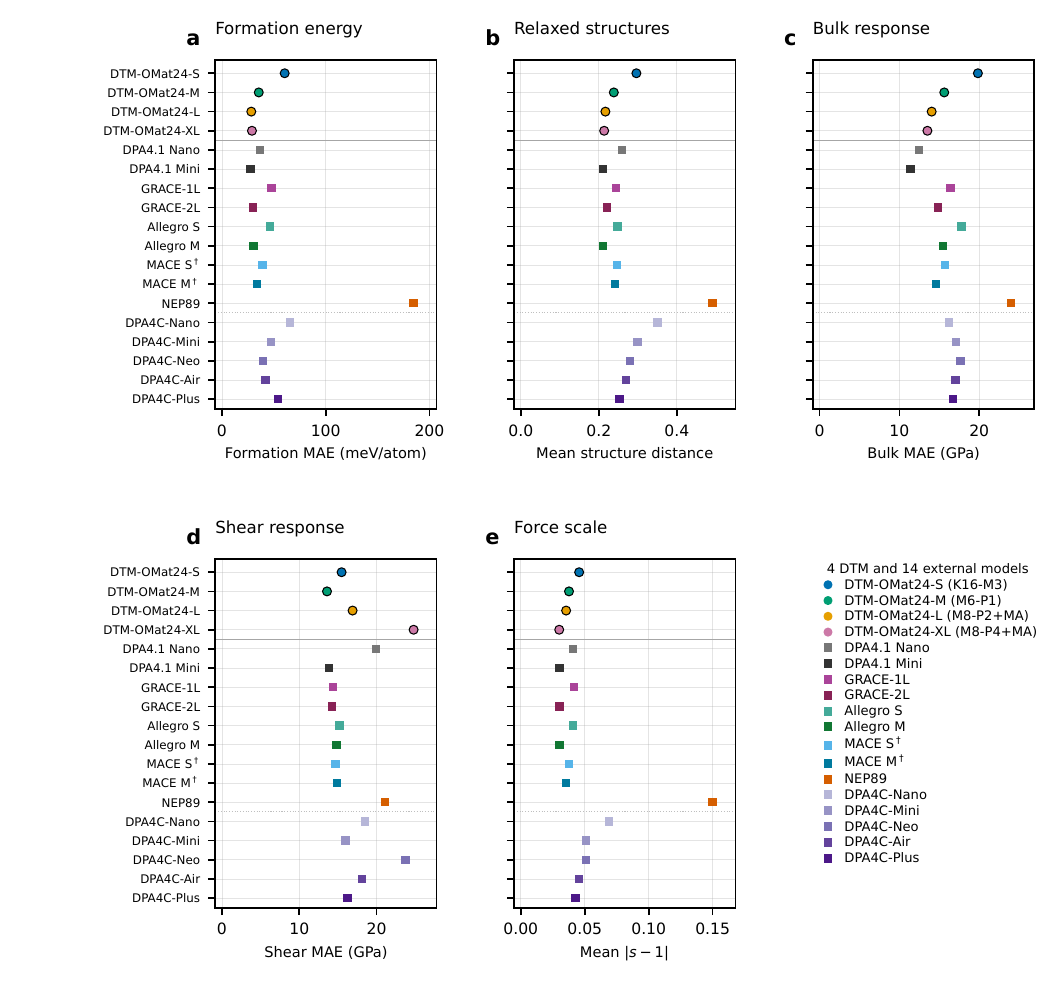}
\caption{\textbf{Energetic and mechanical predictions across models.} 
\textbf{a}, Formation-energy MAE. 
\textbf{b}, Mean structural distance. 
\textbf{c,d}, Bulk- and shear-modulus MAEs at the original strain amplitudes. 
\textbf{e}, Mean $|s-1|$ for the least-squares slope $s$ of predicted versus reference force components, fitted through the origin for each material. DTM circles and external-model squares show means over successful calculations, including finite outliers. Each model attempts 972 equilibrium, 3,953 elasticity and 979 force-scale calculations. For the MACE Small and Medium formation energies (daggers), the N and F references are single-point MACE energies at fixed DFT-selected geometries. Supplementary Figs.~7 and~8 give coverage and strain sensitivity.}\label{fig:properties}\end{figure}

\begin{figure}[p]\centering\includegraphics[width=\linewidth,height=0.86\textheight,keepaspectratio]{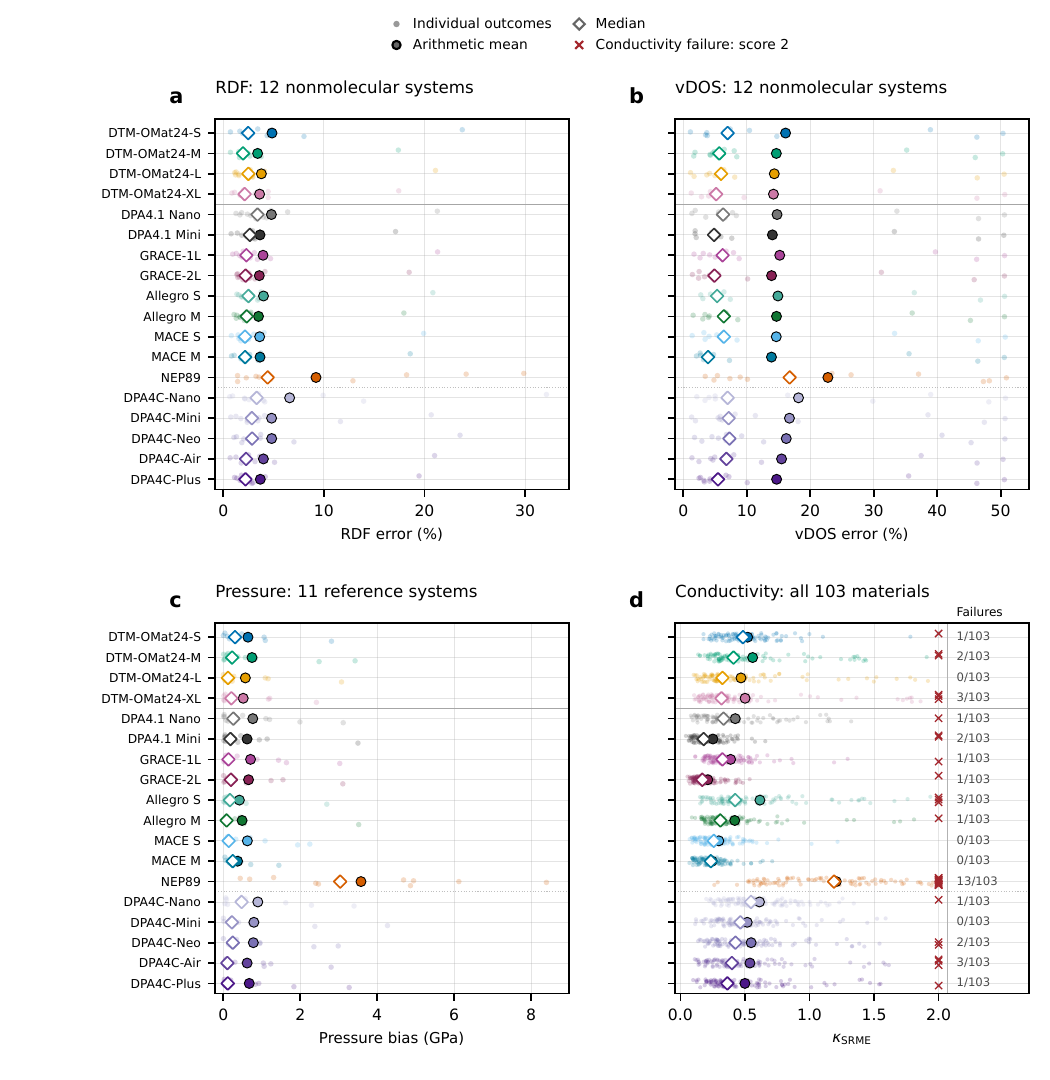}
\caption{\textbf{Finite-temperature predictions and thermal conductivity.} 
\textbf{a,b}, Radial distribution function and vibrational density-of-states errors for 12 nonmolecular systems.
\textbf{c}, Absolute deviations in trajectory-mean pressure for 11 systems with references.
\textbf{d}, Errors in phonon-mode contributions to thermal conductivity ($\kappa_{\rm SRME}$) for 103 materials.
Small points show equally weighted individual outcomes, filled circles the means and open diamonds the medians.
Crosses mark failed conductivity calculations, scored as 2.
Supplementary Fig.~9 includes the five molecular crystals and angular distribution function errors.}\label{fig:dynamics}\end{figure}

\begin{figure}[p]\centering\includegraphics[width=\linewidth,height=0.75\textheight,keepaspectratio]{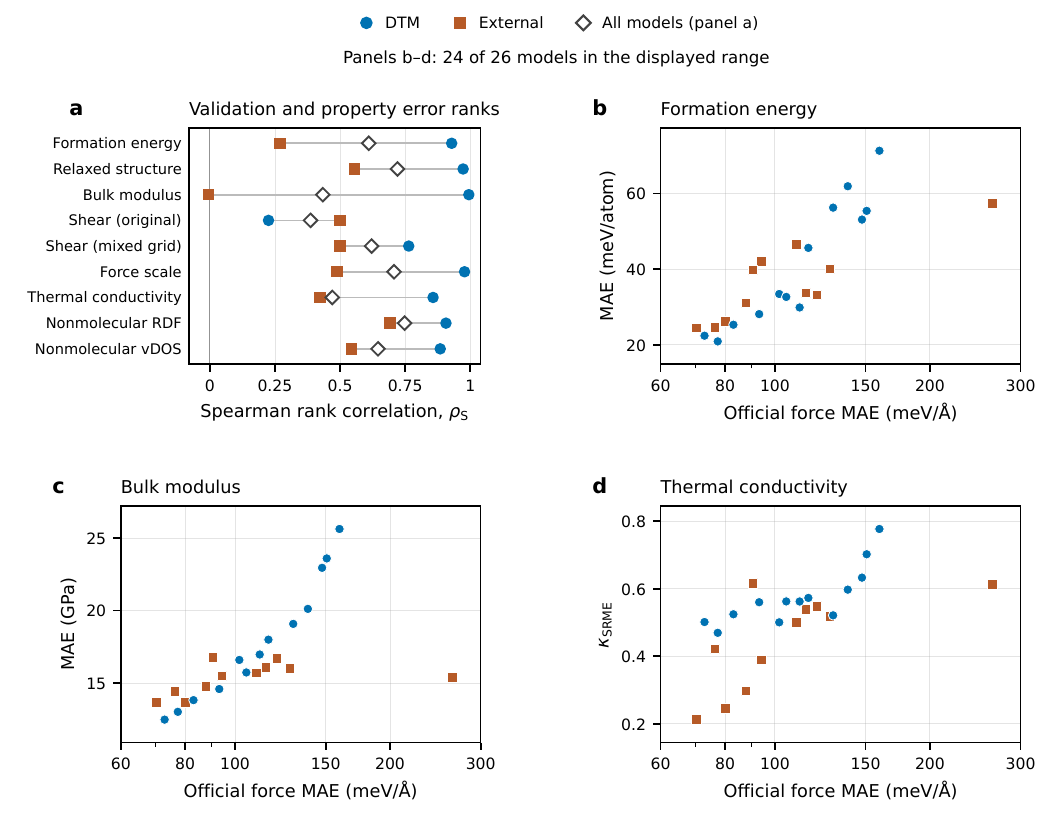}
\caption{\textbf{When validation force errors predict physical performance.} 
\textbf{a}, Spearman correlations of official OMat24 force MAE with property errors for 13 DTM models, 13 external models and all 26 together. For DTM shear-modulus errors, $\rho_{\rm S}=0.225$ at the original strain amplitudes (3,602 common materials) and 0.764 with halved amplitudes for DTM-OMat24-L and DTM-OMat24-XL only (3,603). 
\textbf{b--d}, Formation-energy MAE (698 common materials), bulk-modulus MAE (3,602, original strains) and mean conductivity error ($\kappa_{\rm SRME}$, 103 materials, failures scored as 2) versus official force MAE.
DPA4.1 Nano and Mini, trained on curated OMat24 data, lie above the displayed range (60--300\,meV/\AA{}) and contribute to the external and all-model correlations in \textbf{a}.
For the 11 external models shown in \textbf{b--d}, the formation-energy and bulk-modulus correlations are 0.791 and 0.636 on the same material sets (Supplementary Section~3.7).}
\label{fig:validation-properties}\end{figure}

Within DTM, lower validation force MAE generally identifies more accurate models for formation energies and bulk moduli. 
Figures~\ref{fig:properties} and~\ref{fig:dynamics} compare 4 DTM representatives with 14 external models on static and finite-temperature benchmarks~\cite{matcalc,pota2026,dynmat2026}.
Across all 13 standard-readout DTM models, the Spearman rank correlations between official OMat24 force MAE and errors in formation energy and bulk modulus are 0.929 and 0.995, respectively, on common material sets (Fig.~\ref{fig:validation-properties}a--c).
The relationship is more sensitive to model choice across families. 
Across the 13 external checkpoints, the corresponding correlations are 0.269 and $-0.005$.
For the 11 checkpoints within the force-MAE range shown in Fig.~\ref{fig:validation-properties}b--d, they are 0.791 and 0.636 on the same materials (Supplementary Section~3.7).

The four-stage attention model (DTM-OMat24-XL) improves on the two-stage model (DTM-OMat24-L) in bulk-modulus and force-scaling accuracy, with bulk-modulus MAE falling from 14.0 to 13.5\,GPa (Fig.~\ref{fig:properties}c,e).
On the nonmolecular trajectories, XL also predicts the vibrational density of states more accurately and gives smaller pressure deviations (Fig.~\ref{fig:dynamics}b,c).
L and XL both roughly halve the formation-energy MAE of DTM-OMat24-S and reach accuracy comparable to DPA4.1 Mini~\cite{li2026dpa4}, GRACE-2L and Allegro Medium (Fig.~\ref{fig:properties}a).
L gives a slightly lower formation-energy MAE than XL (28.3 versus 28.9\,meV/atom) and runs 1.85 times as fast on RTX~5090/BCC16.
Among the four representatives, the one-block DTM-OMat24-M gives the lowest radial distribution function error on the nonmolecular trajectories and the lowest mean shear-modulus error under the original strain protocol.
TiPt alone contributes 11.0\,GPa to XL's mean shear-modulus error of 24.8\,GPa (Supplementary Section~3.4).
Figure~\ref{fig:properties} reports means over successful calculations for each model, with completion counts and strain sensitivity in Supplementary Tables~22 and~23.

Thermal conductivity probes harmonic phonon stability and anharmonic coupling~\cite{pota2026,fu2025esen}. 
Across the 13 DTM models, lower validation force MAE generally accompanies lower mean conductivity error ($\rho_{\mathrm S}=0.857$, Fig.~\ref{fig:validation-properties}d).
Across all 103 materials, the four-stage model gives a lower median conductivity error score ($\kappa_{\rm SRME}$) than the two-stage model (0.319 versus 0.326).
It also gives a lower mean on the 100 materials where both models complete the calculation (0.456 versus 0.475). 
In AgBr, AgCl and CuCl, the four-stage model predicts imaginary harmonic modes.
Each failure receives the prescribed score of 2, raising its full-benchmark mean slightly above that of the two-stage model (0.501 versus 0.469).
DPA4.1 Mini, GRACE-2L and MACE Medium give lower mean scores on the same 103 materials (Fig.~\ref{fig:dynamics}d and Supplementary Section~3.6).

\section*{Discussion}

The value of additional local computation depends on which correlations the current representation already exposes.
On HEA16, increasing the base model's angular resolution from $M=4$ to $M=6$ reduced validation force RMSE more than doubling the energy-network width (7.5\% versus 2.8\%).
Increasing $M$ to 8 gave a smaller further gain.
At fixed angular resolution, a first refinement block improved force accuracy on both HEA16 and OMat24.
Even at $M=8$ on HEA16, a second ordinary block gave slightly lower validation force RMSE and higher measured throughput than adding attention to the one-block model.

Refinement changes how much the energy network can learn from a simple readout.
Consider a vector moment that points along one of the principal axes of a second-order moment.
Turning the vector towards another axis leaves the magnitude of each moment unchanged.
An explicit cubic feature can detect this change in alignment.
A refinement block can also make the magnitude of the updated second-order moment depend on the alignment, so that the energy network can detect it through the standard readout.
This overlap helps explain why adding explicit cubic features improved force accuracy more in the base models than after one refinement block.

The refined moments describe the arrangement of neighbours around an atom.
DTM can use this description to weight their contributions in two ways.
Applying a weight after summation scales all contributions to a given moment channel together.
Before summation, the model can give two neighbours different weights when adding their contributions to the same channel.
For both the HEA16- and OMat24-trained controls, weighting the pooled summary ran roughly twice as fast as weighting neighbours individually at equal parameter count.
Individual neighbour weighting gave lower validation force errors in both studies.
A second comparison kept the individual weighting and changed which moments were used to set the weights.
Using the refined moments gave more accurate forces than using the initial moments, with little change in throughput.

DTM shows that a compact local representation can support both fast force evaluation and successive gains in accuracy. 
The same tensor moments that describe the neighbourhood in a base model can be refined to expose additional correlations and then used to reweight individual neighbours. 
Varying where the model spends this computation gives competitive operating points across a broad range of simulation speeds. 
Within DTM, lower force MAE closely tracks improvements in formation energies and bulk moduli.
The four-stage model also improves median thermal-conductivity accuracy, although a few unstable phonon modes raise its mean error because stability depends on how forces change under small displacements around a relaxed crystal.
For materials screening and large-scale dynamics, the useful frontier is therefore the accuracy of the target observable at the throughput the calculation can sustain. 
A connected family of potentials makes it possible to trace how each additional local operation moves that frontier, both for typical predictions and for the materials that dominate the error tail.


\section*{Methods}
\subsection*{Model construction}
Equations~\eqref{eq:init_moments}--\eqref{eq:readout} define the DTM model for radial width $K$, angular-order count $M$, refinement depth $L$ and zero-based attention schedule $\mathcal{J}$. 
Network sizes denote hidden-layer depth $\times$ width.
All DTM models use a $6\,\text{\AA}$ cosine cutoff, fixed 18-component orbital-occupation fingerprints and SiLU activations.
The chemical--radial encoder combines ordered element-pair features with Gaussian radial basis functions. 
Tabulating this network reduces the cost of evaluating neighbour features at deployment~\cite{lu2022dpcompress}. 
In equation~\eqref{eq:init_moments}, $j \in \mathcal{N}_i$ indexes all directed periodic-image edges within the cutoff of atom $i$.

At each attention stage, the query and key projections depend on angular order.
The stage uses a single value projection across all orders.
All three projections are bias-free.
In equation~\eqref{eq:ma}, $\mathbf{g}_{ij}^{(t)}$ and $\mathbf{v}_{ij}^{(t)}$ stack the $K$ channel weights and values for an edge, $\odot$ multiplies them componentwise and $\otimes$ forms the channel--angular outer product. 
The gates are unnormalized across neighbours. Each stage has independent parameters, including a learned residual scale $\eta_{\mathrm{agg}}^{(t)}$ for reaggregation at attention stages. 

Supplementary Information gives the full operators, coupling catalogues, readout definitions and parameter counts.

\subsection*{Datasets, training and checkpoints}
We train the DTM models with TensorAlloy 2.11.2~\cite{chen2020tensoralloy} using mini-batch AdamW.
For a mini-batch of $B$ structures at update $t$, the structure-averaged loss is:
\begin{equation}
\label{eq:total_loss}
\mathcal{L}_{\mathrm{data}}(t) = \frac{1}{B} \sum_{s=1}^B \left[ \lambda_E(t) \ell_{E,s} + \lambda_F(t) \ell_{F,s} + \lambda_\sigma(t) \ell_{\sigma,s} \right].
\end{equation}

The three residuals use the total energy per atom, the $3N_s$ Cartesian force components and the six Voigt stress components of structure $s$.
Each structure therefore contributes one force RMSE to the batch average.
AdamW uses decoupled weight decay of $10^{-3}$. 
Supplementary Eqs.~(S45) and~(S46) specify the stabilized residuals, units and prefactor schedules.

We use seed 42 to split HEA16 into training (90\%), validation (5\%) and test (5\%) sets.
Each HEA16 training run uses 500,000 updates with a batch size of 64, gradient norm clipping at 40 and a cosine learning-rate schedule from $5\times10^{-4}$ to $10^{-6}$.
All reported HEA16 results use the final checkpoints. 
Supplementary Sections~2.1--2.2 give the split counts and training details.

We trained the OMat24 models on the official training set (100,824,585 structures) and evaluated them on the official validation set (1,025,361 structures)~\cite{omat24}.
For all 18 OMat24-trained DTM models, we used a cosine learning-rate schedule from $3\times10^{-4}$ to $10^{-6}$ over 4 million updates.
The training batch size was 256 structures per update.
We selected the final checkpoints for DTM-OMat24-S (K16-M3), -M (M6-P1) and -L (M8-P2+MA), and the stable checkpoint at 3.85 million updates for -XL (M8-P4+MA).
For L and XL, we rescaled an output projection after an update if its quadratic coefficient proxy exceeded the prescribed bound.
Supplementary Sections~3.1 and~3.2 describe the training and validation procedures.

\subsection*{Controlled architecture studies}
The HEA16 survey varies angular resolution, refinement depth, attention, radial width and energy-network width across 25 architectures.
Validation summaries use six terminal runs per architecture (seeds 1--6) and include variation in radial-computation precision.
Test and external evaluations use seeds 1 and 3.

Within each dataset, S, $Q_0$ and A use the same encoder architecture, cutoff, standard readout and four ordinary refinement blocks (Fig.~\ref{fig:hea-refinement}d).
Each block is followed by a residual update to the current moments.
Supplementary Section~1.10 defines the three updates.
All three use $K=32$ and $M=6$.
The HEA16 controls use two seeds (1 and 3), a $3\times256$ radial encoder, a $3\times128$ energy network and 419,993 parameters for each of S, $Q_0$ and A.
At both seeds we also tested models that omit attention or apply it only after the final block.
The three OMat24 controls (S, $Q_0$ and A) each use a $3\times512$ radial encoder and a $3\times256$ energy network and have 975,001 trainable parameters.
All use seed 42.

The HEA16 cubic-readout study pairs C0 and C1 at $K=32$ and either $M=4$ or $M=6$, with or without one ordinary block.
All pairs use a $3\times128$ energy network with attention disabled.
Each pair shares its split, seed (1 or 3), training recipe, precision and terminal-checkpoint rule. 
C1 increases the readout dimension and parameter count. 
The five OMat24 C1 models use seed 42 and terminal checkpoints at 4 million updates.
Of these, four have matched C0 counterparts.
The unrefined K32-M6 model with a $3\times256$ energy network contributes an absolute validation score and a timing measurement.

\subsection*{Accuracy metrics and statistics}
Post-training energy errors weight each structure's per-atom energy error equally.
Force errors pool all Cartesian components across structures.
Stress errors use the six independent ASE Voigt components $(xx,yy,zz,yz,xz,xy)$ in GPa.
We report MAE for both datasets and RMSE for HEA16.

The HEA16 testing subset contains 339,173 atoms. The external alloy datasets are PdCuNi (S1), MoTaVW (S2) and multicomponent alloys (S3)~\cite{song2024unep}. Combined external force scores pool 307,542 Cartesian components from 2,250 structures (102,514 atoms) per checkpoint. Formation-energy comparisons on S1--S3 and the published Materials Project ternary and GNoME alloy sets use model-relaxed elemental references.

All primary OMat24 scores use the same reference labels on the 1,025,361 structures in the final official validation split.
Released models use their configured numerical precision.
Official DTM scores use the trainer's FP32 base operations and BF16 radial computation.
DTM evaluations on the reconstructed DPA-style subset (1,022,745 structures) use full FP32 operations (Supplementary Table~16).

For each structure, we flatten the reference and predicted forces into $\mathbf f,\widehat{\mathbf f}\in\mathbb R^{3N}$ and fit a single least-squares scale $s$,
\begin{equation}\label{eq:methods-force-decomposition}
s = \frac{\mathbf{f}^{\mathsf{T}}\widehat{\mathbf{f}}}{\mathbf{f}^{\mathsf{T}}\mathbf{f}}, \qquad \mathbf{r} = \widehat{\mathbf{f}} - s\mathbf{f}.
\end{equation}

Because $\mathbf f^{\mathsf T}\mathbf r=0$, the squared force error equals $(s-1)^2\|\mathbf f\|^2+\|\mathbf r\|^2$. 
The residual captures errors in force direction and in the relative magnitudes of atomic forces. 
Supplementary Section~2.6 reports its share of the reduction in squared error across matched pairs. 
All structures contribute to the primary accuracy scores. 
For this sensitivity analysis, we examine all neighbour pairs within 6\,\AA{} and exclude a structure if any pair is separated by less than 75\% of the sum of the atoms' covalent radii.
For the force decomposition, we fit the scale for a structure only if the root mean square of its reference force components exceeds $10^{-8}\,\text{eV/\AA}$.
Environment bins use only reference geometries and forces. 
Paired 95\% intervals are percentile bounds from 2,000 whole-structure bootstrap replicates, stratified by S1/S2/S3 on the external sets. 
These intervals show how the estimated gains vary when we resample the structures with the trained models held fixed.
Bins with fewer than 20 contributing structures report counts and point estimates only.

\subsection*{Throughput measurements}
We measured molecular dynamics throughput on the Paratera clusters (\url{https://cloud.paratera.com}) using a single GPU per run, either an NVIDIA A100 PCIe (40\,GB, N32H) or a GeForce RTX~5090 (N56R5), with one MPI rank and one OpenMP thread.
We used TensorMD 2.8.3~\cite{ouyang2025tensormd} for the DTM throughput measurements.
The full periodic cells contain 2,048,000 FCC Cu atoms ($a=3.61\,\text{\AA}$) or 1,024,000 atoms of BCC16 ($a=3.20\,\text{\AA}$), an equiatomic BCC alloy of the 16 HEA16 elements.
For memory-intensive models, we used the same reduced atom count on both GPUs. 
Each of three independent processes ran 100 warmup steps followed by 300 timed NVT steps with a 1\,fs timestep and a four-thermostat Nos\'{e}--Hoover chain at 300\,K with 100\,fs damping.

Throughput is the actual atom count multiplied by 300 and divided by elapsed MD-loop wall time. 
This includes force evaluation, integration and neighbour-list checks. 
Model initialization and warmup precede the timed interval. 
Figure~\ref{fig:frontiers} reports the median and observed range across the three processes.
Supplementary Tables~19--20 give the corresponding rates and actual atom counts.
Supplementary Section~3.3 gives the inference-engine settings and describes the separate FCC Cu comparison with the NEP implementation of Song et al.~\cite{song2026sharednep} (Supplementary Table~21).

\subsection*{Physical and dynamical benchmarks}
The MatCalc~0.5.1 workflows~\cite{matcalc} cover 972 equilibrium, 3,953 elasticity, 1,170 phonon and 979 force-scale materials. 
Equilibrium relaxations start from unseeded 0.1\,\AA{} atomic perturbations and allow up to 500 steps to a force threshold of 0.05\,eV/\AA{}. 
Elastic parent relaxations allow up to 1,000 steps to 0.005\,eV/\AA{}. 
The primary elastic calculations apply normal strains of $\pm(0.005,0.01)$ and shear strains of $\pm(0.03,0.06)$ with fixed internal coordinates. 
Supplementary Section~3.4 gives the half-strain sensitivity analysis. 
Harmonic phonons use 0.015\,\AA{} displacements in supercells spanning at least 20\,\AA{}. Constant-volume heat capacity is evaluated at 300\,K.

Relaxed structural error is the Euclidean distance between structure-level CrystalNN site-fingerprint statistics~\cite{zimmermann2020fingerprints}. 
For force-scale deviation, we fit a single scale $s$ from Eq.~\eqref{eq:methods-force-decomposition} across all frames of each material and average $|s-1|$ over materials.
Static-property means include every finite outcome accepted by each workflow, with success counts reported. 
The 26-model Spearman analysis uses the fixed intersection of jointly successful materials for each static property. 

For anharmonic thermal conductivity~\cite{pota2026}, all 103 materials undergo two-stage self-relaxation and evaluation at 300\,K with 0.03\,\AA{} displacements and no non-analytical correction.
Supplementary Section~3.6 defines the mode-resolved score and its treatment of failures. 
Finite-temperature trajectories follow the protocol adapted from Dyna-Mat~\cite{dynmat2026}. 
At the reference temperature, each trajectory runs for 80,000 NVT steps of 0.25\,fs (20\,ps), with 25\,fs Nos\'{e}--Hoover-chain damping, seed 0 and sampling every 10 steps.
We weight systems equally when averaging errors.
Radial distribution function (RDF) errors use normalized $L_1$ differences. 
Angular distribution function (ADF) and vibrational density of states (vDOS) errors use normalized one-dimensional Wasserstein distances. 
Primary summaries cover 12 nonmolecular systems, 11 of which have reference pressures.
Supplementary Table~24 reports the five molecular crystals separately.
Supplementary Fig.~9 gives the per-system results for all 17 systems.
The molecular-crystal reference trajectories include D3 dispersion corrections. 
The evaluated models use their energies and forces without adding a D3 correction.

\par\medskip
\noindent\textbf{Acknowledgements.}
This work was supported by Advanced Materials-National Science and Technology Major Project (2025ZD0618402) and Yunnan Provincial Key Research and Development Program (202403AA080016).

\noindent\textbf{Author contributions.}
\textbf{Xin Chen} designed the model, implemented the software, and performed model training and analysis. 
\textbf{Lifang Wang} was a principal user of the software and provided valuable feedback. 
\textbf{Bangkai Zhu} conducted tests of DTM models and the software. 
\textbf{Xingyu Gao}, \textbf{Haifeng Liu} and \textbf{Haifeng Song} provided project guidance and theoretical analysis. 
\textbf{Yucheng Ouyang} and \textbf{Ying Liu} contributed to the development of the inference engine TensorMD. 
All authors contributed to writing and revising the manuscript.

\noindent\textbf{Competing interests.}
The authors declare no competing interests.

\clearpage
\clearpage
\bibliographystyle{ncs-unsrtnat}
\addtolength{\bibsep}{-2pt}
\bibliography{references}

@article{mattergen2025,
 author={Zeni, Claudio and Pinsler, Robert and Z{\"u}gner, Daniel and others}, title={A generative model for inorganic materials design}, journal={Nature}, volume={639}, pages={624--632}, year={2025}, doi={10.1038/s41586-025-08628-5}}

@article{chen2018atomic,
 author={Chen, Xin and J{\o}rgensen, Mathias S. and Li, Jun and Hammer, Bj{\o}rk}, title={Atomic Energies from a Convolutional Neural Network}, journal={Journal of Chemical Theory and Computation}, volume={14}, number={7}, pages={3933--3942}, year={2018}, doi={10.1021/acs.jctc.8b00149}}

@article{chen2020tensoralloy,
 author={Chen, Xin and others}, title={{TensorAlloy}: An automatic atomistic neural network program for alloys}, journal={Computer Physics Communications}, volume={250}, pages={107057}, year={2020}, doi={10.1016/j.cpc.2019.107057}}

@article{chen2021machinelearning,
 author={Chen, Xin and Wang, Li-Fang and Gao, Xing-Yu and Zhao, Ya-Fan and Lin, De-Ye and Chu, Wei-Dong and Song, Hai-Feng}, title={Machine learning enhanced empirical potentials for metals and alloys}, journal={Computer Physics Communications}, volume={269}, pages={108132}, year={2021}, doi={10.1016/j.cpc.2021.108132}}

@inproceedings{ouyang2025tensormd,
 author={Ouyang, Yucheng and Chen, Xin and Liu, Ying and others}, title={{TensorMD}: Accelerating Molecular Dynamics with a High-Performance Machine Learning Interatomic Potential}, booktitle={Proceedings of the International Conference for High Performance Computing, Networking, Storage and Analysis}, pages={1631--1645}, year={2025}, doi={10.1145/3712285.3759844}}

@misc{ouyang2026smallfootprint,
 author={Ouyang, Yucheng and Chen, Xin and Liu, Ying and others}, title={Unlocking Multi-Component Bulk-Materials Molecular Dynamics with a Small-Footprint Machine Learning Interatomic Potential}, year={2026}, eprint={2608.16329}, archivePrefix={arXiv}, doi={10.48550/arXiv.2608.16329}, url={https://arxiv.org/abs/2608.16329}}

@article{prome2025,
 author={Gao, Xingyu and Wang, William Yi and Chen, Xin and others}, title={{ProME}: An integrated computational platform for material properties at extremes and its application in multicomponent alloy design}, journal={Materials Genome Engineering Advances}, volume={3}, number={3}, pages={e70029}, year={2025}, doi={10.1002/mgea.70029}}

@article{song2024unep,
 author={Song, Keke and others}, title={General-purpose machine-learned potential for 16 elemental metals and their alloys}, journal={Nature Communications}, volume={15}, pages={10208}, year={2024}, doi={10.1038/s41467-024-54554-x}}

@misc{wu2026torchnep,
 author={Wu, Yong-Chao and Chang, Xiaoya and M{\"a}kinen, Tero and Esfandiarpour, Amin and Shao, Jian-Li and Ala-Nissila, Tapio and Fan, Zheyong and Alava, Mikko}, title={{TorchNEP}: Ultra-Efficient and Accurate Training of Neuroevolution Potentials}, year={2026}, eprint={2606.19557}, archivePrefix={arXiv}, doi={10.48550/arXiv.2606.19557}, url={https://arxiv.org/abs/2606.19557}}

@article{shuang2026comparison,
 author={Shuang, Fei and others}, title={Benchmarking chemically scalable machine-learning interatomic potentials for large-scale simulations of multicomponent alloys}, journal={Physical Review Materials}, volume={10}, pages={073802}, year={2026}, doi={10.1103/1qc9-ypyb}}

@article{grace2026,
 author={Lysogorskiy, Yury and Bochkarev, Anton and Drautz, Ralf}, title={Graph atomic cluster expansion for foundational machine learning interatomic potentials}, journal={npj Computational Materials}, volume={12}, pages={114}, year={2026}, doi={10.1038/s41524-026-01979-1}}

@article{musaelian2023allegro,
 author={Musaelian, Albert and Batzner, Simon and Johansson, Anders and Sun, Lixin and Owen, Cameron J. and Kornbluth, Mordechai and Kozinsky, Boris}, title={Learning local equivariant representations for large-scale atomistic dynamics}, journal={Nature Communications}, volume={14}, pages={579}, year={2023}, doi={10.1038/s41467-023-36329-y}}

@article{wang2024hotpp,
 author={Wang, Junjie and Wang, Yong and Zhang, Haoting and Yang, Ziyang and Liang, Zhixin and Shi, Jiuyang and Wang, Hui-Tian and Xing, Dingyu and Sun, Jian}, title={{E(n)-Equivariant} cartesian tensor message passing interatomic potential}, journal={Nature Communications}, volume={15}, pages={7607}, year={2024}, doi={10.1038/s41467-024-51886-6}}

@article{batatia2025design,
 author={Batatia, Ilyes and others}, title={The design space of {E(3)}-equivariant atom-centred interatomic potentials}, journal={Nature Machine Intelligence}, volume={7}, pages={56--67}, year={2025}, doi={10.1038/s42256-024-00956-x}}

@article{pozdnyakov2020incompleteness,
 author={Pozdnyakov, Sergey N. and others}, title={Incompleteness of Atomic Structure Representations}, journal={Physical Review Letters}, volume={125}, pages={166001}, year={2020}, doi={10.1103/PhysRevLett.125.166001}}

@inproceedings{zhang2018deeppotse,
 author={Zhang, Linfeng and Han, Jiequn and Wang, Han and Saidi, Wissam A. and Car, Roberto and E, Weinan}, title={End-to-end symmetry preserving inter-atomic potential energy model for finite and extended systems}, booktitle={Advances in Neural Information Processing Systems}, volume={31}, year={2018}, pages={4436--4446}, url={https://arxiv.org/abs/1805.09003}}

@inproceedings{schutt2017schnet,
 author={Sch{\"u}tt, Kristof T. and Kindermans, Pieter-Jan and Sauceda, Huziel E. and Chmiela, Stefan and Tkatchenko, Alexandre and M{\"u}ller, Klaus-Robert}, title={{SchNet}: A continuous-filter convolutional neural network for modeling quantum interactions}, booktitle={Advances in Neural Information Processing Systems}, volume={30}, year={2017}, url={https://arxiv.org/abs/1706.08566}}

@article{lu2022dpcompress,
 author={Lu, Denghui and Jiang, Wanrun and Chen, Yixiao and Zhang, Linfeng and Jia, Weile and Wang, Han and Chen, Mohan}, title={{DP Compress}: A Model Compression Scheme for Generating Efficient Deep Potential Models}, journal={Journal of Chemical Theory and Computation}, volume={18}, pages={5559--5567}, year={2022}, doi={10.1021/acs.jctc.2c00102}}

@misc{li2026dpa4,
 author={Li, Tiancheng and Li, Wentao and Peng, Anyang and Xue, Jianming and Zhang, Linfeng and Zhang, Duo and Wang, Han}, title={{DPA4}: Pushing the Accuracy-Cost Frontier of Interatomic Potentials with {EMFA} {SO(2)} Convolution}, year={2026}, eprint={2606.02419}, archivePrefix={arXiv}, doi={10.48550/arXiv.2606.02419}, url={https://arxiv.org/abs/2606.02419}}

@misc{li2026dpa4c,
 author={Li, Tiancheng and Xue, Jianming and Zhang, Linfeng and Zhang, Duo and Wang, Han}, title={Universal Machine-learning Molecular Dynamics at the Speed of Empirical Potentials}, year={2026}, eprint={2608.19041}, archivePrefix={arXiv}, doi={10.48550/arXiv.2608.19041}, url={https://arxiv.org/abs/2608.19041}}

@misc{bu2026wpe,
 author={Bu, Hekai and Fan, Zheyong}, title={Wisevolve Potential Engine: Machine-Learning Molecular Dynamics at 100 Million Atom-Steps per Second on a Single {GPU}}, year={2026}, eprint={2610.17366}, archivePrefix={arXiv}, doi={10.48550/arXiv.2610.17366}, url={https://arxiv.org/abs/2610.17366}}

@inproceedings{song2026sharednep,
 author={Song, Zeyu and Duan, Xiaohui and Zhou, Zikai and Gan, Lin and Yang, Guangwen}, title={Unleashing Shared Memory Potential for {NEP} Machine-Learning Potential in Molecular Dynamics on {GPUs}}, booktitle={Proceedings of the International Conference for High Performance Computing, Networking, Storage and Analysis}, year={2026}, url={https://sc26.supercomputing.org/proceedings/papers/paper_pages/pap477.html}}

@article{liang2026nep89,
 author={Liang, Ting and others}, title={{NEP89}: Universal neuroevolution potential for inorganic and organic materials across 89 elements}, journal={Nature Computational Science}, volume={6}, pages={789--801}, year={2026}, doi={10.1038/s43588-026-01009-6}}

@article{omat24,
 author={Barroso-Luque, Luis and others}, title={The Open Materials 2024 ({OMat24}) inorganic materials dataset and models}, journal={Nature Computational Science}, volume={6}, pages={642--652}, year={2026}, doi={10.1038/s43588-026-00996-w}, eprint={2410.12771}, archivePrefix={arXiv}}

@article{gpumd2025,
 author={Xu, Ke and others}, title={{GPUMD 4.0}: A high-performance molecular dynamics package for versatile materials simulations with machine-learned potentials}, journal={Materials Genome Engineering Advances}, volume={3}, number={3}, pages={e70028}, year={2025}, doi={10.1002/mgea.70028}}

@article{dpa12024,
 author={Zhang, Duo and Bi, Hangrui and Dai, Fu-Zhi and Jiang, Wanrun and Liu, Xinzijian and Zhang, Linfeng and Wang, Han}, title={Pretraining of attention-based deep learning potential model for molecular simulation}, journal={npj Computational Materials}, volume={10}, pages={94}, year={2024}, doi={10.1038/s41524-024-01278-7}}

@article{dpa22024,
 author={Zhang, Duo and Liu, Xinzijian and Zhang, Xiangyu and others}, title={{DPA-2}: a large atomic model as a multi-task learner}, journal={npj Computational Materials}, volume={10}, pages={293}, year={2024}, doi={10.1038/s41524-024-01493-2}}

@article{dpa32026,
 author={Zhang, Duo and Peng, Anyang and Cai, Chun and Li, Wentao and Zhou, Yuanchang and Zeng, Jinzhe and Guo, Mingyu and Zhang, Chengqian and Li, Bowen and Jiang, Hong and Zhu, Tong and Jia, Weile and Zhang, Linfeng and Wang, Han}, title={A graph neural network for the era of large atomistic models}, journal={npj Computational Materials}, volume={12}, pages={276}, year={2026}, doi={10.1038/s41524-026-02146-2}}

@misc{kavanagh2026foundation,
 author={Kavanagh, Se{\'a}n R. and Tan, Chuin Wei and Wang, Menghang and Descoteaux, Marc L. and de Miranda Nascimento, Gabriel and Unneberg, Ulrik and Zichi, Laura and Libbi, Francesco and Rivano, Norma and Glover, Austin and Bharadwaj, Vivek and Johansson, Anders and Witt, William C. and Musaelian, Albert and Kozinsky, Boris}, title={Fast and Accurate Equivariant Foundation Models for Atomistic Simulation}, year={2026}, eprint={2607.28461}, archivePrefix={arXiv}, doi={10.48550/arXiv.2607.28461}, url={https://arxiv.org/abs/2607.28461}}

@inproceedings{liao2024equiformerv2,
 author={Liao, Yi-Lun and Wood, Brandon and Das, Abhishek and Smidt, Tess}, title={{EquiformerV2}: Improved Equivariant Transformer for Scaling to Higher-Degree Representations}, booktitle={International Conference on Learning Representations}, year={2024}, url={https://proceedings.iclr.cc/paper_files/paper/2024/hash/ab12e8f3443c1a789f595b18d8c597b4-Abstract-Conference.html}}

@misc{dynmat2026,
 author={Gawkowski, Miko{\l}aj J. and others}, title={{Dyna-Mat}: End-to-end benchmarking of foundation machine learning interatomic potentials in finite-temperature ensembles}, year={2026}, eprint={2607.03433}, archivePrefix={arXiv}, doi={10.48550/arXiv.2607.03433}, url={https://arxiv.org/abs/2607.03433}}

@article{pota2026,
 author={P{\'o}ta, Bal{\'a}zs and Ahlawat, Paramvir and Cs{\'a}nyi, G{\'a}bor and Simoncelli, Michele}, title={Thermal conductivity predictions with foundation atomistic models}, journal={Nature Communications}, volume={17}, pages={10566}, year={2026}, doi={10.1038/s41467-026-76391-w}, eprint={2408.00755}, archivePrefix={arXiv}}

@misc{matcalc,
 author={Liu, Runze and Liu, Elliott and Riebesell, Janosh and Qi, Ji and Ong, Shyue Ping and Ko, Tsz Wai}, title={{MatCalc}: A {Python} library for calculating materials properties from the potential energy surface ({PES})}, year={2026}, note={Version 0.5.1}, url={https://pypi.org/project/matcalc/0.5.1/}}

@misc{dpa4models,
 author={{DeePMD community}}, title={{DPA4-OMat24} model card and checkpoints (v20260911)}, year={2026}, url={https://huggingface.co/OutisLi/DPA4-OMat24-v20260911/blob/855893b0bdcefa5780d7fdd08f16590aed4f914d/README.md}}

@article{behler2007,
 author={Behler, J{\"o}rg and Parrinello, Michele}, title={Generalized Neural-Network Representation of High-Dimensional Potential-Energy Surfaces}, journal={Physical Review Letters}, volume={98}, pages={146401}, year={2007}, doi={10.1103/PhysRevLett.98.146401}}

@article{chen2022m3gnet,
 author={Chen, Chi and Ong, Shyue Ping}, title={A universal graph deep learning interatomic potential for the periodic table}, journal={Nature Computational Science}, volume={2}, pages={718--728}, year={2022}, doi={10.1038/s43588-022-00349-3}}

@article{macefoundation2025,
 author={Batatia, Ilyes and others}, title={A foundation model for atomistic materials chemistry}, journal={The Journal of Chemical Physics}, volume={163}, pages={184110}, year={2025}, doi={10.1063/5.0297006}}

@article{bartok2013soap,
 author={Bart{\'o}k, Albert P. and Kondor, Risi and Cs{\'a}nyi, G{\'a}bor},
 title={On representing chemical environments}, journal={Physical Review B},
 volume={87}, pages={184115}, year={2013}, doi={10.1103/PhysRevB.87.184115}}

@article{drautz2019ace,
 author={Drautz, Ralf}, title={Atomic cluster expansion for accurate and transferable interatomic potentials},
 journal={Physical Review B}, volume={99}, pages={014104}, year={2019}, doi={10.1103/PhysRevB.99.014104}}

@inproceedings{zaheer2017deepsets,
 author={Zaheer, Manzil and Kottur, Satwik and Ravanbakhsh, Siamak and P{\'o}czos, Barnab{\'a}s and Salakhutdinov, Ruslan and Smola, Alexander J.},
 title={Deep Sets}, booktitle={Advances in Neural Information Processing Systems}, volume={30}, year={2017},
 url={https://papers.neurips.cc/paper_files/paper/2017/hash/f22e4747da1aa27e363d86d40ff442fe-Abstract.html}}

@inproceedings{lee2019settransformer,
 author={Lee, Juho and Lee, Yoonho and Kim, Jungtaek and Kosiorek, Adam and Choi, Seungjin and Teh, Yee Whye},
 title={Set Transformer: A Framework for Attention-based Permutation-Invariant Neural Networks},
 booktitle={Proceedings of the 36th International Conference on Machine Learning},
 series={Proceedings of Machine Learning Research}, volume={97}, pages={3744--3753}, year={2019},
 url={https://proceedings.mlr.press/v97/lee19d.html}}

@inproceedings{fu2025esen,
 author={Fu, Xiang and Wood, Brandon M. and Barroso-Luque, Luis and Levine, Daniel S. and Gao, Meng and Dzamba, Misko and Zitnick, C. Lawrence},
 title={Learning Smooth and Expressive Interatomic Potentials for Physical Property Prediction},
 booktitle={Proceedings of the 42nd International Conference on Machine Learning},
 series={Proceedings of Machine Learning Research}, volume={267}, pages={17875--17893},
 year={2025}, url={https://proceedings.mlr.press/v267/fu25h.html}}

@article{wagstaff2022sets,
 author={Wagstaff, Edward and Fuchs, Fabian B. and Engelcke, Martin and Osborne, Michael A. and Posner, Ingmar},
 title={Universal Approximation of Functions on Sets}, journal={Journal of Machine Learning Research},
 volume={23}, number={151}, pages={1--56}, year={2022}, url={https://jmlr.org/papers/v23/21-0730.html}}

@article{shapeev2016mtp,
 author={Shapeev, Alexander V.}, title={Moment Tensor Potentials: A Class of Systematically Improvable Interatomic Potentials}, journal={Multiscale Modeling \& Simulation}, volume={14}, number={3}, pages={1153--1173}, year={2016}, doi={10.1137/15M1054183}}

@article{zimmermann2020fingerprints,
 author={Zimmermann, Nils E. R. and Jain, Anubhav}, title={Local structure order parameters and site fingerprints for quantification of coordination environment and crystal structure similarity}, journal={RSC Advances}, volume={10}, pages={6063--6081}, year={2020}, doi={10.1039/C9RA07755C}}

@article{cordero2008radii,
 author={Cordero, Beatriz and G{\'o}mez, Ver{\'o}nica and Platero-Prats, Ana E. and Rev{\'e}s, Marc and Echeverr{\'i}a, Jorge and Cremades, Eduard and Barrag{\'a}n, Flavia and Alvarez, Santiago}, title={Covalent radii revisited}, journal={Dalton Transactions}, pages={2832--2838}, year={2008}, doi={10.1039/B801115J}}
\end{document}


\renewcommand{\thesection}{\arabic{section}}
\renewcommand{\theequation}{S\arabic{equation}}
\renewcommand{\thefigure}{\arabic{figure}}
\renewcommand{\thetable}{\arabic{table}}

\renewcommand{\figurename}{Supplementary Fig.}\renewcommand{\tablename}{Supplementary Table}
\maketitle
\begin{figure}[p]\centering
{\footnotesize We generated the supplementary figures in Python using Matplotlib.\par}\medskip
\includegraphics[width=\linewidth,height=0.80\textheight,keepaspectratio]{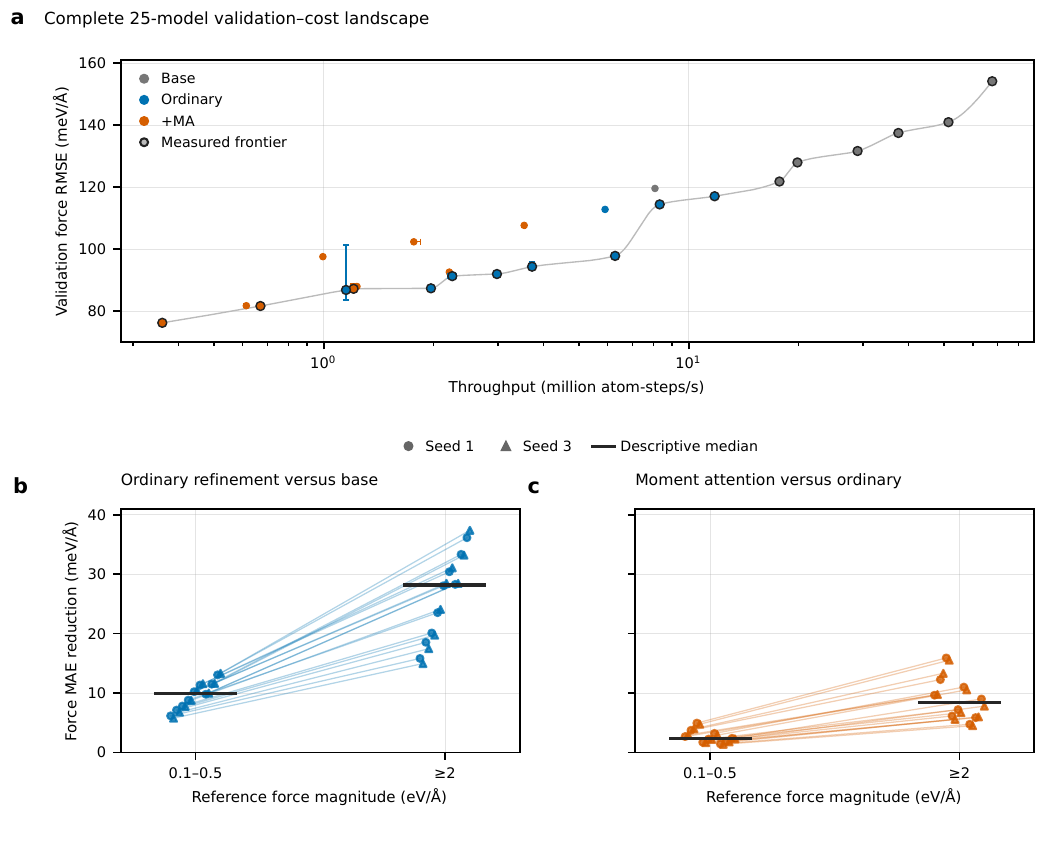}
\caption{\textbf{HEA16 accuracy and throughput, and force errors by reference-force magnitude.}
\textbf{a}, Mean validation force RMSE over six runs and median BCC16 throughput on the RTX~5090 for the seed-1 checkpoint of each of the 25 architectures.
Vertical bars span the six validation errors and horizontal bars span three throughput measurements.
Larger, dark-edged markers identify nondominated points and the faint curve is a visual guide.
\textbf{b,c}, Reductions in testing force MAE from adding ordinary refinement to the base model (\textbf{b}) or moment attention to the corresponding ordinary model (\textbf{c}).
Lines connect the same architecture pair and seed across the two reference-force bands (0.1--0.5 and at least 2\,eV/\AA{}).
Circles and triangles denote seeds 1 and 3, respectively, and black bars mark the median of the 18 comparisons in each band.}\label{fig:si-family-cost-environment}\end{figure}

\begin{figure}[p]\centering\includegraphics[width=\linewidth]{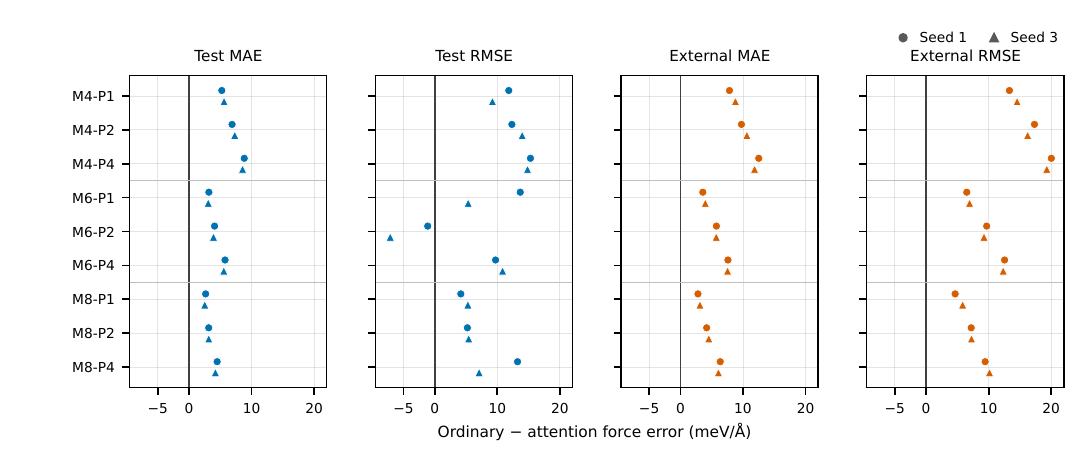}
\caption{\textbf{Moment-attention gains on test and external structures.}
The first two panels show reductions in force MAE and RMSE on the 5,273 HEA16 testing structures and the last two show the corresponding reductions on the 2,250 external alloy structures (S1--S3).
Each panel shows ordinary-minus-attention errors for all nine architecture pairs at seeds 1 and 3, giving 18 comparisons.
Circles denote seed 1 and triangles denote seed 3.
Positive gains favour moment attention.}\label{fig:si-attention-transfer}\end{figure}

\begin{figure}[p]\centering\includegraphics[width=\linewidth,height=0.82\textheight,keepaspectratio]{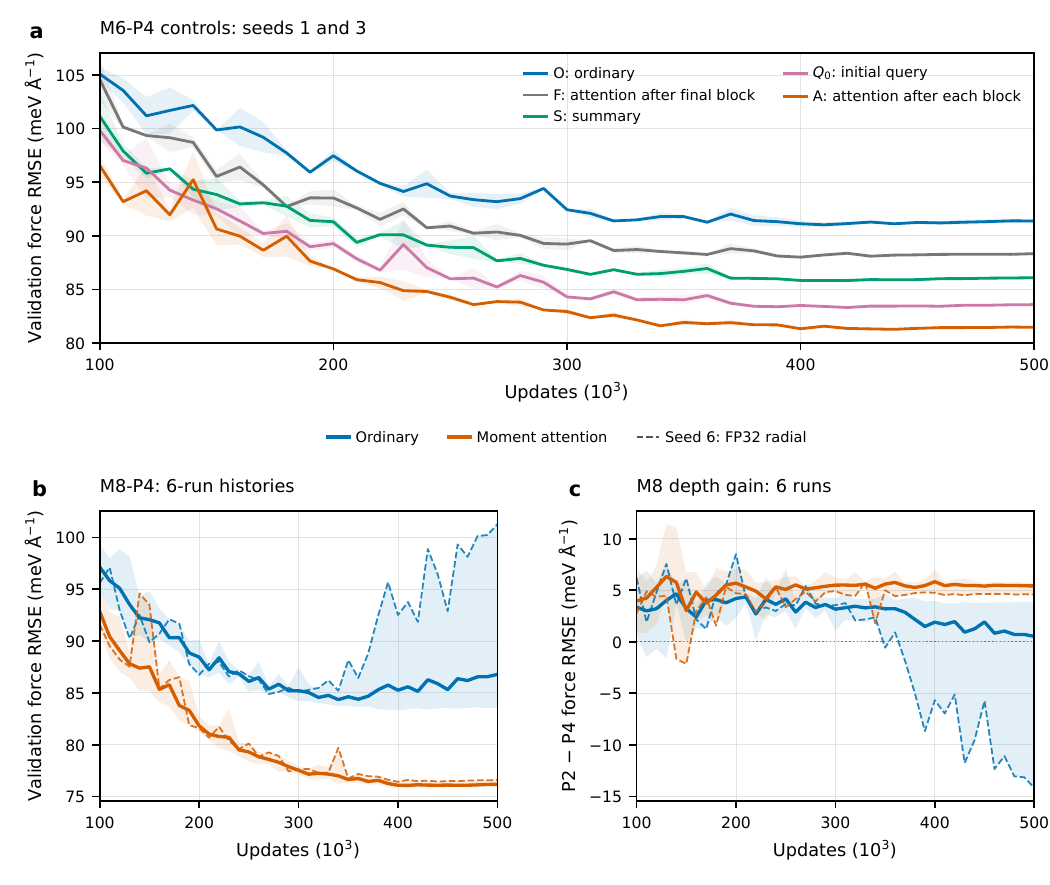}
\caption{\textbf{HEA16 validation force RMSE during training.}
\textbf{a}, Validation histories for the five M6-P4 models O, F, S, $Q_0$ and A (Supplementary Table~\ref{tab:hea16-mechanism}).
\textbf{b}, Validation histories for M8-P4 with ordinary refinement or moment attention.
\textbf{c}, Paired differences in force RMSE between two and four blocks (P2 minus P4), with positive values favouring four blocks.
Solid lines show the mean and shading spans the observed range across seeds 1 and 3 in \textbf{a} and across all six runs in \textbf{b,c}.
Dashed lines identify seed 6, whose ordinary M8-P4 run shows a late increase in validation force RMSE.}\label{fig:si-mechanism}\label{fig:si-family-learning}\end{figure}

\begin{figure}[p]\centering\includegraphics[width=\linewidth,height=0.80\textheight,keepaspectratio]{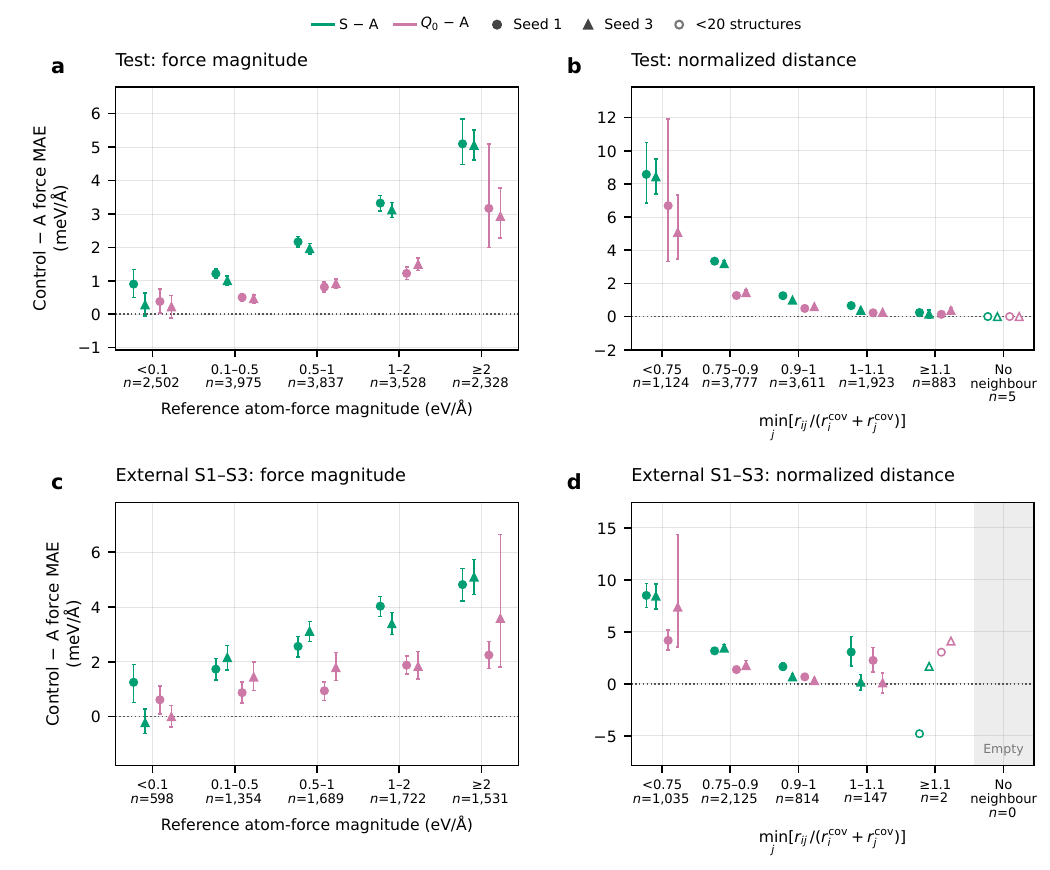}
\caption{\textbf{Environment-resolved paired force gains.}
\textbf{a,b}, HEA16 testing subset.
\textbf{c,d}, External alloy structures (S1--S3).
Panels \textbf{a,c} group atoms by reference-force magnitude and panels \textbf{b,d} by minimum normalized neighbour distance $\min_j[r_{ij}/(r_i^{\rm cov}+r_j^{\rm cov})]$, where $r^{\rm cov}$ is the tabulated covalent radius.
Positive component-MAE differences, S minus A or $Q_0$ minus A, favour A.
Circles and triangles identify seeds 1 and 3.
Bars show pointwise 95\% bootstrap intervals obtained by resampling structures.
Tick labels count the structures contributing atoms to each bin, and a structure may contribute to several bins.
Open markers identify bins with fewer than 20 contributing structures, for which no interval is shown.
Grey regions indicate empty bins.}\label{fig:si-environment-gains}\end{figure}

\begin{figure}[p]\centering\includegraphics[width=\linewidth,height=0.82\textheight,keepaspectratio]{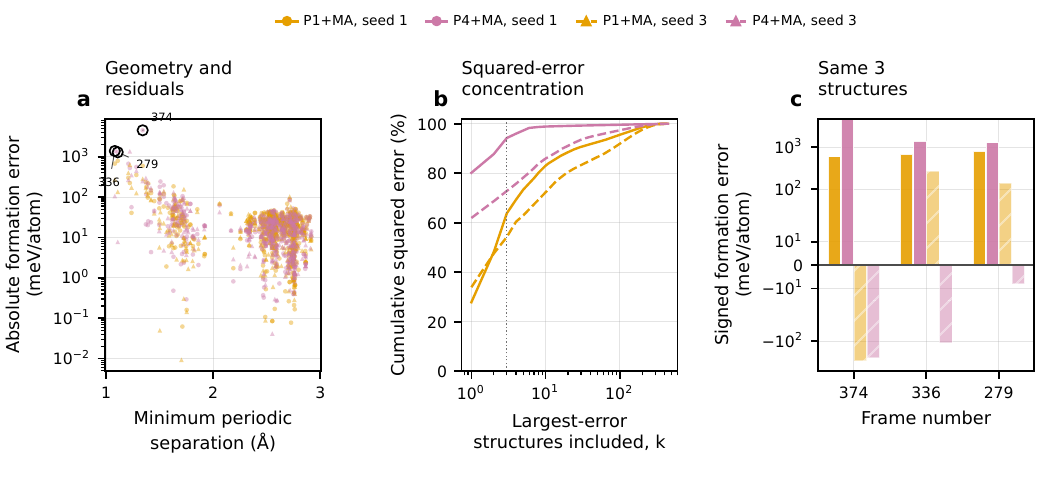}
\caption{\textbf{Concentration of formation-energy errors in the GNoME set.}
We compare standard-readout HEA16 models M8-P1+MA and M8-P4+MA on all 444 GNoME structures at seeds 1 and 3.
\textbf{a}, Absolute formation-energy residuals versus minimum periodic interatomic distance.
\textbf{b}, Cumulative fraction of squared error, with structures added in descending order of absolute error.
\textbf{c}, Signed residuals for both models and seeds on the three structures with the largest absolute errors for M8-P4+MA at seed 1.
Circles and solid lines denote seed 1, and triangles, dashed lines and hatched bars denote seed 3.}\label{fig:si-hea16-gnome}\end{figure}

\begin{figure}[p]\centering\includegraphics[width=\linewidth,height=0.82\textheight,keepaspectratio]{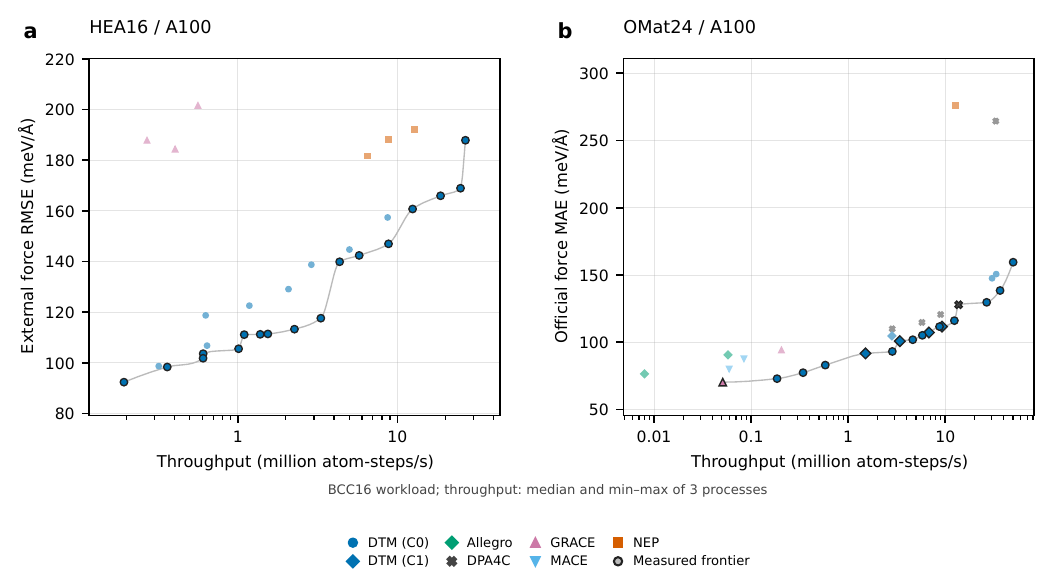}
\caption{\textbf{Force accuracy and BCC16 throughput on the A100.}
\textbf{a}, Force RMSE on the external alloy datasets S1--S3 for the HEA16 benchmark.
\textbf{b}, Force MAE on the official OMat24 validation set.
DTM circles and diamonds denote standard (C0) and cubic (C1) readouts, respectively.
Points and horizontal bars give the median and range of three throughput measurements.
Larger, dark-edged markers identify nondominated models and faint grey curves are visual guides.
Main Fig.~3 gives the RTX~5090 comparison.}\label{fig:si-hea16-timing}\end{figure}

\begin{figure}[p]\centering\includegraphics[width=\linewidth,height=0.83\textheight,keepaspectratio]{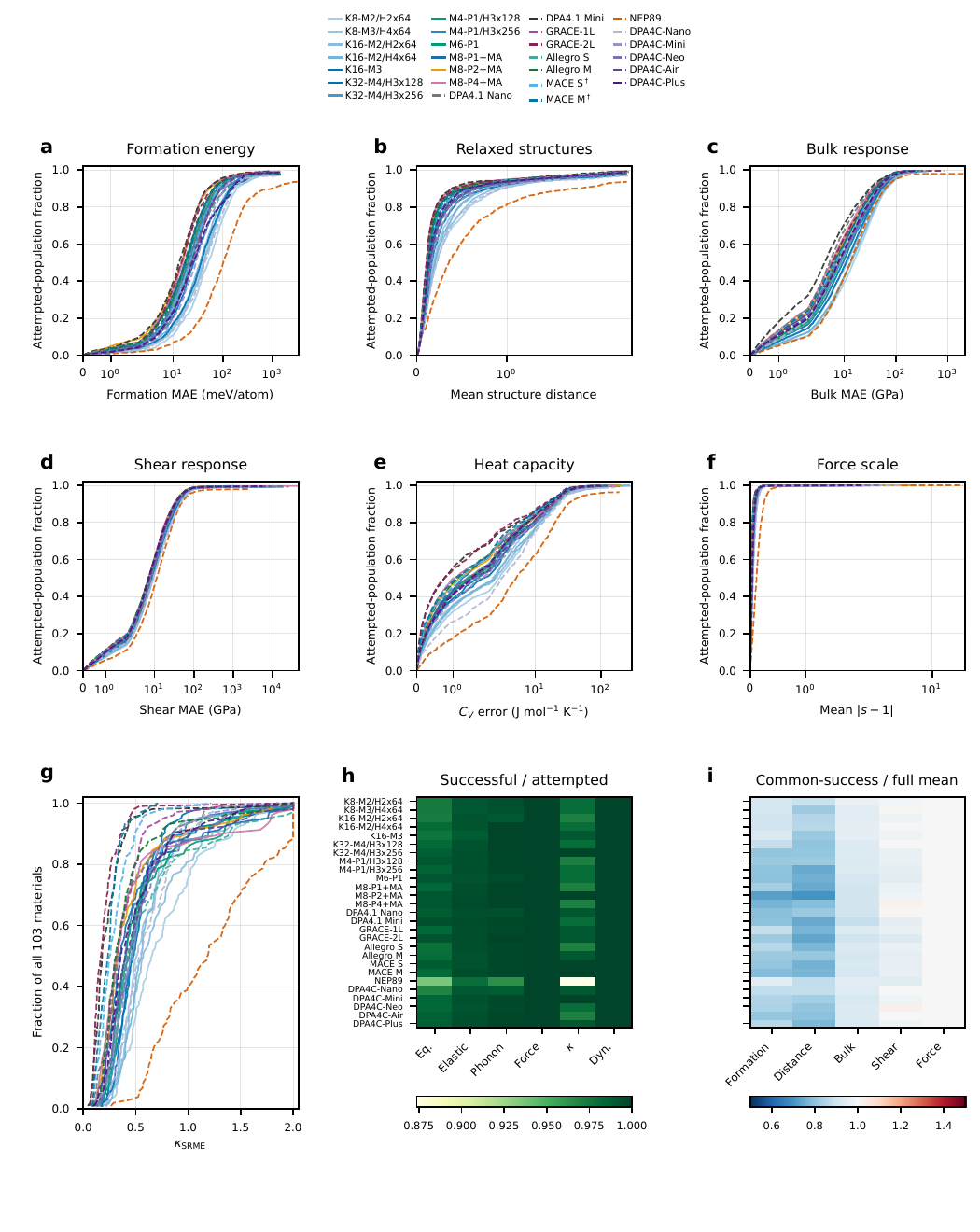}
\caption{\textbf{Property-error distributions and calculation completion.}
\textbf{a--f}, At each error threshold, the curves show the fraction of all attempted materials with a successful result at or below that threshold.
\textbf{g}, Thermal-conductivity scores for all 103 materials, with a score of 2 for each failed calculation.
In \textbf{a--g}, solid curves denote the 13 DTM models and dashed curves the 14 external models.
The dagger marks MACE formation energies that use single-point N and F reference energies at fixed DFT-selected geometries.
\textbf{h}, Completion fraction for each model and benchmark.
\textbf{i}, Each model's mean error on materials completed by all 27 models is divided by its mean error on all materials it completed for that property.
The shared sets comprise 654 materials for equilibrium properties, 3,548 for elasticity and 979 for force scaling.}\label{fig:si-properties}\end{figure}

\begin{figure}[p]\centering\includegraphics[width=\linewidth,height=0.86\textheight,keepaspectratio]{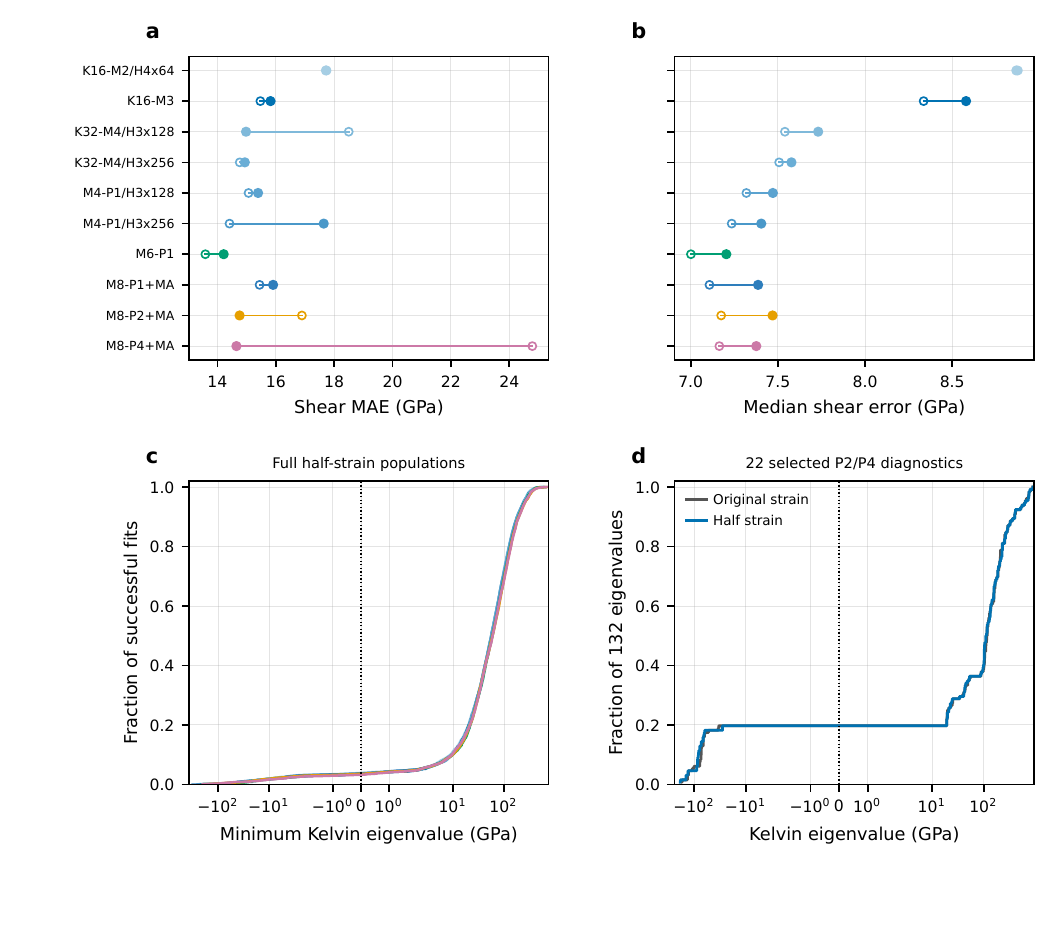}
\caption{\textbf{Elasticity strain sensitivity.}
\textbf{a,b}, Mean and median absolute shear-modulus errors for 10 DTM checkpoints over the successful fits on each strain grid.
Open and filled markers denote the original and halved strain amplitudes, respectively.
\textbf{c}, Cumulative distributions of the minimum Kelvin-stiffness eigenvalue for successful fits at halved strain amplitudes.
\textbf{d}, Cumulative distributions of all six Kelvin-stiffness eigenvalues for 22 selected M8-P2+MA and M8-P4+MA model--material cases, giving 132 eigenvalues per strain grid.
Main Fig.~4 uses the original strain amplitudes for every model.
Supplementary Table~\ref{tab:half-strain} gives the mean bulk-modulus errors.}\label{fig:si-strain}\end{figure}

\begin{figure}[p]\centering\includegraphics[width=\linewidth,height=0.86\textheight,keepaspectratio]{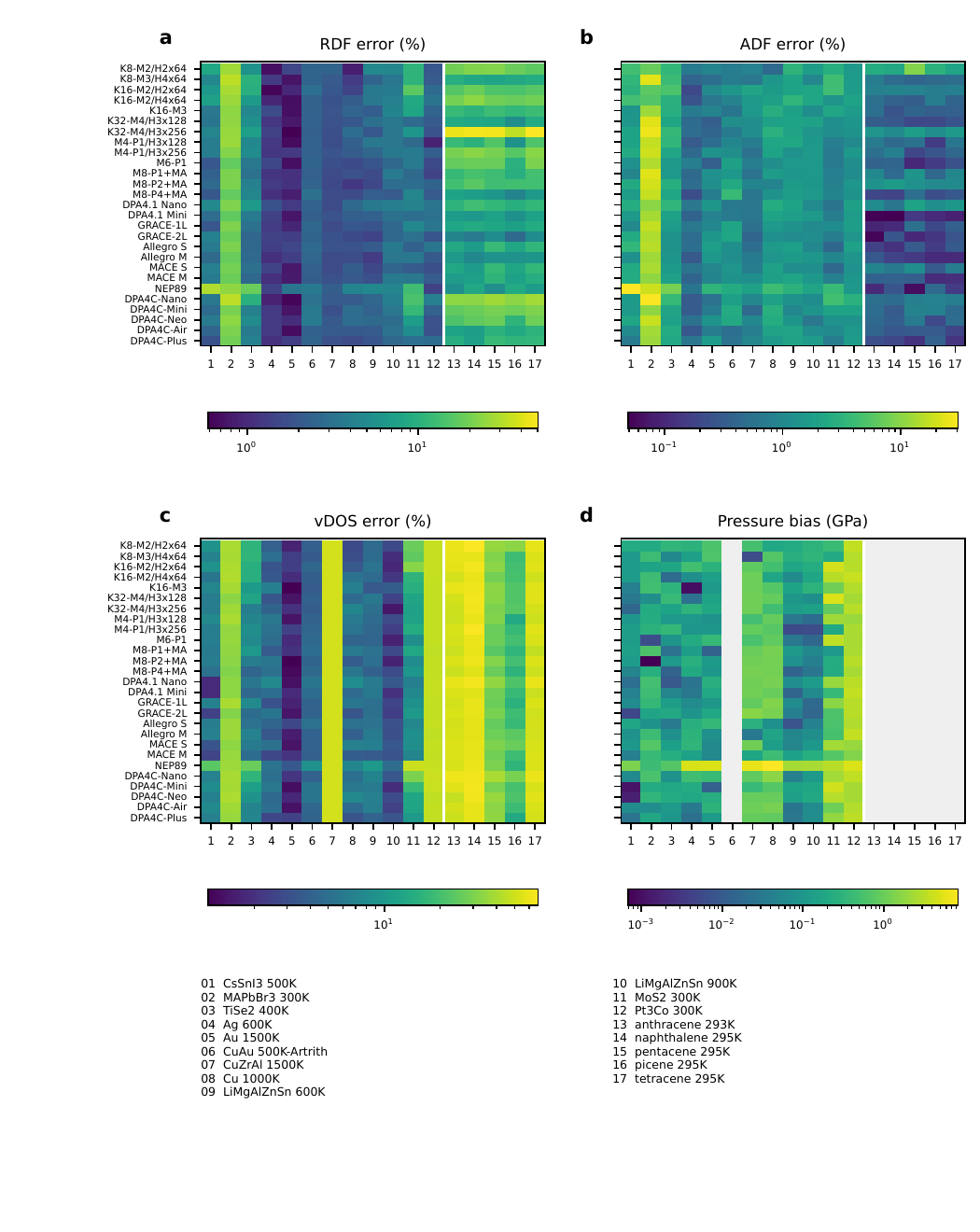}
\caption{\textbf{Structure and dynamics at finite temperature.}
\textbf{a--c}, Errors in the radial distribution function, angular distribution function and vibrational density of states, respectively, for 13 DTM and 14 external models.
\textbf{d}, Absolute difference between predicted and reference trajectory-mean pressures.
Systems 1--12 are nonmolecular and systems 13--17 are molecular crystals with D3 reference trajectories.
Grey cells indicate missing reference pressures.}\label{fig:si-dynamics}\end{figure}

\begingroup\scriptsize
\newcommand{\sinavsection}[2]{\hyperref[#1]{\textbf{\ref*{#1} #2}}\par}
\newcommand{\sinavsubsection}[2]{\hangindent=1em\hspace*{1em}\hyperref[#1]{\ref*{#1} #2}\par}
\noindent\textbf{Contents}\par\medskip
\noindent\begin{minipage}[t]{0.49\textwidth}
\textbf{Sections}\par
\sinavsection{sec:si-tensor-potential}{DeepTensorMoment model architecture}
\sinavsubsection{nav:sub1}{Model specification and periodic geometry}
\sinavsubsection{nav:sub2}{Chemical--radial edge encoding}
\sinavsubsection{nav:sub3}{Harmonic basis and initial moments}
\sinavsubsection{nav:sub4}{Equivariant moment refinement}
\sinavsubsection{nav:sub5}{Moment attention}
\sinavsubsection{nav:sub6}{Standard readout (C0)}
\sinavsubsection{nav:sub7}{Cubic-augmented readout (C1)}
\sinavsubsection{nav:sub8}{Atomic energies, symmetries and derivatives}
\sinavsubsection{nav:sub9}{Parameter allocation and representation size}
\sinavsubsection{nav:sub10}{Matched operator controls and computational cost}
\smallskip
\sinavsection{sec:si-hea16}{HEA16 models and controlled refinement study}
\sinavsubsection{nav:sub11}{HEA16 dataset, architecture study and training}
\sinavsubsection{nav:sub13}{Evaluation datasets and metrics}
\sinavsubsection{nav:sub14}{Evaluation results}
\sinavsubsection{nav:sub15}{Force gains at fixed parameter count}
\sinavsubsection{nav:sub16}{Force-error diagnostics}
\sinavsubsection{nav:sub17}{External alloy benchmarks}
\sinavsubsection{nav:sub18}{Force accuracy and cost of cubic readout}
\sinavsubsection{nav:sub19}{Alloy formation energies with cubic readout}
\sinavsubsection{sec:si-hea16-throughput}{Throughput}
\smallskip
\sinavsection{sec:si-omat24}{OMat24 models, validation and physical evaluation}
\sinavsubsection{nav:sub20}{Models and training data}
\sinavsubsection{nav:sub21}{Validation and controlled comparisons}
\sinavsubsection{nav:sub22}{Inference and throughput}\hspace*{2em}\hyperref[sec:si-sc26-nep]{Additional NEP implementation on FCC Cu}\par
\sinavsubsection{nav:sub23}{Static properties and strain sensitivity}
\sinavsubsection{nav:sub24}{Finite-temperature dynamics}
\sinavsubsection{nav:sub25}{Thermal conductivity}
\sinavsubsection{nav:sub26}{Validation errors and physical predictions}
\smallskip
\end{minipage}\hfill\begin{minipage}[t]{0.49\textwidth}
\textbf{Tables}\par
\hyperref[tab:hea16-architectures]{\ref*{tab:hea16-architectures} HEA16 architectures and accuracy}\par
\hyperref[tab:hea16-mechanism]{\ref*{tab:hea16-mechanism} Matched control validation}\par
\hyperref[tab:hea16-mechanism-test]{\ref*{tab:hea16-mechanism-test} HEA16 control testing errors}\par
\hyperref[tab:hea16-mechanism-external]{\ref*{tab:hea16-mechanism-external} HEA16 control external errors}\par
\hyperref[tab:hea16-mechanism-paired]{\ref*{tab:hea16-mechanism-paired} Paired control improvements}\par
\hyperref[tab:hea16-mechanism-decomposition]{\ref*{tab:hea16-mechanism-decomposition} Force-error decomposition}\par
\hyperref[tab:hea16-mechanism-contact-subset]{\ref*{tab:hea16-mechanism-contact-subset} Force errors with a minimum-distance criterion}\par
\hyperref[tab:hea16-splits]{\ref*{tab:hea16-splits} HEA16 split errors by seed}\par
\hyperref[tab:hea16-formation]{\ref*{tab:hea16-formation} External formation energies}\par
\hyperref[tab:hea16-forces]{\ref*{tab:hea16-forces} External force errors}\par
\hyperref[tab:hea16-cubic-validation]{\ref*{tab:hea16-cubic-validation} HEA16 cubic validation}\par
\hyperref[tab:hea16-cubic-external-forces]{\ref*{tab:hea16-cubic-external-forces} Cubic testing and external errors}\par
\hyperref[tab:hea16-cubic-external-formation]{\ref*{tab:hea16-cubic-external-formation} Cubic formation energies}\par
\hyperref[tab:omat-registry]{\ref*{tab:omat-registry} OMat24 architectures}\par
\hyperref[tab:omat-training-composition]{\ref*{tab:omat-training-composition} OMat24 training composition}\par
\hyperref[tab:benchmarks-official]{\ref*{tab:benchmarks-official} Official validation and reconstructed DPA-style subset}\par
\hyperref[tab:omat24-cubic-validation]{\ref*{tab:omat24-cubic-validation} Paired changes with cubic readout}\par
\hyperref[tab:omat24-sqa-validation]{\ref*{tab:omat24-sqa-validation} OMat24 query controls}\par
\hyperref[tab:throughput-hea16]{\ref*{tab:throughput-hea16} HEA16 throughput}\par
\hyperref[tab:throughput-omat24]{\ref*{tab:throughput-omat24} OMat24 throughput}\par
\hyperref[tab:si-sc26-cu]{\ref*{tab:si-sc26-cu} SC26 NEP throughput on FCC Cu}\par
\hyperref[tab:benchmarks-properties]{\ref*{tab:benchmarks-properties} Static properties and coverage}\par
\hyperref[tab:half-strain]{\ref*{tab:half-strain} Strain sensitivity}\par
\hyperref[tab:benchmarks-dynamics]{\ref*{tab:benchmarks-dynamics} Dynamics and conductivity}\par
\hyperref[tab:si-mae-benchmarks]{\ref*{tab:si-mae-benchmarks} Validation/property correlations}\par
\medskip\textbf{Figures}\par
\hyperref[fig:si-family-cost-environment]{\ref*{fig:si-family-cost-environment} Accuracy, cost and force bands}\par
\hyperref[fig:si-attention-transfer]{\ref*{fig:si-attention-transfer} Attention gains}\par
\hyperref[fig:si-mechanism]{\ref*{fig:si-mechanism} Training histories}\par
\hyperref[fig:si-environment-gains]{\ref*{fig:si-environment-gains} Environment-resolved gains}\par
\hyperref[fig:si-hea16-gnome]{\ref*{fig:si-hea16-gnome} Compressed-structure tail}\par
\hyperref[fig:si-hea16-timing]{\ref*{fig:si-hea16-timing} A100 comparison}\par
\hyperref[fig:si-properties]{\ref*{fig:si-properties} Physical-property distributions}\par
\hyperref[fig:si-strain]{\ref*{fig:si-strain} Strain sensitivity}\par
\hyperref[fig:si-dynamics]{\ref*{fig:si-dynamics} Finite-temperature dynamics}\par
\hyperref[fig:si-mae-physical]{\ref*{fig:si-mae-physical} Structure, shear and force-scale correlations}\par
\hyperref[fig:si-mae-dynamics]{\ref*{fig:si-mae-dynamics} Dynamics correlations}\par
\end{minipage}\endgroup
\clearpage
\section{DeepTensorMoment model architecture}
\label{sec:si-tensor-potential}

\subsection{Model specification and periodic geometry}\label{nav:sub1}

DeepTensorMoment (DTM) represents each atomic neighbourhood through tensor moments of chemical--radial and angular information, following the TensorAlloy and TensorMD formulation~\cite{chen2020tensoralloy,chen2021machinelearning,ouyang2025tensormd,ouyang2026smallfootprint}. 
We use the C0 readout in the primary architecture study.
Dataset-specific widths, training and evaluation are given in Sections~\ref{sec:si-hea16} and~\ref{sec:si-omat24}.

We use the following notation for architectures trained on HEA16 and OMat24.
\begin{equation}
(K,M,L,R,\mathcal J;\,d_{\rm enc}\times w_{\rm enc},\,d_{\rm head}\times w_{\rm head}).
\label{eq:si-architecture-spec}
\end{equation}
Here $K$ is the radial channel width, $M=\ell_{\max}+1$ the number of angular orders, $L$ the number of equivariant moment-refinement blocks (P-blocks), $R$ their low-rank channel width, and $\mathcal J$ the zero-based set of blocks followed by moment attention. 
All refinement blocks in this study use $R=64$.
Multilayer perceptron (MLP) sizes mean hidden-layer depth $\times$ width.
A base model has $L=0$, $R=0$ and $\mathcal J=\varnothing$.
Models with ordinary blocks alone also have $\mathcal J=\varnothing$.
MA denotes moment attention after the listed blocks.
Short labels such as M6-P1 are dataset-specific aliases for these settings. 
The suffix /H3x128 denotes three hidden layers of width 128.
Without a head-width suffix, K32-M4 and M4-P1 use width 256 in the OMat24 series and width 128 in the standard HEA16 series.
K32 is implicit in refined-model aliases when no radial width is shown.
C0 and C1 denote the standard and cubic-augmented readouts, respectively.
Equations~\eqref{eq:si-cubic-projection}--\eqref{eq:si-cubic-invariant} define the cubic projection width $R_{\rm c}$ and contraction families.

A structure contains the atomic numbers $\mathbf Z=(Z_1,\ldots,Z_N)$, Cartesian positions $\mathbf r_i\in\mathbb R^3$, collected as rows in $\mathbf R\in\mathbb R^{N\times3}$, and, for a periodic system, the cell matrix $\mathbf A$, whose rows are the lattice vectors $\mathbf a_1^{\mathsf T},\mathbf a_2^{\mathsf T},\mathbf a_3^{\mathsf T}$.
The atomic numbers are an essential input for a multielement model; the geometry-only shorthand
$(\mathbf R,\mathbf A)\mapsto E_{\mathrm{tot}}$ is understood to be conditioned on $\mathbf Z$.

Let $\mathbf n_e=(n_{e1},n_{e2},n_{e3})\in\mathbb Z^3$ identify the periodic image of neighbour $j$ on a directed edge $e=(i,j,\mathbf n_e)$. 
Its displacement from centre atom $i$, distance, and unit direction are
\begin{equation}
\mathbf r_e
=\mathbf r_j-\mathbf r_i+\sum_{\alpha=1}^{3}n_{e\alpha}\mathbf a_\alpha,
\qquad
r_e=\lVert\mathbf r_e\rVert,
\qquad
\widehat{\mathbf r}_e=\frac{\mathbf r_e}{r_e}.
\label{eq:si-periodic-edge}
\end{equation}
The local environment $\mathcal N_i$ contains all nonzero directed edges centred on $i$ with $r_e<r_{\mathrm c}$. 
Equation~\eqref{eq:si-periodic-edge} is the only place where absolute positions and the cell enter the descriptor: subsequent operations use relative distances, directions, and species.

Let $\mathcal E_i=\{(H_e,\mathbf Y_e)\}_{e\in\mathcal N_i}$ denote the original radial and angular edge features for centre $i$. 
In compact mathematical notation,
\begin{equation}
\begin{aligned}
(\mathbf Z,\mathbf R,\mathbf A)
&\rightarrow \Gamma_{r_{\mathrm c}}
\rightarrow \{\mathcal E_i\}
\rightarrow \mathbf P_i^{[0]}
\rightarrow \overline{\mathbf P}_i \\
&\rightarrow \mathbf G_i
\rightarrow \varepsilon_i
\rightarrow E_{\mathrm{tot}} .
\end{aligned}
\label{eq:si-compact-map}
\end{equation}
Here $\Gamma_{r_{\mathrm c}}$ is the directed periodic neighbour graph, $e$ denotes an edge of this graph, $H_e=(H_{e1},\ldots,H_{eK})$, and $\mathbf Y_e$ collects the angular blocks $\mathbf Y_{e\ell}=\mathbf Y_\ell(\widehat{\mathbf r}_e)$.
The index $k=1,\ldots,K$ labels a learned radial channel, while $\ell=0,\ldots,\ell_{\max}$ is an angular order. 
Throughout, $\mathbf Y$ denotes the fixed Cartesian harmonic basis and $\mathbf P$ the atom-centred moment state, grouped by angular order.
The energy network takes the complete invariant descriptor $\mathbf G$ as input.
The final moment state is $\overline{\mathbf P}_i=\mathbf P_i^{[L]}$, which reduces to the initial moments $\mathbf P_i^{[0]}$ in the base model.
At refinement stage $t$, the moment state evolves as
\begin{equation}
\begin{aligned}
\widetilde{\mathbf P}_i^{[t]}
&=\mathcal P^{[t]}(\mathbf P_i^{[t]}),\\
\mathbf P_i^{[t+1]}
&=\widetilde{\mathbf P}_i^{[t]}
+\begin{cases}
\eta_{\mathrm{agg}}^{[t]}
\boldsymbol\Delta_{i,\mathrm{agg}}^{[t]}
 (\widetilde{\mathbf P}_i^{[t]};\mathcal E_i),&t\in\mathcal J,\\
\mathbf0,&t\notin\mathcal J,
\end{cases}.
\end{aligned}
\label{eq:si-trunk-flow}
\end{equation}
Here $\mathcal P^{[t]}$ is an ordinary P-block, and $\mathcal J\subseteq\{0,\ldots,L-1\}$ selects the stages at which a moment-attention update follows the block and revisits $\mathcal E_i$.
Each stage has independent parameters, and the channel width is fixed throughout the sequence of refinement blocks.

\subsection{Chemical--radial edge encoding}\label{nav:sub2}

We represent each element $Z_i$ by a fixed 18-component orbital-occupation fingerprint $\boldsymbol\chi_i=\boldsymbol\chi(Z_i)\in\mathbb R^{d_\chi}$, with $d_\chi=18$.
Its components are the occupations of
\begin{equation}
(1s,2s,2p,3s,3p,3d,4s,4p,4d,4f,
  5s,5p,5d,5f,6s,6p,6d,7s).
\label{eq:si-orbital-order}
\end{equation}
The radial encoder uses the same element fingerprint that we append to the final atomic descriptor.
To retain the order of the centre and neighbour species while exposing simple chemical correlations, the edge encoder forms
\begin{equation}
\mathbf c_{ij}=
\boldsymbol\chi_i\mathbin{\Vert}\boldsymbol\chi_j
\mathbin{\Vert}(\boldsymbol\chi_i\odot\boldsymbol\chi_j)
\mathbin{\Vert}\lvert\boldsymbol\chi_i-\boldsymbol\chi_j\rvert
\mathbin{\Vert}(\boldsymbol\chi_i+\boldsymbol\chi_j),
\label{eq:si-pair-chemistry}
\end{equation}
where $\Vert$ denotes concatenation and $\odot$ element-wise multiplication. 
Thus $\mathbf c_{ij}\in\mathbb R^{5d_\chi}$ and, in general, $\mathbf c_{ij}\ne\mathbf c_{ji}$.

The distance is expanded in $K_{\mathrm{RBF}}$ fixed Gaussian functions. 
In this equation, $r$, $\mu_\beta$, and $r_{\mathrm c}$ denote their dimensionless numerical values in the code length unit $1\,\text{\AA}$:
\begin{equation}
\phi_\beta(r)=\exp[-\gamma_{\mathrm r}(r-\mu_\beta)^2],
\qquad \beta=1,\ldots,K_{\mathrm{RBF}},
\label{eq:si-gaussian-basis}
\end{equation}
The centres $\mu_\beta$ are uniformly spaced between $0$ and $r_{\mathrm c}$, and the fixed width coefficient is $\gamma_{\mathrm r}=K_{\mathrm{RBF}}/r_{\mathrm c}$.  
A shared SiLU MLP then produces $K$ radial channels,
\begin{equation}
H_{ek}=f_{\mathrm{cut}}(r_e)
\left[
\operatorname{MLP}_{\mathrm r}
\!\left(\boldsymbol\phi(r_e)\mathbin{\Vert}\mathbf c_{ij}\right)
\right]_k,
\label{eq:si-radial-encoder}
\end{equation}
with the cosine envelope
\begin{equation}
f_{\mathrm{cut}}(r)=
\begin{cases}
\tfrac12[1+\cos(\pi r/r_{\mathrm c})], & 0\le r<r_{\mathrm c},\\
0, & r\ge r_{\mathrm c}.
\end{cases}
\label{eq:si-cutoff}
\end{equation}
All edges share the same radial MLP, with element-pair chemistry encoded in
$\mathbf c_{ij}$.

\subsection{Harmonic basis and initial moments}\label{nav:sub3}

For each angular order $\ell$, let $\mathbf Y_\ell(\widehat{\mathbf r}) \in\mathbb R^{2\ell+1}$ denote the real irreducible Cartesian harmonic basis used by the model. 
It can be viewed as a fixed harmonic projection of the degree-$\ell$ Cartesian monomials,
\begin{equation}
\mathbf Y_\ell(\widehat{\mathbf r})
=\boldsymbol\Pi_\ell\boldsymbol\omega^{(\ell)}(\widehat{\mathbf r}),
\qquad
\dim\mathbf Y_\ell=2\ell+1,
\label{eq:si-harmonic-basis}
\end{equation}
where the rows of $\boldsymbol\Pi_\ell$ span homogeneous harmonic (equivalently,
symmetric-traceless) polynomials and are orthonormal in the symmetric Cartesian-tensor metric. 
The resulting basis is not unit normalized on the sphere: for unit directions $\widehat{\mathbf u}$ and $\widehat{\mathbf v}$,
\begin{equation}
\mathbf Y_\ell(\widehat{\mathbf u})^{\mathsf T}
\mathbf Y_\ell(\widehat{\mathbf v})
=\zeta_\ell\operatorname{Leg}_\ell
 (\widehat{\mathbf u}^{\mathsf T}\widehat{\mathbf v}),
\qquad
\zeta_\ell=\frac{2^\ell(\ell!)^2}{(2\ell)!},
\label{eq:si-harmonic-kernel}
\end{equation}
where $\operatorname{Leg}_\ell$ is the Legendre polynomial with $\operatorname{Leg}_\ell(1)=1$. 
In particular, $Y_0=1$, $\mathbf Y_1(\widehat{\mathbf r})= \widehat{\mathbf r}$, and $\|\mathbf Y_\ell\|^2=\zeta_\ell$.
For any $\mathsf O\in O(3)$,
\begin{equation}
\mathbf Y_\ell(\mathsf O\widehat{\mathbf r})
=\mathbf D^{(\ell)}(\mathsf O)
\mathbf Y_\ell(\widehat{\mathbf r}),
\qquad
\mathbf D^{(\ell)}(-\mathbf I)=(-1)^\ell\mathbf I,
\label{eq:si-harmonic-transform}
\end{equation}
where $\mathbf D^{(\ell)}(\mathsf O)^{\mathsf T} \mathbf D^{(\ell)}(\mathsf O)=\mathbf I$. 
Thus squared irrep norms are invariant and the order-$\ell$ block has natural parity $(-1)^\ell$. 
The cumulative angular width through $\ell_{\max}$ is $\sum_{\ell=0}^{\ell_{\max}}(2\ell+1)=(\ell_{\max}+1)^2$.

Earlier work discusses the connection between this harmonic basis and the angular terms of the modified embedded-atom method (MEAM)~\cite{ouyang2025tensormd,ouyang2026smallfootprint}.

The initial atom-centred equivariant moment state is
\begin{equation}
\mathbf P_{ik}^{[0],(\ell)}
=\sum_{e\in\mathcal N_i}
H_{ek}\,
\mathbf Y_\ell(\widehat{\mathbf r}_e)
\in\mathbb R^{2\ell+1}.
\label{eq:si-initial-moment}
\end{equation}
The fixed edge-wise basis $\mathbf Y_\ell$ is evaluated directly in the irreducible representation. 
Radial weighting and neighbour accumulation then produce the atom-centred moment tensor $\mathbf P_{ik}^{[0],(\ell)}$.  
The neighbour sum makes $\mathbf P_i^{[0]}$ independent of neighbour ordering.  
Under a rigid orthogonal transformation, every block $\mathbf P_{ik}^{[0],(\ell)}$ transforms within its order-$\ell$ irreducible space, preserving equivariance.  The stored state contains $d_P=K(\ell_{\max}+1)^2=KM^2$ components per atom, independently of the number of neighbours. 
This finite aggregation is the first compression of the local environment.

\subsection{Equivariant moment refinement}\label{nav:sub4}

A moment-refinement block processes the accumulated state $\mathbf P_i^{[t]}$ at fixed width $K$. 
It combines a linear channel mixer with low-rank tensor products of the current moments. 
Both branches act on the same incoming state. 
For the linear branch,
\begin{equation}
\boldsymbol\Delta_{ik,\mathrm{lin}}^{[t],(\ell)}
=\sum_{k'=1}^{K}
\left(W_{\mathrm{mix},kk'}^{[t],(\ell)}-\delta_{kk'}\right)
\mathbf P_{ik'}^{[t],(\ell)} .
\label{eq:si-same-order-mix}
\end{equation}
Subtracting the identity makes this the increment associated with the channel mixer; the residual connection is added separately below.

Let $\mathcal S$ denote the selected source-order pairs. 
The admissible tensor-product signatures are
\begin{equation}
\begin{aligned}
\mathcal T(\mathcal S)=\bigl\{\,
\tau=(\ell_1,\ell_2\!\rightarrow\!\ell)\ \big|\;&
(\ell_1,\ell_2)\in\mathcal S,\\[-2pt]
&|\ell_1-\ell_2|\le\ell
\le\min(\ell_1+\ell_2,\ell_{\max}),\\[-2pt]
&\ell_1+\ell_2-\ell\equiv0\pmod 2
\,\bigr\}.
\end{aligned}
\label{eq:si-signature-set}
\end{equation}
The reported ordinary blocks use the following source-pair catalogues, indexed by the number of angular orders $M$:
\[
\begin{aligned}
\mathcal S_4={}&\{(1,1),(1,2),(1,3),(2,2),(2,3),(3,3)\},\\
\mathcal S_6={}&\mathcal S_4\cup\{(1,4),(2,4),(2,5),(3,4),(3,5),(4,4)\},\\
\mathcal S_8={}&\mathcal S_6\cup\{(1,5),(1,6),(2,6),(2,7),(3,6),\\
&\hspace{59pt}(3,7),(4,5),(4,6),(5,5)\}.
\end{aligned}
\]
Every target order satisfying Equation~\eqref{eq:si-signature-set} is retained for each listed pair, giving 11, 29 and 59 signatures for $M=4$ ($\ell_{\max}=3$), $M=6$ ($\ell_{\max}=5$) and $M=8$ ($\ell_{\max}=7$), respectively.
No pair has a scalar source. 
All stages use the same coupling catalogue for both readouts, whether or not moment attention follows the block.

Write $\mathcal L_{\mathrm{src}}$ for the orders occurring in $\mathcal S$ and $\mathcal T_\ell$ for the signatures in $\mathcal T(\mathcal S)$ targeting order $\ell$. 
Each source order is projected to channels $\rho=1,\ldots,K_{\mathrm{tp}}$, with $K_{\mathrm{tp}}=R$ in the architecture tables:
\begin{equation}
\mathbf P_{i\rho}^{[t],(\ell),\mathrm{src}}
=\sum_{k=1}^{K}W_{\mathrm{in},k\rho}^{[t],(\ell)}
\mathbf P_{ik}^{[t],(\ell)},
\qquad \ell\in\mathcal L_{\mathrm{src}}.
\label{eq:si-low-rank-source}
\end{equation}
For $\tau=(\ell_1,\ell_2\!\rightarrow\!\ell)$, a fixed coupling tensor constructs the equivariant product
\begin{equation}
\mathbf U_{i\rho}^{[t],\tau}
=\mathcal C_{\ell_1,\ell_2\rightarrow\ell}
\!\left(\mathbf P_{i\rho}^{[t],(\ell_1),\mathrm{src}},
         \mathbf P_{i\rho}^{[t],(\ell_2),\mathrm{src}}\right)
\in\mathbb R^{2\ell+1}.
\label{eq:si-tensor-product}
\end{equation}
The source projections are shared by signatures using the same source order within a layer. 
Both factors use the same low-rank channel $\rho$. 
The parity rule in Equation~\eqref{eq:si-signature-set} retains the natural parity $(-1)^\ell$, making the product $O(3)$-equivariant.

The products targeting each output order are summed, normalized and projected back to width $K$:
\begin{equation}
\boldsymbol\Delta_{ik,\mathrm{bilin}}^{[t],(\ell)}
=\sum_{\rho=1}^{K_{\mathrm{tp}}}
W_{\mathrm{out},\rho k}^{[t],(\ell)}
\frac{1}{\sqrt{|\mathcal T_\ell|}}
\sum_{\tau\in\mathcal T_\ell}\mathbf U_{i\rho}^{[t],\tau},
\qquad |\mathcal T_\ell|>0.
\label{eq:si-p-block-delta}
\end{equation}
For $\mathcal T_\ell=\varnothing$ the bilinear increment is $0$.
The ordinary block in Equation~\eqref{eq:si-trunk-flow} is therefore
\begin{equation}
\widetilde{\mathbf P}_{ik}^{[t],(\ell)}
=\mathbf P_{ik}^{[t],(\ell)}
+\eta_{\mathrm P}^{[t]}
 \boldsymbol\Delta_{ik,\mathrm{lin}}^{[t],(\ell)}
+\eta_{\mathrm P}^{[t]}
 \boldsymbol\Delta_{ik,\mathrm{bilin}}^{[t],(\ell)}.
\label{eq:si-p-block-update}
\end{equation}
Both branches share a learned $\eta_{\mathrm P}^{[t]}$ across angular orders and all components of each irrep, preserving equivariance.
Repeated tensor products combine correlations already retained in the centre's moment state before invariant readout, without revisiting individual neighbours.

\subsection{Moment attention}\label{nav:sub5}\par 
At each selected stage $t\in\mathcal J$, the ordinary refinement block first updates the moments at the central atom.
Moment attention uses these updated moments $\widetilde{\mathbf P}_i^{[t]}$ to reweight the original neighbours and sum their contributions again.
The configuration files call this combined stage an \emph{aggregated P-block}.

Each attention update uses the original edge features $H_{ek}$ and $\mathbf Y_{e\ell}$ from Equations~\eqref{eq:si-radial-encoder} and \eqref{eq:si-harmonic-basis} without accessing neighbouring atoms' hidden states.
In the main text, the pair index $ij$ suppresses the periodic-image index of edge $e=(i,j,\mathbf n_e)$.
The channel index $h$ and edge weight $g_{eh}^{[t]}$ below correspond to $k$ and $g_{ij,k}^{(t)}$ there.

For attention channels $h=1,\ldots,K$, define
\begin{equation}
\begin{aligned}
\mathbf Q_{ih}^{[t],(\ell)}
&=\sum_{k=1}^{K}W_{\mathrm Q,kh}^{[t],(\ell)}
       \widetilde{\mathbf P}_{ik}^{[t],(\ell)},\\
\kappa_{eh}^{[t],(\ell)}
&=\sum_{k=1}^{K}H_{ek}W_{\mathrm K,kh}^{[t],(\ell)},\\
v_{eh}^{[t]}
&=\sum_{k=1}^{K}H_{ek}W_{\mathrm V,kh}^{[t]} .
\end{aligned}
\label{eq:si-attention-projections}
\end{equation}
The query $\mathbf Q$ is equivariant, whereas the radial key $\kappa$ and value $v$ are invariant scalars. 
All projections are bias-free. 
Query and key matrices depend on $\ell$; the value matrix is shared across orders.
For an edge $e$ centred on $i$, the matching score and edge weight are
\begin{equation}
s_{eh}^{[t]}
=\frac{1}{\sqrt{\ell_{\max}+1}}
 \sum_{\ell=0}^{\ell_{\max}}
 \frac{\kappa_{eh}^{[t],(\ell)}
       \bigl(\mathbf Q_{ih}^{[t],(\ell)}\bigr)^{\mathsf T}\mathbf Y_{e\ell}}
      {\sqrt{2\ell+1}},
\qquad
g_{eh}^{[t]}=\tanh s_{eh}^{[t]} .
\label{eq:si-attention-score}
\end{equation}
The factors normalize the sums over irrep components and angular orders.
The $\tanh$ map gives a signed weight $g_{eh}^{[t]}\in(-1,1)$ independently for each edge and channel. It is not normalized over neighbours.  Each channel's score combines all retained
angular orders, and the resulting weight is shared by those orders.

The weighted values form a new irreducible moment tensor,
\begin{equation}
\mathbf P_{ih}^{[t],(\ell),\mathrm{att}}
=\sum_{e\in\mathcal N_i}
 v_{eh}^{[t]}g_{eh}^{[t]}\mathbf Y_{e\ell}.
\label{eq:si-attention-reread}
\end{equation}
An order-specific output projection gives the attention increment in Equation~\eqref{eq:si-trunk-flow}:
\begin{equation}
\boldsymbol\Delta_{ik,\mathrm{agg}}^{[t],(\ell)}
=\sum_{h=1}^{K}W_{\mathrm O,hk}^{[t],(\ell)}
 \mathbf P_{ih}^{[t],(\ell),\mathrm{att}},
\qquad
\mathbf P_{ik}^{[t+1],(\ell)}
=\widetilde{\mathbf P}_{ik}^{[t],(\ell)}
+\eta_{\mathrm{agg}}^{[t]}
 \boldsymbol\Delta_{ik,\mathrm{agg}}^{[t],(\ell)} .
\label{eq:si-attention-update}
\end{equation}
Here $\eta_{\mathrm{agg}}^{[t]}$ is an independently learned scalar, separate from the ordinary block's shared residual scalar.
The keys and values inherit the cutoff envelope from $H_{ek}$.

The query and the corresponding edge-basis vector transform under the same orthogonal matrix $\mathbf D^{(\ell)}$, so their inner product in Equation~\eqref{eq:si-attention-score} is invariant under rotations and reflections.
The edge weights are invariant and the reaggregated moments are equivariant.
The attention update therefore preserves equivariance and retains the same neighbourhood $\mathcal N_i$ at every depth.

Environments with the same initial moment state can produce different weighted sums when their individual edge features differ. 
Reaggregation can therefore change which distinctions are retained at fixed state dimensions. 
Its weights also introduce nonlinear angular dependence. 
The model explicitly represents angular orders up to $\ell_{\max}$.
Nonlinear weighting and energy regression can combine them into more complex angular functions.

\subsection{Standard readout (C0)}\label{nav:sub6}

The standard readout (C0) retains the scalar moments and takes the squared norm of each higher-order moment.
\begin{equation}
G_{ik0}^{\mathrm{base}}=\overline P_{ik,1}^{(0)},
\qquad
G_{ik\ell}^{\mathrm{base}}
=\sum_{q=1}^{2\ell+1}
 \left(\overline P_{ik,q}^{(\ell)}\right)^2,
\quad \ell=1,\ldots,\ell_{\max}.
\label{eq:si-base-invariants}
\end{equation}
The squared norm includes all components without angular averaging.
The superscript $\mathrm{base}$ identifies this standard invariant contraction, which also applies to refined moments.
The C0 descriptor includes the element fingerprint (Equation~\eqref{eq:si-orbital-order}).
\begin{equation}
\begin{aligned}
\mathbf G_i^{\mathrm{base}}
&=\operatorname{vec}\{G_{ik\ell}^{\mathrm{base}}\}_{k,\ell},\\
\mathbf G_i
&=\mathbf G_i^{\mathrm{base}}\mathbin{\Vert}\boldsymbol\chi(Z_i),
\qquad
d_G=K(\ell_{\max}+1)+d_\chi=KM+d_\chi.
\end{aligned}
\label{eq:si-final-descriptor}
\end{equation}
At fixed $K$ and $\ell_{\max}$, the C0 descriptor dimension is independent of neighbour count, refinement depth and the number of attention stages.

For an unrefined base DTM model, substituting Equation~\eqref{eq:si-initial-moment} into the base readout and applying the angular addition theorem gives
\begin{equation}
G_{ik\ell}^{\mathrm{base}}=
\zeta_\ell\sum_{e,e'\in\mathcal N_i}H_{ek}H_{e'k}
\operatorname{Leg}_\ell(\widehat{\mathbf r}_e^{\mathsf T}\widehat{\mathbf r}_{e'}),
\qquad \ell>0.
\label{eq:si-base-correlations}
\end{equation}
The diagonal terms contain radial information, and distinct edges describe angular correlations between pairs of neighbours of the same centre.
Nonlinear regression combines these statistics into dependencies involving more neighbours. 
This density/moment interpretation connects DTM with the broader design space of atom-centred potentials~\cite{batatia2025design}. 
The compact norm readout exposes only part of the correlations available in the moment state. 
For two vector channels $u$ and $v$, their individual squared norms leave $u\cdot v$ unspecified, whereas $\|u+v\|^2$ contains this cross term.
Learned channel transformations before norm reduction can expose such information. Bilinear equivariant products further combine angular orders. 
An ordinary refinement stack therefore learns which invariant correlations reach the shared energy network.

Initial aggregation and invariant readout place different limits on the representation~\cite{pozdnyakov2020incompleteness}.
For a fixed encoder, two environments with the same central element and identical initial moments give the same atomic energy when processed by the same ordinary refinement blocks and energy network.
Refinement can expose correlations within those moments.
Training can change which information the encoder retains.
This argument concerns the energy assigned to the central atom.
Total energies also include the contributions of neighbouring centres, and forces depend on how those energies vary with atomic positions.

\subsection{Cubic-augmented readout (C1)}\label{nav:sub7}\label{sec:si-cubic-readout}

C1 augments the C0 geometric features with scalar contractions of three moment blocks.
A learned matrix $U\in\mathbb R^{K\times R_{\rm c}}$ projects the radial channels of the final moment state,
\begin{equation}
\widehat{\mathbf P}_{i\rho}^{(\ell)}
=\sum_{k=1}^{K}U_{k\rho}\overline{\mathbf P}_{ik}^{(\ell)},
\qquad \rho=1,\ldots,R_{\rm c}.
\label{eq:si-cubic-projection}
\end{equation}
The same matrix is shared across angular orders and contraction families. 
$R_{\rm c}$ is the cubic projection width, separate from the P-block rank $R$. 
All three factors in each cubic invariant use the same projected channel $\rho$.

A family label $f=\ell_1\ell_2\ell_3$ denotes an unordered triple of positive angular orders. 
The implemented catalogue is
\begin{equation}
\mathcal F_{\rm supported}
=\{112,123,134,145,222,224,233,235,244,255,334,345,444,455\}.
\label{eq:si-cubic-families}
\end{equation}
The selected set $\mathcal F$ retains the families whose largest order is at most $\ell_{\max}$. Each triple satisfies the angular-momentum triangle condition and has even total parity. 
To form its scalar, a repeated order is coupled with itself to the remaining order.
For three distinct orders, the two smaller orders couple to the largest.
Let $(a_f,b_f\to c_f)$ denote this canonical coupling. 
The invariant is
\begin{equation}
G_{i\rho f}^{\rm cub}
=\alpha_f\,
\mathcal C_{a_f,b_f\to c_f}
\!\left(\widehat{\mathbf P}_{i\rho}^{(a_f)},
        \widehat{\mathbf P}_{i\rho}^{(b_f)}\right)^{\mathsf T}
\widehat{\mathbf P}_{i\rho}^{(c_f)}.
\label{eq:si-cubic-invariant}
\end{equation}
The fixed trace-coupling tensors use the same Cartesian harmonic basis as Equation~\eqref{eq:si-tensor-product}; their normalization is fixed by that basis. 
There is no division by the moment norms or neighbour count. 
Each $\alpha_f$ is a learned scalar shared across projected channels. 
The controls use $R_{\rm c}=16$, orthogonal initialization of $U$, and $\alpha_f=0.01$ at initialization; these initializations impose no orthogonality or sign constraint during subsequent training.

The augmented descriptor concatenates the standard invariants, cubic features and the element fingerprint:
\begin{equation}
\begin{aligned}
\mathbf G_i^{\rm C1}
&=\mathbf G_i^{\rm base}\mathbin{\Vert}
  \operatorname{vec}\{G_{i\rho f}^{\rm cub}\}_{f,\rho}
  \mathbin{\Vert}\boldsymbol\chi(Z_i),\\
d_G^{\rm C1}&=KM+|\mathcal F|R_{\rm c}+d_\chi.
\end{aligned}
\label{eq:si-cubic-descriptor}
\end{equation}
At $M=4$, the cubic readout uses $\{112,123,222,233\}$ and appends 64 features, while at $M=6$ it uses all 14 listed families and appends 224.
Cubic augmentation changes the readout dimension and first energy-layer input width, but not the stored moment dimension or atomic-energy neighbourhood. 

For the ordinary P-block with an input-independent residual scalar, let $\mathcal L(\mathbf P)$ include the identity and linear update and let $\mathcal B(\mathbf P,\mathbf P)$ denote the scaled bilinear update. 
For each non-scalar output block, the C0 readout gives

\begin{equation}
\begin{aligned}
\widetilde{\mathbf P}&=\mathcal L(\mathbf P)+\mathcal B(\mathbf P,\mathbf P),\\
\|\widetilde{\mathbf P}\|^2
&=\|\mathcal L(\mathbf P)\|^2
+2\left\langle\mathcal L(\mathbf P),\mathcal B(\mathbf P,\mathbf P)\right\rangle
+\|\mathcal B(\mathbf P,\mathbf P)\|^2.
\end{aligned}
\label{eq:si-refinement-readout-expansion}
\end{equation}
The cubic cross term can expose an angular correlation that the separate norms of a vector $v$ and a symmetric traceless tensor $Q$ leave undetermined. 
A $1\otimes1\to2$ update forms the traceless part of $vv^{\mathsf T}$. 
Its inner product with a linear second-order component $Q$ equals $v^{\mathsf T}Qv$, a 112-type invariant that depends on the alignment of $v$ with $Q$.
Refinement followed by C0 can therefore expose a correlation of the same type as an explicit C1 feature.
The two operations use different learned channel projections and can retain different combinations of these correlations.
Equation~\eqref{eq:si-refinement-readout-expansion} applies to the ungated ordinary block at fixed parameters. 
Input-dependent gates, normalization and attention weights fall outside this expansion, and the nonlinear energy network can generate higher body orders. 
The HEA16 cubic-readout controls test how much the explicit contractions add before and after one ordinary block.

\subsection{Atomic energies, symmetries and derivatives}\label{nav:sub8}

The energy network is shared by every atom and receives the complete invariant descriptor as its input,
\begin{equation}
\mathbf x_i^{(0)}=\mathbf G_i.
\label{eq:si-energy-input}
\end{equation}
Each hidden layer applies an affine transformation and a SiLU activation, with a residual connection when the input and output widths match:
\begin{equation}
\mathbf x_i^{(u+1)}=
\begin{cases}
\mathbf x_i^{(u)}
+\operatorname{SiLU}(\mathbf W_u\mathbf x_i^{(u)}+\mathbf b_u),
&\text{with a residual},\\
\operatorname{SiLU}(\mathbf W_u\mathbf x_i^{(u)}+\mathbf b_u),
&\text{otherwise}.
\end{cases}
\label{eq:si-energy-residual}
\end{equation}
Let $\mathbf x_i:=\mathbf x_i^{(L_{\mathrm E})}$ denote the final hidden state. 
A bias-free linear output layer and a trainable element-specific reference energy give
\begin{equation}
\varepsilon_i=E_{Z_i}^{(0)}
+\mathbf w_{\mathrm E}^{\mathsf T}\mathbf x_i,
\qquad
E_{\mathrm{tot}}(\mathbf Z,\mathbf R,\mathbf A)
=\sum_{i=1}^{N}\varepsilon_i.
\label{eq:si-total-energy}
\end{equation}

Relative displacements in Equation~\eqref{eq:si-periodic-edge} give translation invariance, and the lattice-image graph enforces periodicity.
Shared edge and energy networks, neighbour sums and the final atomic sum give invariance to simultaneous permutations of the pairs $(Z_i,\mathbf r_i)$.
The parity-preserving irreducible couplings and scalar contractions give rotational and reflection invariance. 
The finite cutoff makes the atomic map local, while the sum in Equation~\eqref{eq:si-total-energy} makes the total energy extensive.

For the production architecture defined above, forces and stress are conservative derivatives of the same scalar model. 
For the pooled-summary control (S), Section~\ref{sec:si-control-cost} specifies how cutoff crossings affect these derivatives.

Let $\mathbf F$ be the $N\times3$ matrix whose rows are the Cartesian forces.
For a right-handed cell in the row-vector convention specified above, the implemented  stress is
\begin{equation}
\mathbf F
=-\nabla_{\mathbf R}E_{\mathrm{tot}},
\qquad
\boldsymbol\sigma
=\frac{-\mathbf F^{\mathsf T}\mathbf R
+(\nabla_{\mathbf A}E_{\mathrm{tot}})^{\mathsf T}\mathbf A}{V},
\qquad
V=\det\mathbf A>0.
\label{eq:si-energy-derivatives}
\end{equation}
The cell derivative in Eq.~\eqref{eq:si-energy-derivatives} is taken at fixed Cartesian positions $\mathbf R$.
We report stress in the Atomic Simulation Environment (ASE) Voigt order $(xx,yy,zz,yz,xz,xy)$, with pressure $p=-\operatorname{tr}(\boldsymbol\sigma)/3$.
Reference~\cite{chen2020tensoralloy} gives the derivation.

The derivatives include every dependence on geometry: initial moments, queries, radial keys and values, edge weights and reaggregated moments. 
Retaining original edge features for reuse does not detach them from the energy derivative. 
A fused or explicit derivative implementation evaluates this same chain rule.

Neither ordinary refinement nor moment attention reads an updated moment state at a neighbouring centre. 
Each $\varepsilon_i$ therefore depends only on $Z_i$ and $\mathcal N_i(r_{\mathrm c})$ at all depths, a form of strict atomic-energy locality also present in other local equivariant architectures~\cite{musaelian2023allegro}. 
This statement concerns the receptive field of an atomic energy.
The force on an atom collects derivatives of all atomic energies involving that atom, and its dependence can extend across overlapping neighbourhoods.

\subsection{Parameter allocation and representation size}\label{nav:sub9}

\label{sec:si-parameter-accounting}

Parameter count, central-state dimension and execution cost describe different resources. 
Let $n_{\rm enc}$ and $n_{\rm head}$ denote the numbers of trainable parameters in the shared encoder and head, $n_{\rm P}^{(t)}$ the ordinary refinement parameter count, and $n_{\rm MA}^{(t)}$ the attention parameter count.
Let $n_{\rm cub}=0$ for C0 and $n_{\rm cub}=KR_{\rm c}+|\mathcal F|$ for the shared projector and family scales of C1. 
The total number of trainable parameters is
\begin{equation}
 n_\theta=n_{\rm enc}+n_{\rm head}+n_{\rm E0}+n_{\rm cub}
 +\sum_{t=0}^{L-1}n_{\rm P}^{(t)}
 +\sum_{t\in\mathcal J}n_{\rm MA}^{(t)}.
 \label{eq:si-param-total}
\end{equation}
Here $n_{\rm E0}$ counts the trainable elemental reference energies. 
The element fingerprints, harmonic coupling tensors and interpolation samples are not learned parameters.
We count a shared matrix once irrespective of how many edges, atoms or components within an angular order use it.
The central state contains $KM^2$ components per atom. 
The C0 geometric readout has $KM$ invariants; C1 adds $|\mathcal F|R_{\rm c}$. 
These dimensions do not increase when a fixed-width refinement stage is added.

For a dense network with layer widths $d_0,\ldots,d_{H+1}$, the matrix and bias count is
\begin{equation}
 n_{\rm MLP}=\sum_{r=0}^{H}(d_rd_{r+1}+b_rd_{r+1}),
 \label{eq:si-mlp-count}
\end{equation}
where $b_r=1$ for a layer with a bias. 
Residual connections add no parameters in the stated MLP implementation. 
The radial network input has 16 Gaussian features and $5\times18$ chemical features. 
Every radial linear layer has a bias. 
The head input has $KM+18$ features for C0 and $KM+|\mathcal F|R_{\rm c}+18$ for C1. 
All hidden layers have biases, and the final scalar linear layer has no bias. 
The supplied implementation allocates 113 learned elemental energy-offset entries, including the entry at index 0.
This allocated table is counted consistently even when some entries are unused by a dataset.

At fixed backbone and hidden-layer widths, cubic augmentation adds
\begin{equation}
\Delta n_\theta^{\rm C1-C0}
=KR_{\rm c}+|\mathcal F|+|\mathcal F|R_{\rm c}w_{\rm head},
\label{eq:si-cubic-parameter-increment}
\end{equation}
where $w_{\rm head}$ is the first hidden-layer width. 
The last term is the extra input-layer weight count; no new bias is required. 
For $K=32$, $R_{\rm c}=16$ and $w_{\rm head}=128$, the increments are 8,708 parameters at $M=4$ and 29,198 at $M=6$, with or without one ordinary P-block.
Therefore, C0/C1 comparisons are not fixed-parameter comparisons.

An ordinary block contains a $K\times K$ same-order map per angular order, $K\times R$ source projections, $R\times K$ target projections and a residual scalar.
If the coupling catalogue uses $n_{\rm src}$ source orders and $n_{\rm dst}$ target orders, its count is
\begin{equation}
 n_{\rm P}=MK^2+(n_{\rm src}+n_{\rm dst})KR+1.
 \label{eq:si-p-count}
\end{equation}
For the catalogues at $M=4,6,8$, the source orders are $1,\ldots,M-1$ and the target orders are $0,\ldots,M-1$.
Thus the catalogue contributes $(2M-1)KR$ projection parameters. 
The coupling tensors are fixed.

Each moment-attention update contains $3M$ order-specific matrices, one shared value matrix (all $K\times K$) and one residual scalar.
\begin{equation}
 n_{\rm MA}=(3M+1)K^2+1.
 \label{eq:si-ma-count}
\end{equation}
The three controls use projection matrices of the same dimensions and one learned residual scale per update.
At $K=32$ and $M=6$, each moment-attention update has 19,457 trainable parameters.
At $K=32$ and $M=8$, it has 25,601.
At $M=8$, four attention updates contribute 102,404 trainable parameters while preserving the stored central-state width.

\paragraph{Trainable parameter counts.}
We count trainable parameters using the TensorAlloy 2.11.2 module definitions, excluding the 16 fixed Gaussian centres and their fixed width.
The architecture tables report the totals for the complete model families, and Supplementary Table~\ref{tab:hea16-mechanism} gives the matched-control counts.
Radial tabulation storage depends on the grid and chemical-pair coverage and is separate from trainable parameters~\cite{lu2022dpcompress}.

For the four highlighted standard-readout OMat24 models, the allocations are
\[
\begin{array}{lrrrr}
\text{Model}&\text{Encoder}&\text{Energy head}&\text{Refinement}&\text{Attention}\\
\text{K16-M3 (DTM-OMat24-S)}&588{,}304&16{,}832&0&0\\
\text{M6-P1 (DTM-OMat24-M)}&596{,}512&185{,}856&28{,}673&0\\
\text{M8-P2+MA (DTM-OMat24-L)}&596{,}512&666{,}624&77{,}826&51{,}202\\
\text{M8-P4+MA (DTM-OMat24-XL)}&596{,}512&666{,}624&155{,}652&102{,}404
\end{array}
\]
The encoder and energy head together account for approximately 100\% of the trainable parameters in the base model and 83.0\% in the deepest model.
Adding 113 elemental references to the listed allocations gives the totals in Supplementary Table~\ref{tab:omat-registry}.
\FloatBarrier

\subsection{Matched operator controls and computational cost}\label{nav:sub10}

\label{sec:si-control-cost}

The pooled-summary control (S), initial-moment control ($Q_0$) and standard moment attention (A) implement two constructions for permutation-invariant set functions~\cite{zaheer2017deepsets,lee2019settransformer}.
The dimension of the pooled state limits the capacity of such representations~\cite{wagstaff2022sets}. 
A uses the edge-resolved moment attention of Section~\ref{nav:sub5}. 
S retains the post-block query in Equation~\eqref{eq:si-attention-projections} and obtains its keys and values by projecting the initial moments $\mathbf P_i^{\rm init}=\mathbf P_i^{[0]}$.

\begin{equation}
\begin{aligned}
\overline{\mathbf K}_{ih}^{[t],(\ell)}&=\sum_k W_{\mathrm K,kh}^{[t],(\ell)}\mathbf P_{ik}^{\mathrm{init},(\ell)},\\
\overline{\mathbf V}_{ih}^{[t],(\ell)}&=\sum_k W_{\mathrm V,kh}^{[t]}\mathbf P_{ik}^{\mathrm{init},(\ell)},\\
\overline{s}_{ih}^{[t]}&=\frac{1}{\max(z_i,1)\sqrt M}\sum_{\ell=0}^{M-1}\frac{\bigl(\mathbf Q_{ih}^{[t],(\ell)}\bigr)^{\mathsf T}\overline{\mathbf K}_{ih}^{[t],(\ell)}}{\sqrt{2\ell+1}},\\
\boldsymbol\Delta_{ik,\mathrm S}^{[t],(\ell)}&=\sum_h W_{\mathrm O,hk}^{[t],(\ell)}\tanh\!\bigl(\overline{s}_{ih}^{[t]}\bigr)\overline{\mathbf V}_{ih}^{[t],(\ell)} .
\end{aligned}
\label{eq:si-summary-control}
\end{equation}
Here $z_i=|\mathcal N_i(r_{\mathrm c})|$ counts physical neighbours inside the cutoff, excluding padding and skin-only neighbours. 
We hold this integer fixed when differentiating the energy on a given neighbour graph. 
At a cutoff crossing, the crossing edge's radial contribution vanishes smoothly, but the change in $z_i$ rescales the gate acting on the remaining pooled contributions. 
If these are nonzero, the energy can jump. Forces and stress for S are therefore derivatives on the current graph. 
We use S for static-error and execution-cost comparisons. 
Both A and $Q_0$ omit the neighbour-count divisor.

The residual update retains the learned scalar in Equation~\eqref{eq:si-attention-update}.
Let $W$ denote either the order-specific key matrix $W_{\mathrm K}^{[t],(\ell)}$ or the shared value matrix $W_{\mathrm V}^{[t]}$ in Equation~\eqref{eq:si-summary-control}.
Substituting the initial moments from Equation~\eqref{eq:si-initial-moment} and exchanging the channel and neighbour sums gives
\begin{equation}
 \sum_k W_{kh}\mathbf P_{ik}^{[0],(\ell)}
 = \sum_k W_{kh}\sum_{e\in\mathcal N_i}H_{ek}\mathbf Y_{e\ell}
 = \sum_{e\in\mathcal N_i}\left(\sum_k H_{ek}W_{kh}\right)\mathbf Y_{e\ell}.
 \label{eq:si-pooled-projections}
\end{equation}
S can therefore compute its pooled keys and values from the stored initial moments and complete the update at the central atom.
For scalar values and scores, the distinction from edge weighting is
\begin{equation}
 \sum_j v_j\tanh(s_j)
 \ne \tanh\!\left(\frac1z\sum_j s_j\right)\sum_j v_j
 \label{eq:si-nonlinear-pooling}
\end{equation}
in general.

At each attention stage, $Q_0$ forms its query from the initial moments $\mathbf P_i^{[0]}$.
As in A, the stage computes its keys and values from the original radial features and adds the update to the current moment state.
Coordinate and parameter derivatives through the initial moments remain active.
Each stage has its own projection matrices, so initial-query computations cannot generally be replaced by a shared projection.
A draws its query from the corresponding post-block state.

For a fixed $K,M$ and coupling catalogue, ordinary blocks operate on centre states. 
The leading forward operation count per neighbour-resolved attention update is
\begin{equation}
 C_{\rm MA}=O\!\left(NK^2M^2+E[(M+1)K^2+KM^2]\right),
 \label{eq:si-refresh-work}
\end{equation}
where $E$ is the directed edge count. 
The order-specific key maps and shared value map contribute edge projections.
Contractions, gates and reaggregation contribute the remaining edge work.
S computes its pooled update from the central moments, so its work scales with the number of centres at fixed widths after the initial aggregation.
The two edge-resolved variants have the same leading operation classes.
Their runtime also depends on intermediate reuse, kernel layout and the derivative evaluation needed for forces and stress. 

\FloatBarrier

\section{HEA16 models and controlled refinement study}
\label{sec:si-hea16}
\mergedsubsection{HEA16 dataset, architecture study and training}\label{nav:sub11}\label{nav:sub12}

HEA16 comprises 105,464 UNEP-v1 structures containing Ag, Al, Au, Cr, Cu, Mg, Mo, Ni, Pb, Pd, Pt, Ta, Ti, V, W and Zr. 
A fixed split (seed 42) assigns 94,918 structures to training, 5,273 to validation and 5,273 to testing. 
The training set contains 7,015 unary and 87,903 binary structures. 

The 25-architecture study uses a 6\,\AA{} cosine cutoff, 16 Gaussian inputs, a shared $3\times256$ radial encoder and the standard invariant readout. 
The 21 core models use 32 radial channels and a $3\times128$ energy network.
For each of the settings $M=4,6,8$, we compare one base model with six models containing one, two or four rank-64 blocks, with or without moment attention after every block.
The grid also includes four radial-width and energy-head controls (Supplementary Table~\ref{tab:hea16-family-endpoints}).
In checkpoint names, $M=\ell_{\max}+1$. 
In the external model comparisons, five of the six representatives use a $3\times128$ energy network.
K16-M3 uses $4\times64$.

For structure $s$ with $N_s$ atoms, the energy error per atom $\delta e_s$, Cartesian force errors $\Delta F_{sa\alpha}$ and Voigt stress errors $\Delta\sigma_{sc}$ enter main Eq.~(5) through
\begin{equation}
\begin{aligned}
\ell_{E,s}&=\sqrt{\delta e_s^2+\epsilon},\\
\ell_{F,s}&=\sqrt{\frac{1}{3N_s}\sum_{a,\alpha}(\Delta F_{sa\alpha})^2+\epsilon},\\
\ell_{\sigma,s}&=\sqrt{\frac{1}{6}\sum_{c=1}^{6}(\Delta\sigma_{sc})^2+\epsilon}.
\end{aligned}
\label{eq:si-training-objective}
\end{equation}
We use $\epsilon=10^{-8}$ with energy errors in eV/atom, force errors in eV/\AA{} and stress errors in eV/\AA{}\textsuperscript{3}.
At update $t=0,\ldots,T-1$, the loss prefactors for energy, force and stress follow

\begin{equation}
\lambda_x(t)=\lambda_{x,0}^{1-p(t)}\lambda_{x,T}^{p(t)},\qquad
p(t)=\min\!\left(1,\max\!\left(0,\frac{t}{T-1}\right)\right).
\label{eq:si-loss-schedule}
\end{equation}

The initial and final weights are 0.1 and 100 for energy, 100 and 10 for force, and 160 and 40 for stress. 
We evaluated validation errors every 10,000 updates during the 500,000-update training schedule described in Methods.

All 150 fits reached the terminal checkpoint. 
The architecture survey reports the arithmetic mean and range of the six terminal validation RMSEs.
All six runs used FP32 base operations.
Seeds 1--5 used BF16 radial computation and seed 6 used FP32 radial computation, so the six-run range includes this precision difference.
The post-training test and external evaluations use the terminal checkpoints at seeds 1 and 3. 
Throughput uses the seed-1 checkpoint.

\subsection{Evaluation datasets and metrics}\label{nav:sub13}
\label{sec:si-hea16-evaluation}
\label{sec:si-mechanism-heldout}
The testing subset contains 5,273 structures, 339,173 atoms and 1,017,519 Cartesian force components. 
The five external datasets used for formation energies are the published Materials Project ternary, GNoME, PdCuNi, MoTaVW and multicomponent-alloy sets associated with Fig.~2d, Fig.~2e and Figs.~S1--S3 of Ref.~\cite{song2024unep}.
They contain 482, 444, 123, 1,215 and 912 structures, respectively. 
For forces, we use the PdCuNi (S1), MoTaVW (S2) and multicomponent-alloy (S3) sets, which contain 36,189, 156,891 and 114,462 Cartesian force components, respectively.
Each of the 307,542 components from the 2,250 structures contributes equally to the combined S1--S3 force score.
Validation scores recorded during training use the precision of that run.
For post-training evaluations on the HEA16 splits and external alloys, we use FP32 base and radial operations, disable TF32 and accumulate errors in float64.

\paragraph{Formation energies.}
Formation energies use the published fixed geometries and model-specific relaxed elemental references, separately from the total-energy and force evaluations above:
\begin{equation}
e_{\rm f}^{(m)}(\mathcal X)=
\frac{E^{(m)}(\mathcal X)-\sum_Z N_Z\mu_Z^{(m)}}{N}.
\label{eq:si-formation-energy}
\end{equation}
The reference $\mu_Z^{(m)}$ is obtained from the model's canonical elemental phase in a $3\times3\times3$ supercell, relaxed with zero-pressure hydrostatic \texttt{ExpCellFilter}, force tolerance 0.01\,eV/\AA{} and at most 500 steps per optimizer attempt. 
For the NEP elemental references, we used BFGS when the initial relaxation did not converge.
Section~\ref{sec:si-cubic-external} gives the elemental-reference protocol for the cubic-readout models.
The published DFT formation values have their own reference convention. 
Formation errors weight structures equally.

\paragraph{Paired comparisons.}
Paired comparisons preserve structure and atom identities. 
For structure $s$, let $e_m(s)$ be the RMS of its $3N_s$ Cartesian force errors. 
The per-structure gains $e_{\rm S}(s)-e_{\rm A}(s)$ and $e_{Q_0}(s)-e_{\rm A}(s)$ are positive when A is more accurate. 
Population gains instead subtract component-pooled errors. 
We resample whole structures jointly across models in 2,000 paired bootstrap replicates and recompute the pooled score in each replicate.
For S1--S3 combined, we resample structures separately within each dataset.
The intervals condition on each fitted checkpoint.
Source-trajectory groups are unidentified, and the two seeds represent two fitted models.

\subsection{Evaluation results}\label{nav:sub14}
\label{sec:si-family-heldout}

Supplementary Table~\ref{tab:hea16-family-endpoints} gives validation, testing and external force RMSE for all 25 architectures and identifies the six named HEA16 models.

\formatHEATable{\begingroup\footnotesize\setlength{\tabcolsep}{2.5pt}
\begin{longtable}{@{}llrrrrrr@{}}
\caption{HEA16 architectures, trainable parameters and force RMSEs (meV/\AA{}).
Configuration columns give $K/M/L$; attention positions $\mathcal J$; and energy-network depth by width.
All use a $3\times256$ radial encoder, 6\,\AA{} cutoff and $R=64$ for refinement.
Validation averages six terminal runs.
The seed-1 and seed-3 results at 500,000 updates include all structures in the testing subset and the external alloy datasets (S1--S3).}\label{tab:hea16-architectures}\label{tab:hea16-family-endpoints}\\
\toprule
Model & $K/M/L$; $\mathcal J$; head & Params & Val. & Test s1 & Test s3 & Ext. s1 & Ext. s3\\
\midrule\endfirsthead
\multicolumn{8}{l}{\textit{Supplementary Table \thetable{} (continued)}}\\
\toprule
Model & $K/M/L$; $\mathcal J$; head & Params & Val. & Test s1 & Test s3 & Ext. s1 & Ext. s3\\
\midrule\endhead\bottomrule\endfoot
K16-M3 & 16/3/0; --; $4\times64$ & 180,033 & 154.097 & 154.784 & 154.916 & 187.856 & 187.760\\
K16-M4 & 16/4/0; --; $4\times64$ & 181,057 & 140.897 & 141.314 & 141.751 & 168.922 & 170.676\\
K16-M4/H3x128 & 16/4/0; --; $3\times128$ & 206,977 & 137.416 & 136.649 & 138.344 & 165.935 & 164.541\\
K32-M4/H3x256 & 32/4/0; --; $3\times256$ & 336,785 & 127.905 & 128.189 & 128.431 & 157.388 & 157.267\\
K32-M4 & 32/4/0; --; $3\times128$ & 219,281 & 131.603 & 132.925 & 133.330 & 160.720 & 159.914\\
M4-P1 & 32/4/1; --; $3\times128$ & 237,714 & 117.007 & 119.236 & 117.184 & 142.400 & 143.078\\
M4-P1+MA & 32/4/1; 0; $3\times128$ & 251,027 & 107.613 & 107.408 & 107.968 & 129.083 & 128.537\\
M4-P2 & 32/4/2; --; $3\times128$ & 256,147 & 114.380 & 113.932 & 115.558 & 139.888 & 139.819\\
M4-P2+MA & 32/4/2; 0,1; $3\times128$ & 282,773 & 102.311 & 101.604 & 101.584 & 122.574 & 123.606\\
M4-P4 & 32/4/4; --; $3\times128$ & 293,013 & 112.761 & 111.933 & 112.057 & 138.733 & 137.626\\
M4-P4+MA & 32/4/4; 0,1,2,3; $3\times128$ & 346,265 & 97.527 & 96.636 & 97.230 & 118.725 & 118.340\\
K32-M6 & 32/6/0; --; $3\times128$ & 227,473 & 121.761 & 120.952 & 121.791 & 146.965 & 147.718\\
M6-P1 & 32/6/1; --; $3\times128$ & 256,146 & 97.769 & 105.593 & 97.760 & 117.617 & 118.171\\
M6-P1+MA & 32/6/1; 0; $3\times128$ & 275,603 & 92.545 & 91.928 & 92.419 & 111.124 & 111.226\\
M6-P2 & 32/6/2; --; $3\times128$ & 284,819 & 94.296 & 93.553 & 93.329 & 113.284 & 113.406\\
M6-P2+MA & 32/6/2; 0,1; $3\times128$ & 323,733 & 87.161 & 94.679 & 100.440 & 103.609 & 104.155\\
M6-P4 & 32/6/4; --; $3\times128$ & 342,165 & 91.243 & 90.887 & 91.868 & 111.239 & 110.692\\
M6-P4+MA & 32/6/4; 0,1,2,3; $3\times128$ & 419,993 & 81.706 & 81.167 & 81.028 & 98.705 & 98.370\\
K32-M8 & 32/8/0; --; $3\times128$ & 235,665 & 119.520 & 118.890 & 119.671 & 144.715 & 144.246\\
M8-P1 & 32/8/1; --; $3\times128$ & 274,578 & 91.921 & 91.611 & 92.171 & 111.411 & 110.881\\
M8-P1+MA & 32/8/1; 0; $3\times128$ & 300,179 & 87.909 & 87.444 & 86.880 & 106.775 & 105.042\\
M8-P2 & 32/8/2; --; $3\times128$ & 313,491 & 87.308 & 87.188 & 86.841 & 105.547 & 105.985\\
M8-P2+MA & 32/8/2; 0,1; $3\times128$ & 364,693 & 81.603 & 81.969 & 81.418 & 98.347 & 98.736\\
M8-P4 & 32/8/4; --; $3\times128$ & 391,317 & 86.786 & 89.389 & 82.846 & 101.800 & 101.939\\
M8-P4+MA & 32/8/4; 0,1,2,3; $3\times128$ & 493,721 & 76.185 & 76.156 & 75.750 & 92.365 & 91.811\\
\end{longtable}\endgroup
}


Across nine architecture pairs evaluated at seeds 1 and 3, adding moment attention lowered force MAE on the HEA16 testing subset in all 18 comparisons and lowered force RMSE in 16 (Supplementary Fig.~\ref{fig:si-attention-transfer}).
The two exceptions were the M6-P2 models at seeds 1 and 3.
Among test structures whose squared force error increased when attention was added, a 48-atom Pt--V structure alone accounted for 96.1\% of the increase at seed 1 and 97.5\% at seed 3.
Its minimum interatomic distance is 1.5614\,\AA{}, and its minimum normalized neighbour distance is 0.5403.
All reported errors on the full testing subset include this structure.

Supplementary Fig.~\ref{fig:si-family-cost-environment}a compares validation force RMSE with throughput, and Supplementary Fig.~\ref{fig:si-family-learning}b,c shows the late increase in validation force RMSE in one of the six ordinary M8-P4 runs.

\subsection{Force gains at fixed parameter count}\label{nav:sub15}\label{sec:si-mechanism}

We compared the pooled-summary control (S), the initial-moment control ($Q_0$) and standard moment attention (A) at the same refinement depth (four blocks) and parameter count (419,993).
Section~\ref{sec:si-control-cost} defines the three updates.
We also evaluated the ordinary model without attention (O) and a model with attention only after the final block (F) to test attention placement.
At seeds 1 and 3, S, $Q_0$ and F copied the initial values of their shared parameters from A and trained with fresh optimizer states and the same sampling order.
Supplementary Table~\ref{tab:hea16-mechanism} gives the validation errors for all five models, and Supplementary Fig.~\ref{fig:si-mechanism}a shows their validation histories.
For the throughput measurements in Supplementary Table~\ref{tab:throughput-hea16}, S uses a fused Tensor Core implementation with TF32 arithmetic, while $Q_0$ and A use the same SIMT implementation of neighbour reaggregation.

{
\footnotesize
\setlength{\tabcolsep}{3.5pt}
\begin{longtable}{@{}llrrrrrrr@{}}
\caption{Matched HEA16 mechanism controls at K32/M6/P4/R64.
We evaluated validation errors after 500,000 training updates.
Energy errors are in meV/atom, force errors in meV/\AA{} and stress errors in GPa.
Mean rows average the validation errors for seeds 1 and 3.
Counts include only trainable parameters.
Section~\ref{sec:si-control-cost} defines S, $Q_0$ and A.
Section~\ref{sec:si-mechanism} describes the HEA16 controls, including O and F, and their initialization.}\label{tab:hea16-mechanism}\\
\toprule
Model & Seed & \shortstack{Trainable\\parameters} & \multicolumn{3}{c}{RMSE} & \multicolumn{3}{c}{MAE}\\
\cmidrule(lr){4-6}\cmidrule(lr){7-9}
 & & & Energy & Force & Stress & Energy & Force & Stress\\
\midrule
\endfirsthead
\multicolumn{9}{l}{\textit{Continued}}\\
\toprule
Model & Seed & \shortstack{Trainable\\parameters} & \multicolumn{3}{c}{RMSE} & \multicolumn{3}{c}{MAE}\\
 & & & Energy & Force & Stress & Energy & Force & Stress\\
\midrule
\endhead
\bottomrule
\endfoot
O & 1 & 342,165 & 8.862 & 91.557 & 0.4299 & 3.595 & 57.488 & 0.1673\\
 & 3 &  & 7.742 & 91.214 & 0.4363 & 3.381 & 57.405 & 0.1662\\
 & Mean &  & 8.302 & 91.386 & 0.4331 & 3.488 & 57.447 & 0.1668\\
\midrule
F & 1 & 361,622 & 9.283 & 88.335 & 0.4038 & 3.395 & 55.648 & 0.1600\\
 & 3 &  & 8.129 & 88.333 & 0.4016 & 3.356 & 55.773 & 0.1582\\
 & Mean &  & 8.706 & 88.334 & 0.4027 & 3.376 & 55.711 & 0.1591\\
\midrule
S & 1 & 419,993 & 8.541 & 85.967 & 0.4009 & 3.296 & 54.464 & 0.1510\\
 & 3 &  & 8.511 & 86.203 & 0.4022 & 3.238 & 54.329 & 0.1487\\
 & Mean &  & 8.526 & 86.085 & 0.4016 & 3.267 & 54.397 & 0.1498\\
\midrule
$Q_0$ & 1 & 419,993 & 7.768 & 83.654 & 0.3696 & 3.190 & 52.950 & 0.1485\\
 & 3 &  & 8.103 & 83.497 & 0.3821 & 3.159 & 53.095 & 0.1509\\
 & Mean &  & 7.936 & 83.576 & 0.3759 & 3.175 & 53.023 & 0.1497\\
\midrule
A & 1 & 419,993 & 9.575 & 81.509 & 0.3618 & 4.365 & 51.914 & 0.1473\\
 & 3 &  & 9.994 & 81.413 & 0.3874 & 3.164 & 51.789 & 0.1503\\
 & Mean &  & 9.785 & 81.461 & 0.3746 & 3.765 & 51.852 & 0.1488\\
\end{longtable}
}

\paragraph{Testing on HEA16 and external alloys.}
At both seeds, A gives the lowest force MAE and RMSE of the three equal-parameter models on the HEA16 testing subset and on S1--S3 combined (Supplementary Tables~\ref{tab:hea16-mechanism-test} and~\ref{tab:hea16-mechanism-external}).
Rankings differ on individual external datasets.
Supplementary Table~\ref{tab:hea16-mechanism-paired} gives the paired force-RMSE gains, their bootstrap intervals and the number of structures for which A improves on each control.

{
\footnotesize
\setlength{\tabcolsep}{4pt}
\begin{longtable}{@{}llrrrrrr@{}}
\caption{HEA16 testing-subset errors of the K32/M6/P4/R64 controls at terminal step 500000. Each checkpoint uses the same 5,273 structures, 339,173 atoms and 1,017,519 Cartesian force components. Energy errors are total-energy errors per atom in meV/atom; forces are in meV/\AA{} and stress in GPa. Seeds are reported separately.}\label{tab:hea16-mechanism-test}\\
\toprule
Model & Seed & \multicolumn{2}{c}{Energy} & \multicolumn{2}{c}{Force} & \multicolumn{2}{c}{Stress}\\
\cmidrule(lr){3-4}\cmidrule(lr){5-6}\cmidrule(lr){7-8}
 & & MAE & RMSE & MAE & RMSE & MAE & RMSE\\
\midrule
\endfirsthead
\multicolumn{8}{l}{\textit{Continued}}\\
\toprule
Model & Seed & \multicolumn{2}{c}{Energy} & \multicolumn{2}{c}{Force} & \multicolumn{2}{c}{Stress}\\
\cmidrule(lr){3-4}\cmidrule(lr){5-6}\cmidrule(lr){7-8}
 & & MAE & RMSE & MAE & RMSE & MAE & RMSE\\
\midrule
\endhead
\bottomrule
\endfoot
O & 1 & 3.865 & 6.958 & 57.312 & 90.887 & 0.1807 & 0.5526\\
O & 3 & 3.940 & 7.446 & 57.132 & 91.868 & 0.1729 & 0.5802\\
S & 1 & 3.476 & 6.715 & 54.161 & 86.355 & 0.1602 & 0.5217\\
S & 3 & 4.309 & 7.013 & 53.975 & 85.512 & 0.1599 & 0.4797\\
$Q_0$ & 1 & 3.456 & 5.541 & 52.744 & 96.003 & 0.1556 & 0.4741\\
$Q_0$ & 3 & 3.584 & 6.122 & 52.797 & 85.694 & 0.1568 & 0.4911\\
A & 1 & 4.046 & 6.492 & 51.558 & 81.167 & 0.1609 & 0.4923\\
A & 3 & 3.303 & 5.788 & 51.570 & 81.028 & 0.1554 & 0.4877\\
\end{longtable}
}

{
\footnotesize
\setlength{\tabcolsep}{4pt}
\begin{longtable}{@{}llrrrr@{}}
\caption{External force MAE/RMSE of the matched HEA16 controls, in meV/\AA. S1, S2 and S3 contain 123, 1,215 and 912 structures and 36,189, 156,891 and 114,462 Cartesian force components, respectively. The final column includes all 2,250 external alloy structures. Results are reported for both seeds.}\label{tab:hea16-mechanism-external}\\
\toprule
Model & Seed & S1 & S2 & S3 & S1--S3 combined\\
\midrule
\endfirsthead
\multicolumn{6}{l}{\textit{Continued}}\\
\toprule
Model & Seed & S1 & S2 & S3 & S1--S3 combined\\
\midrule
\endhead
\bottomrule
\endfoot
O & 1 & 21.670/29.801 & 66.845/103.189 & 94.496/135.542 & 71.820/111.239\\
O & 3 & 21.942/30.477 & 66.381/102.871 & 94.055/134.620 & 71.452/110.692\\
S & 1 & 20.574/28.388 & 63.116/97.576 & 89.504/128.095 & 67.931/105.161\\
S & 3 & 20.138/27.842 & 62.860/96.918 & 89.188/127.109 & 67.632/104.385\\
$Q_0$ & 1 & 19.894/27.630 & 61.152/93.372 & 86.962/124.503 & 65.903/101.522\\
$Q_0$ & 3 & 19.716/27.254 & 61.591/161.505 & 87.331/124.627 & 66.244/138.473\\
A & 1 & 19.672/27.068 & 59.520/90.847 & 84.779/120.968 & 64.232/98.705\\
A & 3 & 20.346/27.886 & 58.558/89.744 & 85.024/121.302 & 63.912/98.370\\
\end{longtable}
}

{
\footnotesize
\setlength{\tabcolsep}{4pt}
\begin{longtable}{@{}lllrll@{}}
\caption{Paired force-RMSE gains of A relative to each control.
We report $\mathrm{RMSE}_{\rm control}-\mathrm{RMSE}_{\rm A}$ in meV/\AA{}, so positive gains favour A.
The 95\% paired bootstrap intervals follow Section~\ref{sec:si-hea16-evaluation}.
The Improved column counts structures with lower force RMSE under A than under the control.}\label{tab:hea16-mechanism-paired}\\
\toprule
Control & Seed & Population & Gain & 95\% interval & Improved\\
\midrule
\endfirsthead
\multicolumn{6}{l}{\textit{Continued}}\\
\toprule
Control & Seed & Population & Gain & 95\% interval & Improved\\
\midrule
\endhead
\bottomrule
\endfoot
O & 1 & Test & 9.720 & [8.389, 10.880] & 4376/5273\\
O & 1 & S1--S3 combined & 12.534 & [11.604, 13.537] & 1880/2250\\
O & 3 & Test & 10.840 & [9.020, 14.150] & 4332/5273\\
O & 3 & S1--S3 combined & 12.322 & [11.294, 13.438] & 1867/2250\\
S & 1 & Test & 5.188 & [3.080, 8.215] & 3678/5273\\
S & 1 & S1--S3 combined & 6.456 & [5.862, 7.065] & 1551/2250\\
S & 3 & Test & 4.484 & [3.881, 5.394] & 3599/5273\\
S & 3 & S1--S3 combined & 6.016 & [5.280, 6.720] & 1510/2250\\
$Q_0$ & 1 & Test & 14.836 & [1.514, 36.935] & 3219/5273\\
$Q_0$ & 1 & S1--S3 combined & 2.816 & [2.246, 3.477] & 1403/2250\\
$Q_0$ & 3 & Test & 4.666 & [1.838, 9.709] & 3248/5273\\
$Q_0$ & 3 & S1--S3 combined & 40.103 & [2.516, 96.124] & 1359/2250\\
\end{longtable}
}

\subsection{Force-error diagnostics}\label{nav:sub16}\label{sec:si-force-decomposition}

We compared nine pairs of ordinary and attention models at seeds 1 and 3.
For each model and structure, we fitted a single scale factor to the reference forces (main Eq.~(6)).
The minimum-distance criterion retained 4,149 test and 1,215 external structures.
Each pair within 6\,\AA{} had to be separated by at least 75\% of the sum of the atoms' covalent radii.
Of these structures, 3,778 test and 796 external structures also met the force threshold for fitting the scale, with a root mean square reference force component above $10^{-8}$\,eV/\AA{}.
We divided the decrease in summed squared residuals by the decrease in total squared force error.
The residual accounted for 89.4--95.6\% of the improvement on the 3,778 eligible test structures across 18 comparisons (\path{paired-force-decomposition.csv} in the Source Data).

Supplementary Table~\ref{tab:hea16-mechanism-decomposition} gives the residual's share of total squared force error for each individual control checkpoint.
For this calculation we divided each checkpoint's summed squared residual by its total squared force error on all structures above the reference-force threshold, irrespective of interatomic distance.

{
\footnotesize
\setlength{\tabcolsep}{4pt}
\begin{longtable}{@{}llrr@{}}
\caption{Residual share of total squared force error for each control checkpoint. The test and S1--S3 populations include all 4,848 and 1,770 structures, respectively, that meet the reference-force threshold.}\label{tab:hea16-mechanism-decomposition}\\
\toprule
Model & Seed & Test residual (\%) & S1--S3 residual (\%)\\
\midrule
\endfirsthead
\multicolumn{4}{l}{\textit{Continued}}\\
\toprule
Model & Seed & Test residual (\%) & S1--S3 residual (\%)\\
\midrule
\endhead
\bottomrule
\endfoot
O & 1 & 96.09 & 96.35\\
O & 3 & 94.84 & 96.25\\
S & 1 & 95.40 & 96.41\\
S & 3 & 95.74 & 95.80\\
$Q_0$ & 1 & 88.32 & 96.76\\
$Q_0$ & 3 & 94.68 & 71.92\\
A & 1 & 96.38 & 96.95\\
A & 3 & 96.82 & 96.69\\
\end{longtable}
}

\paragraph{Force magnitude and neighbour distance.}\label{sec:si-environment-gains}
We grouped atoms in the full HEA16 testing subset and external alloy sets by reference-force magnitude and minimum normalized neighbour distance.
For each atom, we divided its distance to each neighbour within 6\,\AA{} by the sum of their covalent radii and took the smallest ratio~\cite{cordero2008radii}.
The neighbourhood includes periodic images and excludes zero-shift self edges.
The distance bins start at 0, 0.75, 0.9, 1.0 and 1.1, and the force bins at 0, 0.1, 0.5, 1 and 2\,eV/\AA{}.
Each bin includes its lower boundary and excludes its upper boundary.
The final bin is unbounded, and a separate distance category retains atoms without neighbours within 6\,\AA{}.
For control $c\in\{\mathrm S,Q_0\}$, the reduction in squared force error for atom $i$ is

\begin{equation}
d_i^{c,\mathrm A}=\|\mathbf F_i^{c}-\mathbf F_i^{\rm DFT}\|^2-\|\mathbf F_i^{\mathrm A}-\mathbf F_i^{\rm DFT}\|^2.
\label{eq:si-environment-gain}
\end{equation}

We computed force MAE and RMSE over the Cartesian components in each bin.
For the structure-balanced gain, we averaged $d_i^{c,\mathrm A}$ over the contributing atoms within each structure and then weighted those structures equally.
The paired bootstrap in Section~\ref{sec:si-hea16-evaluation} uses the same resampling weights across bins.
We report pointwise intervals only for bins with at least 20 contributing structures.

On the HEA16 testing subset and on S1--S3 combined, the absolute force-MAE gain of A over S was larger for atoms with reference-force magnitudes of at least 2\,eV/\AA{} than for those in the 0.1--0.5\,eV/\AA{} bin at both seeds (Supplementary Fig.~\ref{fig:si-environment-gains}).
The same ordering held for the 18 comparisons of ordinary refinement with the base models and the 18 comparisons of attention with ordinary refinement (Supplementary Fig.~\ref{fig:si-family-cost-environment}b,c).
Atoms with a minimum normalized neighbour distance below 0.75 comprised 5.67\% of the test atoms and accounted for 36--40\% of the S-to-A reduction in squared force error across seeds.
On S1--S3 combined, these atoms comprised 14.12\% of the population and accounted for 42--46\% of the reduction.
After excluding every structure containing such a pair, A still gave lower force MAE and RMSE than S and $Q_0$ on both subsets at both seeds (Supplementary Table~\ref{tab:hea16-mechanism-contact-subset}).

{
\footnotesize
\setlength{\tabcolsep}{4pt}
\begin{longtable}{@{}lllrrrl@{}}
\caption{Paired force-error gains after excluding structures with unusually short interatomic distances.
Positive gains favour A and are in meV/\AA.
We exclude a structure if any neighbour pair within 6\,\AA{} is separated by less than 75\% of the sum of their covalent radii.
The subsets contain 4,149 test and 1,215 external structures.
The full populations remain the primary comparison.
The last column gives exploratory 95\% paired structure-bootstrap intervals for the subset RMSE gain; the external bootstrap preserves S1/S2/S3 strata.}\label{tab:hea16-mechanism-contact-subset}\\
\toprule
Population & Control & Seed & \multicolumn{3}{c}{Gain (meV/\AA)} & 95\% interval\\
\cmidrule(lr){4-6}
 & & & Full MAE & Subset MAE & Subset RMSE & RMSE gain\\
\midrule
\endfirsthead
\multicolumn{7}{l}{\textit{Continued}}\\
\toprule
Population & Control & Seed & \multicolumn{3}{c}{Gain (meV/\AA)} & 95\% interval\\
\cmidrule(lr){4-6}
 & & & Full MAE & Subset MAE & Subset RMSE & RMSE gain\\
\midrule
\endhead
\bottomrule
\endfoot
Test & S & 1 & 2.603 & 1.778 & 3.013 & [2.788, 3.247]\\
Test & S & 3 & 2.405 & 1.554 & 2.609 & [2.391, 2.815]\\
Test & $Q_0$ & 1 & 1.187 & 0.716 & 1.242 & [1.066, 1.449]\\
Test & $Q_0$ & 3 & 1.227 & 0.770 & 1.168 & [0.925, 1.401]\\
S1--S3 combined & S & 1 & 3.699 & 1.987 & 3.239 & [2.762, 3.743]\\
S1--S3 combined & S & 3 & 3.720 & 1.782 & 2.752 & [2.107, 3.401]\\
S1--S3 combined & $Q_0$ & 1 & 1.671 & 1.009 & 1.684 & [1.274, 2.116]\\
S1--S3 combined & $Q_0$ & 3 & 2.332 & 0.758 & 1.272 & [0.771, 1.762]\\
\end{longtable}
}

\subsection{External alloy benchmarks}\label{nav:sub17}
The external alloy comparison evaluates DTM-HEA16-XXS through DTM-HEA16-XL at seeds 1 and 3 alongside released TorchNEP and GRACE models~\cite{wu2026torchnep,shuang2026comparison}. 
NEP44/66/88 use the released seed-0 TorchNEP parameters, radial/angular cutoffs of 6/5\,\AA{} and a 1--2\,\AA{} ZBL term. 
The GRACE comparison uses the released GRACE-FS-S, GRACE-FS-M-Aug and GRACE-FS-L models; M-Aug uses augmented training data. 
The DTM architectures and training settings are specified above. 
All models are evaluated on the same published alloy structures.

Supplementary Table~\ref{tab:hea16-training} gives the energy, force and stress errors for the six DTM architectures used in the external comparison.
\formatHEATable*{\begingroup\footnotesize\setlength{\tabcolsep}{2.5pt}
\begin{longtable}{@{}llrrrrrr@{}}
\caption{Prediction MAE/RMSE on the HEA16 training, validation and testing subsets for the same 12 checkpoints.
Energy errors are in meV/atom, force errors in meV/\AA{} and stress errors in GPa.
Post-training evaluations use the fixed splits and common archived inference settings; their validation scores are distinct from the trainer-validation means in Table~\ref{tab:hea16-family-endpoints}.
Seeds are placed side by side; each cell is MAE/RMSE.}\label{tab:hea16-splits}\label{tab:hea16-training}\label{tab:hea16-evaluation}\label{tab:hea16-test}\\
\toprule
Model & Subset & E s1 & F s1 & Stress s1 & E s3 & F s3 & Stress s3\\
\midrule\endfirsthead
\multicolumn{8}{l}{\textit{Supplementary Table \thetable{} (continued)}}\\
\toprule
Model & Subset & E s1 & F s1 & Stress s1 & E s3 & F s3 & Stress s3\\
\midrule\endhead\bottomrule\endfoot
K16-M3 & Training & 7.363/12.982 & 95.771/151.829 & 0.290/0.738 & 6.771/12.621 & 95.913/152.145 & 0.289/0.732\\
K32-M4 & Training & 4.856/8.865 & 81.329/128.539 & 0.227/0.544 & 4.707/8.400 & 80.920/127.903 & 0.223/0.526\\
M4-P1 & Training & 4.078/7.182 & 71.339/113.020 & 0.197/0.464 & 4.296/7.388 & 71.740/113.578 & 0.203/0.480\\
M6-P1 & Training & 3.262/5.556 & 60.274/94.175 & 0.164/0.365 & 3.838/6.020 & 60.423/94.390 & 0.164/0.374\\
M8-P1+MA & Training & 2.853/4.989 & 54.213/84.156 & 0.141/0.310 & 3.435/6.709 & 54.433/84.418 & 0.148/0.342\\
M8-P4+MA & Training & 2.857/4.352 & 46.667/71.617 & 0.121/0.251 & 2.478/4.036 & 46.711/71.693 & 0.123/0.283\\
K16-M3 & Validation & 7.884/15.067 & 96.409/153.749 & 0.310/0.859 & 7.495/16.547 & 96.572/153.738 & 0.304/0.811\\
K32-M4 & Validation & 5.531/11.383 & 82.547/132.079 & 0.257/0.761 & 5.330/11.347 & 82.138/130.992 & 0.251/0.710\\
M4-P1 & Validation & 4.729/11.095 & 72.590/116.552 & 0.218/0.563 & 4.940/11.720 & 72.971/117.234 & 0.229/0.666\\
M6-P1 & Validation & 3.921/9.623 & 61.436/97.882 & 0.184/0.491 & 4.460/9.642 & 61.654/97.745 & 0.185/0.516\\
M8-P1+MA & Validation & 3.433/8.050 & 55.446/87.237 & 0.157/0.388 & 3.865/9.773 & 55.682/87.540 & 0.163/0.431\\
M8-P4+MA & Validation & 3.440/7.726 & 48.430/75.952 & 0.139/0.359 & 3.022/7.335 & 48.465/75.992 & 0.139/0.346\\
K16-M3 & Testing & 8.394/15.748 & 96.887/154.784 & 0.322/0.982 & 7.950/16.304 & 96.978/154.916 & 0.329/1.047\\
K32-M4 & Testing & 5.636/11.371 & 82.825/132.925 & 0.269/0.931 & 5.627/11.272 & 82.415/133.330 & 0.273/1.000\\
M4-P1 & Testing & 4.814/9.419 & 72.676/119.236 & 0.225/0.664 & 5.050/10.378 & 73.017/117.184 & 0.236/0.683\\
M6-P1 & Testing & 3.876/7.502 & 61.516/105.593 & 0.190/0.591 & 4.433/8.296 & 61.578/97.760 & 0.192/0.692\\
M8-P1+MA & Testing & 3.420/6.575 & 55.313/87.444 & 0.170/0.567 & 4.008/8.545 & 55.507/86.880 & 0.172/0.542\\
M8-P4+MA & Testing & 3.243/5.549 & 48.280/76.156 & 0.141/0.405 & 3.010/5.559 & 48.340/75.750 & 0.147/0.489\\
\end{longtable}\endgroup
}

Supplementary Tables~\ref{tab:hea16-formation} and~\ref{tab:hea16-forces} report formation-energy and force MAE/RMSE across the datasets introduced in Section~\ref{sec:si-hea16-evaluation}, with seeds reported separately. 
The force table reports S1, S2 and S3 both individually and combined.
\formatHEATable{{\footnotesize
\setlength{\tabcolsep}{3.8pt}
\begin{longtable}{@{}lrrrrr@{}}
\caption{Published-structure formation-energy MAE/RMSE (meV/atom). Equal structure weights and independently relaxed elemental references are used. All configurations are retained.}\label{tab:hea16-formation}\\
\toprule
Checkpoint & MP ternaries & GNoME & S1 & S2 & S3\\
\midrule
\endfirsthead
\multicolumn{6}{l}{\textit{Continued}}\\
\toprule
Checkpoint & MP ternaries & GNoME & S1 & S2 & S3\\
\midrule
\endhead
\bottomrule
\endfoot
K16-M3; s1 & 38.793/66.302 & 53.277/151.107 & 5.511/5.940 & 10.401/16.368 & 12.185/16.228\\
K16-M3; s3 & 36.869/59.296 & 36.654/59.055 & 4.956/5.611 & 12.294/20.495 & 13.738/17.657\\
K32-M4; s1 & 35.455/65.194 & 29.181/41.347 & 3.334/3.717 & 8.675/15.090 & 11.628/15.354\\
K32-M4; s3 & 35.983/71.143 & 31.834/51.902 & 3.153/3.617 & 7.448/13.108 & 9.727/12.925\\
M4-P1; s1 & 31.006/60.319 & 28.600/59.254 & 3.906/4.328 & 6.950/12.416 & 9.992/13.126\\
M4-P1; s3 & 30.976/57.509 & 29.431/58.301 & 3.557/4.041 & 7.779/11.001 & 11.929/15.733\\
M6-P1; s1 & 30.086/57.748 & 29.245/73.178 & 2.675/3.044 & 5.869/9.155 & 9.737/12.440\\
M6-P1; s3 & 31.146/65.338 & 47.752/224.637 & 1.648/2.021 & 6.197/10.172 & 10.058/12.631\\
M8-P1+MA; s1 & 25.483/53.322 & 29.754/71.289 & 2.254/2.664 & 4.865/8.223 & 9.326/12.116\\
M8-P1+MA; s3 & 26.787/52.691 & 26.012/51.740 & 2.355/2.796 & 5.178/6.894 & 9.760/12.390\\
M8-P4+MA; s1 & 25.595/53.044 & 41.124/239.530 & 3.781/4.044 & 4.974/6.814 & 9.596/11.780\\
M8-P4+MA; s3 & 24.289/49.448 & 29.296/81.967 & 3.815/3.994 & 4.916/6.757 & 9.242/11.523\\
NEP44 & 60.164/107.145 & 60.776/84.519 & 3.828/4.559 & 14.146/29.443 & 17.030/23.330\\
NEP66 & 61.567/111.532 & 57.991/78.117 & 3.283/4.185 & 14.921/33.585 & 17.340/23.975\\
NEP88 & 66.190/115.673 & 57.499/80.447 & 3.157/3.712 & 12.780/28.847 & 16.897/23.513\\
GRACE-FS-S & 46.158/76.749 & 64.232/180.280 & 2.751/3.395 & 22.683/37.903 & 13.574/19.734\\
GRACE-FS-M-Aug & 43.177/68.613 & 65.533/198.964 & 2.624/3.237 & 16.181/28.547 & 11.402/15.841\\
GRACE-FS-L & 51.799/93.674 & 68.354/194.857 & 2.880/3.505 & 16.417/32.679 & 12.286/17.352\\
\end{longtable}
}
}
\formatHEATable{{
\footnotesize
\setlength{\tabcolsep}{3.8pt}
\begin{longtable}{@{}lrrrr@{}}
\caption{Cartesian-component force MAE/RMSE (meV/\AA) for each external alloy dataset and for S1--S3 combined.}\label{tab:hea16-forces}\\
\toprule
Checkpoint & S1 & S2 & S3 & S1--S3 combined\\
\midrule
\endfirsthead
\multicolumn{5}{l}{\textit{Continued}}\\
\toprule
Checkpoint & S1 & S2 & S3 & S1--S3 combined\\
\midrule
\endhead
\bottomrule
\endfoot
K16-M3; s1 & 33.080/45.496 & 119.853/181.018 & 154.291/221.925 & 122.459/187.857\\
K16-M3; s3 & 33.581/46.264 & 119.110/180.292 & 155.108/222.466 & 122.443/187.760\\
K32-M4; s1 & 30.669/42.957 & 97.412/150.662 & 134.932/194.184 & 103.522/160.720\\
K32-M4; s3 & 30.146/41.549 & 96.217/150.394 & 134.029/192.772 & 102.515/159.914\\
M4-P1; s1 & 26.875/36.998 & 86.061/133.379 & 118.670/172.238 & 91.233/142.400\\
M4-P1; s3 & 26.576/36.832 & 86.039/133.344 & 120.321/173.789 & 91.801/143.078\\
M6-P1; s1 & 22.346/31.136 & 70.932/109.199 & 99.708/143.243 & 75.925/117.618\\
M6-P1; s3 & 23.259/32.369 & 71.097/110.727 & 99.842/142.772 & 76.167/118.171\\
M8-P1+MA; s1 & 21.182/29.628 & 63.355/98.641 & 90.933/130.454 & 68.657/106.775\\
M8-P1+MA; s3 & 20.948/29.097 & 62.999/96.040 & 90.917/129.368 & 68.441/105.042\\
M8-P4+MA; s1 & 18.601/25.885 & 54.366/84.936 & 79.393/113.234 & 59.472/92.364\\
M8-P4+MA; s3 & 19.226/26.789 & 54.443/83.380 & 80.147/113.542 & 59.866/91.811\\
NEP44 & 43.436/58.673 & 120.001/178.887 & 162.219/232.787 & 126.704/192.090\\
NEP66 & 40.392/53.898 & 115.375/172.830 & 162.040/230.519 & 123.919/188.035\\
NEP88 & 39.425/53.707 & 111.497/165.632 & 156.928/223.980 & 119.925/181.676\\
GRACE-FS-S & 36.667/50.395 & 128.947/195.836 & 163.040/236.204 & 130.777/201.566\\
GRACE-FS-M-Aug & 34.888/49.494 & 117.780/181.325 & 147.328/213.261 & 119.023/184.359\\
GRACE-FS-L & 34.318/47.416 & 117.026/178.797 & 153.005/224.200 & 120.684/187.832\\
\end{longtable}
}
}

\Needspace{6\baselineskip}\paragraph{Compressed-structure tail.}
Across all 444 GNoME structures, increasing attention depth from one to four blocks raises formation-energy MAE and RMSE at both seeds (Supplementary Table~\ref{tab:hea16-formation} and Supplementary Fig.~\ref{fig:si-hea16-gnome}).
At seed 1, three structures contribute 94.24\% of the four-block model's squared error.
The same three structures account for 2.46\% at seed 3, where the three largest errors contribute 72.80\%.

\subsection{Force accuracy and cost of cubic readout}\label{nav:sub18}

We compared standard (C0) and cubic-augmented (C1) readouts at $M=4$ and $M=6$, with or without one ordinary refinement block.
We trained each C0/C1 pair with the same recipe at seeds 1 and 3 and evaluated the terminal checkpoints after 500,000 updates.
C1 lowered validation force MAE and RMSE in all eight paired comparisons (Supplementary Table~\ref{tab:hea16-cubic-validation}).

{\footnotesize
\setlength{\tabcolsep}{3.4pt}
\begin{longtable}{@{}llrrrrrr@{}}
\caption{Terminal validation errors for the HEA16 readout controls.
All models use 32 radial channels, a $3\times128$ energy network and no moment attention.
C0/C1 denotes standard/cubic-augmented readout; P1 denotes one ordinary refinement block.
Energy errors are in meV/atom, force errors in meV/\AA{} and stress errors in GPa.
Seeds 1 and 3 are reported individually; all runs use 500,000 updates.}\label{tab:hea16-cubic-validation}\\
\toprule
 & & \multicolumn{2}{c}{Energy} & \multicolumn{2}{c}{Force} & \multicolumn{2}{c}{Stress}\\
Model & Seed & MAE & RMSE & MAE & RMSE & MAE & RMSE\\
\midrule
\endfirsthead
\multicolumn{8}{l}{\textit{Continued}}\\
\toprule
Model & Seed & MAE & RMSE & MAE & RMSE & MAE & RMSE\\
\midrule
\endhead
\bottomrule
\endfoot
M4-C0 & 1 & 5.455 & 11.399 & 82.699 & 132.243 & 0.256552 & 0.758025\\
M4-C0 & 3 & 5.232 & 11.446 & 82.277 & 131.116 & 0.250754 & 0.711371\\
M4-C1 & 1 & 5.169 & 11.636 & 78.207 & 125.676 & 0.240631 & 0.683434\\
M4-C1 & 3 & 5.036 & 10.503 & 77.977 & 125.217 & 0.235469 & 0.633766\\
M4-C0-P1 & 1 & 4.661 & 11.062 & 72.735 & 116.717 & 0.217759 & 0.564986\\
M4-C0-P1 & 3 & 4.740 & 11.550 & 73.116 & 117.340 & 0.227851 & 0.661868\\
M4-C1-P1 & 1 & 4.630 & 10.249 & 72.400 & 116.260 & 0.218419 & 0.592963\\
M4-C1-P1 & 3 & 4.673 & 9.962 & 72.450 & 116.392 & 0.222356 & 0.621198\\
M6-C0 & 1 & 4.690 & 9.588 & 76.497 & 122.265 & 0.222029 & 0.596060\\
M6-C0 & 3 & 4.833 & 10.770 & 76.660 & 121.880 & 0.228932 & 0.618927\\
M6-C1 & 1 & 4.162 & 9.393 & 68.202 & 108.695 & 0.197992 & 0.527978\\
M6-C1 & 3 & 4.233 & 9.470 & 68.259 & 108.774 & 0.203873 & 0.551378\\
M6-C0-P1 & 1 & 3.834 & 9.580 & 61.575 & 98.025 & 0.185401 & 0.496788\\
M6-C0-P1 & 3 & 3.860 & 9.497 & 61.811 & 97.950 & 0.187498 & 0.522746\\
M6-C1-P1 & 1 & 3.900 & 8.657 & 60.191 & 95.394 & 0.183421 & 0.576204\\
M6-C1-P1 & 3 & 3.861 & 9.363 & 60.200 & 96.227 & 0.176911 & 0.471859\\
\end{longtable}
}

C1 also lowered force MAE and RMSE on the HEA16 testing subset and on the three external alloy datasets combined.
At both angular resolutions and both seeds, the force-error reductions on both populations were smaller after one refinement block than in the base models (Supplementary Table~\ref{tab:hea16-cubic-external-forces}).
Main Fig.~2c shows the force RMSEs on the HEA16 testing subset.
Adding C1 reduced throughput for both the base and one-block models on BCC16 (Supplementary Table~\ref{tab:throughput-hea16}).

\begingroup
\footnotesize
\setlength{\tabcolsep}{3pt}
\begin{longtable}{@{}llrrrrrrrr@{}}
\caption{Force errors (meV/\AA{}) of standard- and cubic-readout models on the testing subset and the external alloy structures.
All models have K32 and a $3\times128$ energy head at terminal step 500000; P1 denotes one rank-64 ordinary refinement block.
C0/C1 denotes the absence/presence of explicit cubic invariants.
Seeds are listed separately.
Test scores cover all 5,273 testing structures, and external scores cover all 2,250 structures in S1--S3.
The evaluator and metric definitions match Section~\ref{sec:si-mechanism-heldout}.}\label{tab:hea16-cubic-external-forces}\\
\toprule
 & & \multicolumn{2}{c}{Test MAE} & \multicolumn{2}{c}{Test RMSE} & \multicolumn{2}{c}{External MAE} & \multicolumn{2}{c}{External RMSE}\\
\cmidrule(lr){3-4}\cmidrule(lr){5-6}\cmidrule(lr){7-8}\cmidrule(lr){9-10}
Model & Seed & C0 & C1 & C0 & C1 & C0 & C1 & C0 & C1\\
\midrule
\endfirsthead
\toprule
 & & \multicolumn{2}{c}{Test MAE} & \multicolumn{2}{c}{Test RMSE} & \multicolumn{2}{c}{External MAE} & \multicolumn{2}{c}{External RMSE}\\
\cmidrule(lr){3-4}\cmidrule(lr){5-6}\cmidrule(lr){7-8}\cmidrule(lr){9-10}
Model & Seed & C0 & C1 & C0 & C1 & C0 & C1 & C0 & C1\\
\midrule
\endhead
\bottomrule
\endfoot
M4 & 1 & 82.825 & 78.195 & 132.925 & 125.650 & 103.522 & 98.594 & 160.720 & 153.860\\
M4 & 3 & 82.415 & 77.996 & 133.330 & 126.226 & 102.515 & 98.789 & 159.914 & 154.124\\
M4-P1 & 1 & 72.676 & 72.427 & 119.236 & 116.745 & 91.233 & 91.146 & 142.400 & 141.848\\
M4-P1 & 3 & 73.017 & 72.368 & 117.184 & 116.241 & 91.801 & 90.411 & 143.078 & 141.257\\
M6 & 1 & 76.284 & 67.898 & 120.952 & 109.045 & 94.090 & 84.557 & 146.965 & 132.189\\
M6 & 3 & 76.448 & 68.029 & 121.791 & 108.908 & 94.636 & 84.658 & 147.718 & 132.632\\
M6-P1 & 1 & 61.516 & 60.004 & 105.593 & 97.005 & 75.925 & 74.728 & 117.617 & 116.091\\
M6-P1 & 3 & 61.578 & 59.829 & 97.760 & 95.744 & 76.167 & 74.657 & 118.171 & 115.144\\
\end{longtable}
\endgroup

\subsection{Alloy formation energies with cubic readout}\label{nav:sub19}\label{sec:si-cubic-external}

We evaluated the same 16 checkpoints on the published geometries of the five external alloy datasets using Eq.~\eqref{eq:si-formation-energy}.
For each checkpoint, we relaxed the elemental references with the hydrostatic cell filter and QuasiNewton using the convergence settings in Section~\ref{sec:si-hea16-evaluation}.
These calculations used TensorMD 2.8.1 in FP32 with 0.001\,\AA{} radial interpolation.
Across the 40 comparisons of C0 and C1 at the same $M$, refinement depth, seed and dataset, C1 lowered formation-energy MAE in 20 and RMSE in 24 (Supplementary Table~\ref{tab:hea16-cubic-external-formation}).
For M6-P1 at seed 1, an Al$_8$Mg$_{12}$Pd$_8$ structure with a minimum interatomic distance of 1.2223\,\AA{} accounted for 98.68\% of the cubic-readout model's squared formation-energy error on GNoME.

\begingroup
\footnotesize
\setlength{\tabcolsep}{3.5pt}
\begin{longtable}{@{}lrrrrr@{}}
\caption{Cubic-readout formation-energy MAE/RMSE (meV/atom) for the matched terminal C0/C1 checkpoints at seeds 1 and 3. Architecture labels match Table~\ref{tab:hea16-cubic-external-forces}. MP ternaries, GNoME, S1, S2 and S3 contain 482, 444, 123, 1,215 and 912 fixed published geometries, respectively. Each predictor uses its own relaxed elemental references and the common evaluation protocol in Section~\ref{sec:si-cubic-external}. All structures enter each score with equal weight, including the large GNoME tail.}\label{tab:hea16-cubic-external-formation}\\
\toprule
Checkpoint & MP ternaries & GNoME & S1 & S2 & S3\\
\midrule
\endfirsthead
\toprule
Checkpoint & MP ternaries & GNoME & S1 & S2 & S3\\
\midrule
\endhead
\bottomrule
\endfoot
M4-C0; s1 & 35.455/65.194 & 29.181/41.347 & 3.334/3.717 & 8.675/15.090 & 11.628/15.354\\
M4-C1; s1 & 42.984/92.142 & 30.775/49.507 & 2.486/2.911 & 7.954/11.342 & 10.784/13.975\\
M4-C0; s3 & 35.983/71.143 & 31.834/51.902 & 3.153/3.617 & 7.448/13.108 & 9.727/12.925\\
M4-C1; s3 & 32.089/60.471 & 32.042/61.697 & 4.070/4.522 & 7.680/12.586 & 11.523/15.553\\
M4-C0-P1; s1 & 31.006/60.319 & 28.600/59.254 & 3.906/4.328 & 6.950/12.416 & 9.992/13.126\\
M4-C1-P1; s1 & 34.743/73.130 & 28.661/57.477 & 4.074/4.513 & 7.188/15.608 & 11.390/14.580\\
M4-C0-P1; s3 & 30.976/57.509 & 29.431/58.301 & 3.557/4.041 & 7.779/11.001 & 11.929/15.733\\
M4-C1-P1; s3 & 40.776/86.870 & 24.460/53.966 & 3.897/4.568 & 7.882/12.000 & 12.020/15.176\\
M6-C0; s1 & 37.888/77.534 & 32.081/49.070 & 1.972/2.513 & 7.663/12.414 & 12.956/16.544\\
M6-C1; s1 & 29.534/53.126 & 25.225/48.942 & 1.887/2.363 & 6.450/9.200 & 9.980/13.354\\
M6-C0; s3 & 38.067/77.371 & 27.854/40.676 & 2.172/2.892 & 8.064/13.090 & 12.112/15.955\\
M6-C1; s3 & 29.252/62.457 & 32.082/111.916 & 1.537/1.966 & 5.967/8.811 & 9.933/13.148\\
M6-C0-P1; s1 & 30.086/57.748 & 29.245/73.178 & 2.675/3.044 & 5.869/9.155 & 9.737/12.440\\
M6-C1-P1; s1 & 28.888/55.758 & 101.145/1379.526 & 1.746/2.085 & 6.022/9.041 & 10.103/12.760\\
M6-C0-P1; s3 & 31.146/65.338 & 47.752/224.637 & 1.648/2.021 & 6.197/10.172 & 10.058/12.631\\
M6-C1-P1; s3 & 28.481/60.329 & 32.707/104.183 & 2.167/2.551 & 5.933/9.210 & 9.581/11.873\\
\end{longtable}
\endgroup

\FloatBarrier

\subsection{Throughput}\label{sec:si-hea16-throughput}
Supplementary Table~\ref{tab:throughput-hea16} in Section~\ref{sec:si-moving} reports throughput for HEA16-trained models on FCC Cu and BCC16 using the RTX~5090 and A100 GPUs.

\section{OMat24 models, validation and physical evaluation}\label{sec:si-omat24}
\label{sec:si-omat}
\subsection{Models and training data}\label{nav:sub20}\label{sec:si-omat-training}
We trained all 18 DTM architectures on the full official OMat24 training set.
Of these, 13 use the standard readout (C0) and 5 use the cubic-augmented readout (C1).
Supplementary Table~\ref{tab:omat-registry} gives the architecture and trainable parameter count of each model.
Of the five named standard-readout models, we used DTM-OMat24-S, -M, -L and -XL for the main-text physical benchmarks.

\formatOMatTable{{\footnotesize
\setlength{\tabcolsep}{3.3pt}
\begin{longtable}{@{}lrrrlllr@{}}
\caption{OMat24 architectures, readouts and trainable parameter counts.
Parentheses identify the five named OMat24 checkpoints with standard readout.
Every model uses a 6\,\AA{} cutoff and a $3\times512$ radial encoder; all refinement blocks use $R=64$.
Energy-network sizes give hidden-layer depth by width, and $\mathcal J$ lists the blocks followed by moment attention.
C0 and C1 denote the standard and cubic-augmented readouts, respectively.
Counts include 113 elemental offsets and are architecture-derived for C0 or recorded trainer totals for C1.
We selected the checkpoint at 4 million updates for each model except DTM-OMat24-XL (M8-P4+MA), for which we selected a stable checkpoint at 3.85 million updates.
Validation and deployment use the same selected checkpoint for each model.}\label{tab:omat-registry}\label{tab:benchmarks-models}\\
\toprule
Model & $K$ & $M$ & $L$ & $\mathcal J$ & Energy MLP & Readout & \shortstack{Trainable\\parameters}\\
\midrule
\endfirsthead
\multicolumn{8}{l}{\textit{Continued}}\\
\toprule
Model & $K$ & $M$ & $L$ & $\mathcal J$ & Energy MLP & Readout & \shortstack{Trainable\\parameters}\\
\midrule
\endhead
\bottomrule
\endfoot
K8-M2/H2x64 & 8 & 2 & 0 & -- & $2\times64$ & C0 & 590,777\\
K8-M3/H4x64 & 8 & 3 & 0 & -- & $4\times64$ & C0 & 599,609\\
K16-M2/H2x64 & 16 & 2 & 0 & -- & $2\times64$ & C0 & 595,905\\
K16-M2/H4x64 & 16 & 2 & 0 & -- & $4\times64$ & C0 & 604,225\\
K16-M3 & 16 & 3 & 0 & -- & $4\times64$ & C0 & 605,249\\
K32-M4/H3x128 & 32 & 4 & 0 & -- & $3\times128$ & C0 & 648,593\\
M4+C1/H3x128 & 32 & 4 & 0 & -- & $3\times128$ & C1 & 657,301\\
K32-M4/H3x256 & 32 & 4 & 0 & -- & $3\times256$ & C0 & 766,097\\
M4+C1/H3x256 & 32 & 4 & 0 & -- & $3\times256$ & C1 & 782,997\\
M4-P1/H3x128 & 32 & 4 & 1 & -- & $3\times128$ & C0 & 667,026\\
M4-P1+C1/H3x128 & 32 & 4 & 1 & -- & $3\times128$ & C1 & 675,734\\
M4-P1/H3x256 & 32 & 4 & 1 & -- & $3\times256$ & C0 & 784,530\\
M6+C1/H3x256 & 32 & 6 & 0 & -- & $3\times256$ & C1 & 840,351\\
M6-P1 & 32 & 6 & 1 & -- & $3\times256$ & C0 & 811,154\\
M6-P1+C1/H3x256 & 32 & 6 & 1 & -- & $3\times256$ & C1 & 869,024\\
M8-P1+MA & 32 & 8 & 1 & 0 & $3\times512$ & C0 & 1,327,763\\
M8-P2+MA & 32 & 8 & 2 & 0,1 & $3\times512$ & C0 & 1,392,277\\
M8-P4+MA & 32 & 8 & 4 & 0,1,2,3 & $3\times512$ & C0 & 1,521,305\\
\end{longtable}
}
}

For DTM-OMat24-L and -XL, we rescaled an output projection after an optimizer update if its quadratic coefficient proxy exceeded the cap for that layer and angular order.
The model configurations give the caps and metric scales.

Supplementary Table~\ref{tab:omat-training-composition} lists the 11 source groups in the official training set.
On each pass, we shuffled the shards and the structures within each shard, retaining the original source proportions.
We interleaved up to 32 shards at a time, except for M8-P1+MA, which used up to 64 shards and a batch size of 128.
The four highlighted models used seed 42 and were validated every 50,000 updates.
The gradient-norm clipping threshold was 10 for K16-M3 and 1 for the three refined models.

\begin{table}[htbp]\centering\small
\caption{Composition of the official OMat24 training set.
Fractions give the percentage of training structures in each source group.}\label{tab:omat-training-composition}
\begin{tabular}{lrr}\toprule
Source group & Structures & Fraction (\%) \\ \midrule
aimd-from-PBE-1000-npt & 21,269,486 & 21.0955 \\
aimd-from-PBE-1000-nvt & 20,256,650 & 20.0910 \\
aimd-from-PBE-3000-npt & 6,076,290 & 6.0266 \\
aimd-from-PBE-3000-nvt & 7,839,846 & 7.7757 \\
rattled-1000 & 11,388,510 & 11.2954 \\
rattled-1000-subsampled & 3,879,741 & 3.8480 \\
rattled-300 & 6,319,139 & 6.2675 \\
rattled-300-subsampled & 3,464,007 & 3.4357 \\
rattled-500 & 6,922,197 & 6.8656 \\
rattled-500-subsampled & 3,975,416 & 3.9429 \\
rattled-relax & 9,433,303 & 9.3562 \\
\midrule Total & 100,824,585 & 100 \\
\bottomrule\end{tabular}
\end{table}

\subsection{Validation and controlled comparisons}\label{nav:sub21}\label{sec:si-omat-validation}

Supplementary Table~\ref{tab:benchmarks-official} reports energy, force and stress MAEs for the 18 DTM endpoints and 17 released predictors on the corrected official validation set and the reconstructed DPA-style subset.
Energy error is the mean absolute per-atom energy residual across structures. 
Force error pools Cartesian residuals across atoms.
Stress error pools the six symmetric ASE components $(xx,yy,zz,yz,xz,xy)$ in\,GPa.
The 11 official source databases contribute 57,684,726 force components and 6,152,166 stress components.
We scored all finite predictions, including extreme errors, directly against the reference labels without a validation-fitted energy shift.

We reconstructed the DPA-style subset from the recovered selection rules because the published structure-ID list was unavailable. 
We retained structures with $-25\le E/N\le 0$\,eV/atom and maximum absolute Cartesian force component at most 50\,eV/\AA{}, applied no stress filter and excluded four single-atom periodic structures, as inferred from the recovered procedure.
These rules removed 2,616 structures. For released predictors, both cohorts used the same saved predictions at each model's configured precision, with float64 error reduction. 
For DTM, official values retain the trainer's rounding and FP32-base/BF16-radial precision with TF32 disabled. 
Evaluations on the reconstructed DPA-style subset used FP32 for both base and radial operations with float64 error reduction.
The DTM comparison therefore includes both the population and precision changes.

The released predictors comprised DPA4.1, DPA4C, MACE, GRACE, Allegro and NEP89. Both DPA families used release v20260911. 
The reconstructed force MAEs for DPA4.1 Nano and Mini agree with the authors' release scores to the reported precision~\cite{dpa4models}. 
The excluded structures account for the extreme full-cohort means for these two models.
We evaluated NEP89 against the official OMat24 PBE(+U) labels.
Source Data also give its scores against D3(BJ)-consistent labels.

\formatOMatTable{\par
\begingroup
\footnotesize
\setlength{\tabcolsep}{3pt}
\begin{longtable}{@{}lrrrrrrr@{}}
\caption{OMat24 validation MAEs on the official validation set and the reconstructed DPA-style subset.
The official cohort contains 1,025,361 structures and 19,228,242 atoms.
The reconstructed subset contains 1,022,745 structures and 19,120,236 atoms.
$E$ is energy per atom (meV/atom), $F$ is Cartesian-component force (meV/\AA{}) and $S_6$ is six-component stress (GPa).
Trainable parameter counts come from DTM architecture or training records and from released checkpoints or published model counts for the other predictors.}\label{tab:benchmarks-official}\label{tab:benchmarks-curated}\\
\toprule
Model & \multicolumn{3}{c}{Official set} & \multicolumn{3}{c}{Reconstructed subset} & Trainable\\
\cmidrule(lr){2-4}\cmidrule(lr){5-7}
 & $E$ & $F$ & $S_6$ & $E$ & $F$ & $S_6$ & parameters\\
\midrule
\endfirsthead
\multicolumn{8}{l}{\textit{Continued}}\\
\toprule
Model & \multicolumn{3}{c}{Official set} & \multicolumn{3}{c}{Reconstructed subset} & Trainable\\
\cmidrule(lr){2-4}\cmidrule(lr){5-7}
 & $E$ & $F$ & $S_6$ & $E$ & $F$ & $S_6$ & parameters\\
\midrule
\endhead
M8-P4+MA & 14.625 & 73.000 & 0.615 & 13.110 & 71.464 & 0.608 & 1,521,305\\
M8-P2+MA & 15.177 & 77.445 & 0.631 & 13.766 & 75.841 & 0.621 & 1,392,277\\
M8-P1+MA & 15.756 & 83.070 & 0.647 & 14.276 & 81.441 & 0.639 & 1,327,763\\
M6-P1+C1/H3x256 & 17.719 & 91.731 & 0.683 & 16.078 & 89.869 & 0.677 & 869,024\\
M6-P1 & 17.960 & 93.210 & 0.687 & 16.571 & 91.485 & 0.679 & 811,154\\
M6+C1/H3x256 & 18.855 & 100.866 & 0.710 & 17.207 & 98.936 & 0.702 & 840,351\\
M4-P1/H3x256 & 19.273 & 101.953 & 0.719 & 17.585 & 100.164 & 0.711 & 784,530\\
M4-P1+C1/H3x128 & 20.187 & 104.745 & 0.733 & 18.581 & 102.910 & 0.725 & 675,734\\
M4-P1/H3x128 & 20.352 & 105.207 & 0.734 & 18.745 & 103.503 & 0.725 & 667,026\\
M4+C1/H3x256 & 20.075 & 107.304 & 0.736 & 18.396 & 105.443 & 0.726 & 782,997\\
K32-M4/H3x256 & 20.600 & 111.689 & 0.747 & 18.970 & 109.683 & 0.740 & 766,097\\
M4+C1/H3x128 & 21.383 & 111.721 & 0.757 & 19.691 & 109.840 & 0.747 & 657,301\\
K32-M4/H3x128 & 22.007 & 116.132 & 0.768 & 20.436 & 114.208 & 0.759 & 648,593\\
K16-M3 & 25.433 & 129.733 & 0.829 & 23.825 & 127.653 & 0.820 & 605,249\\
K8-M3/H4x64 & 28.326 & 138.521 & 0.874 & 26.747 & 136.212 & 0.869 & 599,609\\
K16-M2/H4x64 & 29.165 & 147.620 & 0.897 & 27.524 & 145.438 & 0.887 & 604,225\\
K16-M2/H2x64 & 30.349 & 150.726 & 0.918 & 28.762 & 148.152 & 0.910 & 595,905\\
K8-M2/H2x64 & 33.891 & 159.569 & 0.964 & 32.167 & 157.058 & 0.951 & 590,777\\
DPA4.1 Nano & 2266.289 & 52461.796 & 64.515 & 18.348 & 94.723 & 0.700 & 480,650\\
DPA4.1 Mini & 191.790 & 13611.734 & 12.706 & 14.018 & 69.986 & 0.613 & 655,390\\
DPA4.1 Neo & 14.914 & 92.108 & 0.617 & 12.047 & 58.647 & 0.575 & 1,125,244\\
DPA4.1 Air & 13.352 & 63.288 & 0.577 & 10.771 & 52.162 & 0.551 & 5,148,371\\
DPA4.1 Plus & 12.609 & 57.510 & 0.558 & 9.956 & 46.584 & 0.532 & 8,849,084\\
DPA4C-Nano & 62.605 & 264.475 & 1.109 & 33.002 & 140.068 & 0.903 & 29,809\\
DPA4C-Mini & 25.531 & 127.918 & 0.803 & 22.870 & 116.344 & 0.781 & 145,513\\
DPA4C-Neo & 23.176 & 120.670 & 0.775 & 20.726 & 111.335 & 0.755 & 342,089\\
DPA4C-Air & 21.770 & 114.791 & 0.755 & 19.349 & 105.667 & 0.734 & 433,673\\
DPA4C-Plus & 19.798 & 110.031 & 0.724 & 17.121 & 98.672 & 0.701 & 1,456,961\\
GRACE-2L & 15.666 & 70.392 & 0.630 & 13.871 & 68.644 & 0.618 & 21,764,956\\
GRACE-1L & 21.669 & 94.243 & 0.719 & 19.722 & 90.735 & 0.702 & 4,461,497\\
MACE M & 18.045 & 80.157 & 0.711 & 16.314 & 78.398 & 0.701 & 9,063,204\\
MACE S & 19.596 & 87.763 & 0.748 & 17.865 & 85.944 & 0.738 & 8,222,244\\
Allegro M & 18.606 & 76.496 & 0.671 & 16.938 & 75.143 & 0.662 & 7,165,605\\
Allegro S & 21.564 & 90.635 & 0.742 & 19.826 & 88.873 & 0.733 & 1,397,029\\
NEP89 & 348.733 & 276.048 & 2.025 & 346.510 & 265.177 & 2.005 & 976,331\\
\bottomrule
\end{longtable}
\endgroup
}

\paragraph{Cubic-readout comparisons.}
In the four matched comparisons, cubic readout reduced force MAE (Supplementary Table~\ref{tab:omat24-cubic-validation}).
At $M=4$ with three hidden layers of width 128, the force-MAE reduction was 3.8\% before refinement and 0.44\% after one block.
These are single-seed comparisons.

\begin{table}[!htbp]\centering
\caption{Effect of cubic readout on mean absolute errors for the official OMat24 validation set.
Entries are C1 minus C0 for separately trained endpoints under the same recipe.
P1 denotes one ordinary refinement block.
H3$\times$128 and H3$\times$256 denote three hidden energy-network layers of width 128 and 256, respectively.}\label{tab:omat24-cubic-validation}
\begin{tabular}{@{}lrrr@{}}\toprule
Architecture & \shortstack{$\Delta$ energy MAE\\(meV/atom)} & \shortstack{$\Delta$ force MAE\\(meV/\AA{})} & \shortstack{$\Delta$ stress MAE\\(GPa)}\\\midrule
K32-M4/H3$\times$128 & $-0.624$ & $-4.411$ & $-0.011581$\\
K32-M4/H3$\times$256 & $-0.525$ & $-4.385$ & $-0.011191$\\
M4-P1/H3$\times$128 & $-0.165$ & $-0.462$ & $-0.001140$\\
M6-P1/H3$\times$256 & $-0.241$ & $-1.479$ & $-0.004174$\\
\bottomrule\end{tabular}
\end{table}

The fifth cubic model, the unrefined K32-M6/H3$\times$256, had no matched C0 checkpoint.
 Supplementary Table~\ref{tab:benchmarks-official} lists its absolute MAEs as M6+C1/H3$\times$256.

\paragraph{Summary and query controls.}
We trained S, $Q_0$ and A on the same structures, using seed 42 for initialization and the same sampling order.
After the initial $Q_0$ run became unstable, we resumed training from the checkpoint at 2.55 million updates.
A reduced force MAE by 5.87\% relative to S and by 2.50\% relative to $Q_0$ (Supplementary Table~\ref{tab:omat24-sqa-validation}).

\begin{table}[!htbp]\centering
\caption{Mean absolute errors for the seed-42 controls on the official OMat24 validation set.
S and A use terminal checkpoints at 4,000,000 updates.
$Q_0$ uses the resumed checkpoint selected at 3,950,000 updates.}\label{tab:omat24-sqa-validation}
\begin{tabular}{@{}lrrr@{}}\toprule
Control & Energy (meV/atom) & Force (meV/\AA{}) & 6-component stress (GPa)\\\midrule
S & 16.063 & 82.926 & 0.647028\\
$Q_0$ & 15.804 & 80.056 & 0.643586\\
A & 15.548 & 78.056 & 0.636060\\
\bottomrule\end{tabular}
\end{table}

\FloatBarrier

\subsection{Inference and throughput}\label{nav:sub22}\label{sec:si-moving}

The HEA16 and OMat24 cluster timings follow the protocol in Methods, ``Throughput measurements''.
LAMMPS built each lattice, used seeded type selections for the equiatomic alloy and initialized velocities with seed 2026. 
GPUMD loaded prepared extended-XYZ cells with seeded occupations and velocities. 
These realizations differ between engines. 
Both engines recreated the thermostat after warmup and continued particle positions and velocities. 
LAMMPS also continued its timestep count. 
The native thermostats used $3N-3$ degrees of freedom in LAMMPS and $3N$ in GPUMD. 
Both engines wrote thermodynamic output every 20 steps. LAMMPS used a 1\,\AA{} neighbour skin, checked for rebuilding at every step, sorted atoms every 50 steps in 2\,\AA{} bins and used \texttt{timer full nosync}. 
Its supported pair styles ran through Kokkos on the GPU.

DTM used TensorMD 2.8.3 in FP32 Speed mode, with 0.005\,\AA{} radial interpolation and automatic tiling and force scatter. 
The architecture survey requested Tensor Core backward kernels. 
Moment-attention models used their SIMT reaggregation path. 
The OMat24 cubic models used the full Cu and BCC16 cells and the same per-platform builds used for the standard readouts. 
Cubic base models used fused Tensor Core force evaluation with a tile size of 32,768. 
P1 models evaluated the U and V force derivatives separately with a tile size of 8,192. 
The S, $Q_0$ and A controls used automatic backward dispatch.
All three were timed on the same cells of 2,048,000 FCC Cu atoms and 1,024,000 BCC16 atoms on each GPU (Supplementary Tables~\ref{tab:throughput-hea16} and~\ref{tab:throughput-omat24}).

DPA4C used the 5 November 2026 development build of DeePMD-Kit 3.2.0, its CUDA2 runtime and 0.005\,\AA{} radial tables.
GRACE used version 0.6.1 LAMMPS pair styles, with the FP32 implementation for the universal models. 
MACE 0.3.16 ran through ML-IAP, and Allegro 0.8.3 with NequIP 0.19.1 through \texttt{pair\_allegro}. 
NEP used GPUMD 5.9.1~\cite{gpumd2025} on both GPUs and GPUMD-WPE 1.0.0~\cite{bu2026wpe} on RTX 5090 with identical potential parameters. 
Main Fig.~3 uses the WPE measurements. Supplementary Fig.~\ref{fig:si-hea16-timing} shows the A100 comparison.

Supplementary Tables~\ref{tab:throughput-hea16} and~\ref{tab:throughput-omat24} give the actual atom counts for each model and the median and observed range for each timing. 
For example, the OMat24 production models with moment attention used 256,000 Cu atoms and 128,000 BCC16 atoms on each GPU. 
On BCC16, cubic augmentation of the K32-M4/H3$\times$128 and K32-M4/H3$\times$256 bases reduced throughput by 25.5\% and 22.2\% on A100 and by 16.5\% and 10.6\% on RTX~5090, respectively.
For M4-P1/H3$\times$128 and M6-P1/H3$\times$256 the corresponding reductions were 51.4\% and 47.1\% on A100 and 42.5\% and 36.9\% on RTX 5090. 
These paired measurements include the wider energy-network input and its force kernels.

\begin{landscape}\formatHEATable{\begingroup\footnotesize\setlength{\tabcolsep}{2.5pt}
\begin{longtable}{@{}lrrrrrr@{}}
\caption{HEA16 throughput on Paratera N32H (A100) and N56R5 (RTX~5090).
Each entry is median [minimum, maximum] million atom-steps/s from three processes.
Cu and BCC16 are column groups; their actual atom counts are retained.
Superscript w identifies GPUMD-WPE; other NEP rows use GPUMD.
A dash denotes an unmeasured case.
Separate groups identify the architecture comparisons and matched S/$Q_0$/A controls.}\label{tab:throughput-hea16}\\
\toprule
Model & \shortstack{Cu\\atoms} & \shortstack{Cu\\A100} & \shortstack{Cu\\RTX~5090} & \shortstack{BCC16\\atoms} & \shortstack{BCC16\\A100} & \shortstack{BCC16\\RTX~5090}\\
\midrule\endfirsthead
\multicolumn{7}{l}{\textit{Supplementary Table \thetable{} (continued)}}\\
\toprule
Model & \shortstack{Cu\\atoms} & \shortstack{Cu\\A100} & \shortstack{Cu\\RTX~5090} & \shortstack{BCC16\\atoms} & \shortstack{BCC16\\A100} & \shortstack{BCC16\\RTX~5090}\\
\midrule\endhead\bottomrule\endfoot
\multicolumn{7}{l}{\textit{DTM trained on HEA16}}\\
K16-M3 & 2,048,000 & 28.849 [28.576, 28.859] & 60.291 [60.198, 60.528] & 1,024,000 & 26.717 [26.710, 26.721] & 67.775 [67.770, 67.997]\\
K16-M4/H4x64 & 2,048,000 & 24.476 [24.278, 24.524] & 46.064 [45.946, 46.070] & 1,024,000 & 24.892 [24.868, 24.928] & 51.426 [51.406, 51.448]\\
K16-M4/H3x128 & 2,048,000 & 18.488 [18.486, 18.490] & 34.098 [34.020, 34.283] & 1,024,000 & 18.662 [18.654, 18.677] & 37.501 [37.472, 37.621]\\
K32-M4/H3x256 & 2,048,000 & 9.655 [9.648, 9.659] & 20.064 [20.062, 20.118] & 1,024,000 & 8.691 [8.677, 8.693] & 19.837 [19.835, 19.860]\\
K32-M4/H3x128 & 2,048,000 & 14.294 [14.287, 14.294] & 28.935 [28.910, 29.017] & 1,024,000 & 12.463 [12.463, 12.469] & 28.992 [28.940, 28.995]\\
M4+C1 & 2,048,000 & 10.306 [10.306, 10.306] & 25.648 [25.634, 25.678] & 1,024,000 & 9.247 [9.214, 9.249] & 23.978 [23.973, 24.070]\\
M4-P1 & 2,048,000 & 6.478 [6.457, 6.479] & 13.284 [13.283, 13.356] & 1,024,000 & 5.762 [5.755, 5.776] & 11.767 [11.741, 11.973]\\
M4-P1+C1 & 2,048,000 & 2.761 [2.760, 2.766] & 7.044 [7.031, 7.078] & 1,024,000 & 2.790 [2.789, 2.798] & 7.137 [7.118, 7.140]\\
M4-P1+MA & 256,000 & 1.842 [1.842, 1.843] & 3.027 [3.025, 3.038] & 128,000 & 2.077 [2.076, 2.079] & 3.540 [3.525, 3.563]\\
M4-P2 & 2,048,000 & 4.690 [4.683, 4.696] & 9.441 [9.398, 9.461] & 1,024,000 & 4.347 [4.340, 4.357] & 8.322 [8.296, 8.355]\\
M4-P2+MA & 256,000 & 0.985 [0.984, 0.985] & 1.556 [1.551, 1.584] & 128,000 & 1.181 [1.180, 1.182] & 1.764 [1.748, 1.844]\\
M4-P4 & 2,048,000 & 3.029 [3.028, 3.029] & 6.023 [6.022, 6.037] & 1,024,000 & 2.890 [2.887, 2.893] & 5.896 [5.893, 5.916]\\
M4-P4+MA & 256,000 & 0.515 [0.515, 0.515] & 0.805 [0.803, 0.809] & 128,000 & 0.628 [0.628, 0.629] & 0.994 [0.993, 1.001]\\
K32-M6 & 2,048,000 & 9.622 [9.618, 9.636] & 17.436 [17.410, 17.459] & 1,024,000 & 8.804 [8.802, 8.806] & 17.713 [17.699, 17.766]\\
M6+C1 & 2,048,000 & 4.628 [4.617, 4.629] & 11.587 [11.563, 11.663] & 1,024,000 & 4.042 [4.040, 4.043] & 11.427 [11.421, 11.486]\\
M6-P1 & 2,048,000 & 3.435 [3.435, 3.437] & 5.745 [5.678, 5.903] & 1,024,000 & 3.314 [3.314, 3.315] & 6.281 [6.271, 6.287]\\
M6-P1+C1 & 2,048,000 & 1.580 [1.580, 1.581] & 3.678 [3.677, 3.678] & 1,024,000 & 1.632 [1.631, 1.632] & 3.929 [3.928, 3.935]\\
M6-P1+MA & 256,000 & 0.929 [0.929, 0.930] & 1.666 [1.652, 1.697] & 128,000 & 1.097 [1.096, 1.097] & 2.208 [2.207, 2.213]\\
M6-P2 & 2,048,000 & 2.319 [2.316, 2.319] & 4.134 [4.133, 4.134] & 1,024,000 & 2.264 [2.263, 2.267] & 3.721 [3.710, 3.754]\\
M6-P2+MA & 256,000 & 0.512 [0.512, 0.513] & 0.984 [0.984, 0.985] & 128,000 & 0.606 [0.606, 0.606] & 1.207 [1.206, 1.210]\\
M6-P4 & 2,048,000 & 1.430 [1.429, 1.430] & 2.415 [2.411, 2.416] & 1,024,000 & 1.381 [1.381, 1.382] & 2.249 [2.192, 2.314]\\
M6-P4+MA & 256,000 & 0.270 [0.270, 0.270] & 0.511 [0.511, 0.511] & 128,000 & 0.319 [0.319, 0.319] & 0.613 [0.613, 0.614]\\
K32-M8 & 2,048,000 & 5.037 [5.032, 5.038] & 7.823 [7.822, 7.835] & 1,024,000 & 5.006 [5.005, 5.007] & 8.078 [8.073, 8.079]\\
M8-P1 & 2,048,000 & 1.554 [1.554, 1.554] & 2.870 [2.841, 2.873] & 1,024,000 & 1.542 [1.540, 1.544] & 2.981 [2.981, 2.982]\\
M8-P1+MA & 256,000 & 0.570 [0.570, 0.571] & 1.002 [1.002, 1.003] & 128,000 & 0.641 [0.640, 0.641] & 1.233 [1.231, 1.236]\\
M8-P2 & 2,048,000 & 1.007 [1.007, 1.007] & 1.944 [1.944, 1.945] & 1,024,000 & 1.011 [1.010, 1.012] & 1.966 [1.966, 1.967]\\
M8-P2+MA & 256,000 & 0.317 [0.317, 0.317] & 0.571 [0.571, 0.571] & 128,000 & 0.361 [0.361, 0.362] & 0.671 [0.671, 0.674]\\
M8-P4 & 2,048,000 & 0.596 [0.596, 0.600] & 1.158 [1.158, 1.158] & 1,024,000 & 0.605 [0.605, 0.606] & 1.151 [1.149, 1.151]\\
M8-P4+MA & 256,000 & 0.167 [0.167, 0.167] & 0.303 [0.303, 0.303] & 128,000 & 0.193 [0.193, 0.193] & 0.361 [0.361, 0.362]\\
\multicolumn{7}{l}{\textit{External models in the alloy comparison}}\\
GRACE-FS-L & 2,048,000 & 0.262 [0.262, 0.262] & 0.587 [0.587, 0.587] & 1,024,000 & 0.270 [0.269, 0.270] & 0.608 [0.608, 0.608]\\
GRACE-FS-M-Aug & 2,048,000 & 0.386 [0.386, 0.386] & 0.907 [0.907, 0.907] & 1,024,000 & 0.405 [0.405, 0.405] & 0.959 [0.959, 0.959]\\
GRACE-FS-S & 2,048,000 & 0.509 [0.508, 0.512] & 1.130 [1.129, 1.130] & 1,024,000 & 0.562 [0.562, 0.562] & 1.247 [1.246, 1.248]\\
NEP44 & 2,048,000 & 15.356 [15.343, 15.359] & 33.985 [33.960, 34.052] & 1,024,000 & 12.763 [12.762, 12.763] & 31.720 [31.715, 31.747]\\
NEP44$^{\mathrm w}$ & 2,048,000 & -- & 103.546 [103.511, 103.588] & 1,024,000 & -- & 51.447 [51.421, 51.448]\\
NEP66 & 2,048,000 & 11.100 [11.100, 11.102] & 25.084 [25.054, 25.183] & 1,024,000 & 8.827 [8.819, 8.833] & 22.568 [22.566, 22.604]\\
NEP66$^{\mathrm w}$ & 2,048,000 & -- & 77.335 [77.135, 77.595] & 1,024,000 & -- & 39.370 [39.358, 39.374]\\
NEP88 & 2,048,000 & 7.916 [7.907, 7.918] & 19.162 [19.154, 19.217] & 1,024,000 & 6.529 [6.529, 6.529] & 16.795 [16.768, 16.841]\\
NEP88$^{\mathrm w}$ & 2,048,000 & -- & 80.934 [80.828, 81.207] & 1,024,000 & -- & 30.093 [30.090, 30.108]\\
\multicolumn{7}{l}{\textit{Matched HEA16 controls: K32/M6/P4, $3\times128$ energy network}}\\
S & 2,048,000 & 0.715 [0.715, 0.716] & 1.295 [1.294, 1.296] & 1,024,000 & 0.708 [0.708, 0.708] & 1.300 [1.299, 1.301]\\
$Q_0$ & 2,048,000 & 0.283 [0.282, 0.283] & 0.550 [0.549, 0.551] & 1,024,000 & 0.336 [0.336, 0.336] & 0.651 [0.647, 0.652]\\
A & 2,048,000 & 0.286 [0.285, 0.286] & 0.572 [0.572, 0.573] & 1,024,000 & 0.334 [0.334, 0.334] & 0.669 [0.668, 0.670]\\
\end{longtable}\endgroup
}
\formatOMatTable{\begingroup\footnotesize\setlength{\tabcolsep}{2.5pt}
\begin{longtable}{@{}lrrrrrr@{}}
\caption{OMat24 throughput on Paratera N32H (A100) and N56R5 (RTX~5090).
Each entry is median [minimum, maximum] million atom-steps/s from three processes.
Cu and BCC16 are column groups; their actual atom counts are retained.
Superscript w identifies GPUMD-WPE; other NEP rows use GPUMD.
A dash denotes an unmeasured case.
Separate groups identify the architecture comparisons and matched S/$Q_0$/A controls.}\label{tab:throughput-omat24}\label{tab:benchmarks-throughput}\\
\toprule
Model & \shortstack{Cu\\atoms} & \shortstack{Cu\\A100} & \shortstack{Cu\\RTX~5090} & \shortstack{BCC16\\atoms} & \shortstack{BCC16\\A100} & \shortstack{BCC16\\RTX~5090}\\
\midrule\endfirsthead
\multicolumn{7}{l}{\textit{Supplementary Table \thetable{} (continued)}}\\
\toprule
Model & \shortstack{Cu\\atoms} & \shortstack{Cu\\A100} & \shortstack{Cu\\RTX~5090} & \shortstack{BCC16\\atoms} & \shortstack{BCC16\\A100} & \shortstack{BCC16\\RTX~5090}\\
\midrule\endhead\bottomrule\endfoot
\multicolumn{7}{l}{\textit{DTM trained on OMat24}}\\
K8-M2/H2x64 & 2,048,000 & 48.132 [47.607, 49.353] & 87.761 [87.726, 88.188] & 1,024,000 & 50.044 [49.993, 50.045] & 105.450 [105.446, 106.122]\\
K8-M3/H4x64 & 2,048,000 & 35.447 [35.442, 35.459] & 73.581 [73.268, 73.862] & 1,024,000 & 36.713 [36.290, 36.781] & 88.732 [88.575, 88.806]\\
K16-M2/H2x64 & 2,048,000 & 39.805 [39.764, 39.819] & 73.411 [73.360, 73.637] & 1,024,000 & 33.475 [33.473, 33.512] & 87.087 [86.984, 87.113]\\
K16-M2/H4x64 & 2,048,000 & 35.864 [35.857, 35.931] & 64.889 [64.743, 65.010] & 1,024,000 & 30.368 [30.325, 30.416] & 77.157 [77.102, 77.373]\\
K16-M3 & 2,048,000 & 29.117 [29.112, 29.129] & 60.607 [60.440, 60.637] & 1,024,000 & 26.701 [26.626, 26.701] & 68.813 [68.013, 69.168]\\
K32-M4/H3x128 & 2,048,000 & 14.480 [14.474, 14.488] & 28.497 [28.444, 28.672] & 1,024,000 & 12.419 [12.357, 12.471] & 29.059 [29.032, 29.091]\\
M4+C1/H3x128 & 2,048,000 & 10.478 [10.472, 10.479] & 25.147 [24.976, 25.319] & 1,024,000 & 9.258 [9.257, 9.259] & 24.268 [24.233, 24.369]\\
K32-M4/H3x256 & 2,048,000 & 9.561 [9.559, 9.568] & 20.066 [20.055, 20.102] & 1,024,000 & 8.735 [8.735, 8.737] & 19.931 [19.917, 20.014]\\
M4+C1/H3x256 & 2,048,000 & 7.281 [7.280, 7.284] & 17.751 [17.724, 17.865] & 1,024,000 & 6.795 [6.795, 6.802] & 17.820 [17.737, 17.937]\\
M4-P1/H3x128 & 2,048,000 & 6.400 [6.397, 6.417] & 13.182 [13.180, 13.193] & 1,024,000 & 5.811 [5.807, 5.818] & 11.959 [11.878, 12.319]\\
M4-P1+C1/H3x128 & 2,048,000 & 2.772 [2.767, 2.772] & 6.944 [6.931, 6.959] & 1,024,000 & 2.822 [2.821, 2.823] & 6.876 [6.872, 6.882]\\
M4-P1/H3x256 & 2,048,000 & 5.085 [5.066, 5.090] & 9.992 [9.911, 10.178] & 1,024,000 & 4.627 [4.621, 4.634] & 11.287 [11.278, 11.352]\\
M6+C1/H3x256 & 2,048,000 & 3.781 [3.781, 3.782] & 9.633 [9.627, 9.696] & 1,024,000 & 3.406 [3.405, 3.412] & 9.521 [9.499, 9.570]\\
M6-P1 & 2,048,000 & 3.012 [3.010, 3.013] & 5.746 [5.740, 5.749] & 1,024,000 & 2.851 [2.845, 2.856] & 5.914 [5.912, 5.918]\\
M6-P1+C1/H3x256 & 2,048,000 & 1.441 [1.440, 1.443] & 3.512 [3.507, 3.517] & 1,024,000 & 1.509 [1.509, 1.511] & 3.734 [3.732, 3.740]\\
M8-P1+MA & 256,000 & 0.519 [0.519, 0.519] & 0.957 [0.957, 0.958] & 128,000 & 0.581 [0.581, 0.582] & 1.166 [1.165, 1.169]\\
M8-P2+MA & 256,000 & 0.300 [0.299, 0.300] & 0.552 [0.552, 0.553] & 128,000 & 0.343 [0.342, 0.343] & 0.650 [0.649, 0.650]\\
M8-P4+MA & 256,000 & 0.160 [0.160, 0.160] & 0.296 [0.295, 0.296] & 128,000 & 0.185 [0.185, 0.185] & 0.351 [0.351, 0.352]\\
\multicolumn{7}{l}{\textit{External models in the universal-potential comparison}}\\
Allegro M & 4,000 & 0.006 [0.006, 0.006] & 0.013 [0.013, 0.013] & 7,168 & 0.008 [0.008, 0.008] & 0.017 [0.017, 0.017]\\
Allegro S & 16,384 & 0.039 [0.039, 0.039] & 0.062 [0.062, 0.062] & 8,192 & 0.058 [0.058, 0.058] & 0.092 [0.092, 0.092]\\
GRACE-1L medium & 2,048,000 & 0.177 [0.177, 0.177] & 0.408 [0.408, 0.408] & 1,024,000 & 0.205 [0.205, 0.205] & 0.458 [0.458, 0.459]\\
GRACE-2L medium & 16,384 & 0.042 [0.042, 0.042] & 0.082 [0.082, 0.082] & 8,192 & 0.051 [0.051, 0.051] & 0.100 [0.100, 0.100]\\
DPA4C-Nano & 2,048,000 & 32.691 [32.492, 33.101] & 81.653 [81.617, 81.820] & 1,024,000 & 33.188 [33.153, 33.318] & 72.185 [72.050, 72.266]\\
DPA4C-Mini & 2,048,000 & 13.907 [13.898, 13.913] & 39.663 [39.580, 39.786] & 1,024,000 & 13.736 [13.702, 13.813] & 30.213 [30.161, 30.270]\\
DPA4C-Neo & 2,048,000 & 8.777 [8.759, 8.780] & 23.427 [23.371, 23.480] & 1,024,000 & 8.968 [8.968, 8.988] & 17.446 [17.418, 17.501]\\
DPA4C-Air & 2,048,000 & 6.733 [6.718, 6.734] & 16.177 [16.166, 16.235] & 1,024,000 & 5.754 [5.745, 5.758] & 11.520 [11.519, 11.576]\\
DPA4C-Plus & 2,048,000 & 3.670 [3.669, 3.675] & 9.091 [9.082, 9.213] & 1,024,000 & 2.842 [2.842, 2.842] & 5.790 [5.782, 5.829]\\
NEP89 & 2,048,000 & 15.292 [15.274, 15.298] & 33.713 [33.707, 33.751] & 1,024,000 & 12.758 [12.758, 12.759] & 31.654 [31.528, 31.764]\\
NEP89$^{\mathrm w}$ & 2,048,000 & -- & 101.614 [101.451, 101.720] & 1,024,000 & -- & 52.496 [52.476, 52.509]\\
MACE Medium & 4,000 & 0.046 [0.046, 0.046] & 0.072 [0.072, 0.072] & 8,192 & 0.059 [0.059, 0.059] & 0.095 [0.094, 0.095]\\
MACE Small & 16,384 & 0.071 [0.071, 0.071] & 0.115 [0.115, 0.116] & 8,192 & 0.084 [0.084, 0.084] & 0.141 [0.141, 0.141]\\
\multicolumn{7}{l}{\textit{Matched OMat24 controls: K32/M6/P4, $3\times256$ energy network}}\\
\multicolumn{7}{@{}p{0.93\linewidth}@{}}{S and A were scored and timed at their terminal checkpoints.
$Q_0$ was scored at the resumed 3.95-million-update checkpoint and timed at the original 4-million-update checkpoint.
Its timing measures the cost of the operator.}\\*
S & 2,048,000 & 0.685 [0.685, 0.685] & 1.263 [1.263, 1.264] & 1,024,000 & 0.680 [0.679, 0.680] & 1.265 [1.262, 1.272]\\
$Q_0$ & 2,048,000 & 0.281 [0.281, 0.281] & 0.551 [0.551, 0.551] & 1,024,000 & 0.326 [0.325, 0.326] & 0.649 [0.649, 0.649]\\
A & 2,048,000 & 0.284 [0.284, 0.284] & 0.566 [0.564, 0.566] & 1,024,000 & 0.329 [0.329, 0.329] & 0.670 [0.670, 0.670]\\
\end{longtable}\endgroup
}\end{landscape}

\subsubsection*{Additional NEP implementation on FCC Cu}\phantomsection\label{sec:si-misc}\label{sec:si-sc26-nep}

On a separate RTX~5090 workstation, we tested the \texttt{loop\_trans} implementation of Song et al.~\cite{song2026sharednep} on the full FCC Cu cell, using the timing intervals specified in Methods.
A three-thermostat chain ran continuously through warmup and timing, and we recorded the pressure tensor at the run boundaries.

We compiled for the RTX~5090 (SM120) and used the authors' generator to set the model dimensions.
We enabled the existing global pressure calculation and distributed the one-time setup of the NEP89 coefficients across GPU blocks to meet the launch limit.
Removing unused NEP88 buffers allowed the full cell to fit in GPU memory.

\begin{table}[!htb]
\centering
\caption{Median FCC Cu throughput in million atom-steps/s. The \texttt{loop\_trans} measurements used the workstation. The GPUMD and GPUMD-WPE measurements used the Paratera RTX~5090 (Supplementary Tables~\ref{tab:throughput-hea16} and~\ref{tab:throughput-omat24}).}
\label{tab:si-sc26-cu}
\begin{tabular}{@{}lrrr@{}}
\toprule
Model & SC26 \texttt{loop\_trans} & GPUMD & GPUMD-WPE \\
\midrule
NEP44 & 62.068 & 33.985 & 103.546 \\
NEP66 & 53.621 & 25.084 & 77.335 \\
NEP88 & 45.312 & 19.162 & 80.934 \\
NEP89 & 27.718 & 33.713 & 101.614 \\
\bottomrule
\end{tabular}
\end{table}

The tested force kernel omits short-range Ziegler--Biersack--Littmark (ZBL) repulsion. 
All pairs in the Cu replays lay outside the ZBL cutoff. 
A separate BCC16 check contained pairs within the cutoff, where the missing term affects forces. We therefore report the workstation comparison only for FCC Cu. 
GPUMD-WPE gave the highest throughput for each of the four Cu models.

\FloatBarrier

\subsection{Static properties and strain sensitivity}\label{nav:sub23}\label{sec:si-omat-static}

We evaluated the \href{https://huggingface.co/datasets/materialyze/matcalc-bench}{MatCalc benchmark collection}~\cite{matcalc} under the run conditions in Methods, ``Physical and dynamical benchmarks.'' 
The 13 DTM checkpoints and 14 external calculators, including DPA4.1 Nano and Mini and all 5 DPA4C models, attempted the same material lists.
The force-scale dataset contained 9,308 frames. 
We relaxed the canonical elemental reference phases separately for each model. 
For N and F with MACE Small and Medium we used single-point energies at the fixed DFT-selected geometries. 
DPA4.1 Nano and Mini used uncompressed exports qualified against their official-validation checkpoints.

Supplementary Table~\ref{tab:benchmarks-properties} reports means and counts for finite successful outcomes. 
Paired comparisons use the intersection of the relevant models' successful materials. 
Phonon relaxations targeted a maximum force of 0.05\,eV/\AA{} within 1,000 steps. 
For released models, acceptance required both this terminal force criterion and a completed, finite heat-capacity result. 
For DTM, we retained the original recorded completions with finite heat capacity.
Completion ranged from 96.3\% to 100.0\% across models. 
Heat-capacity distributions use these accepted populations. 
Imaginary-mode stability was recorded separately.

\formatOMatTable{
\par
\begingroup
\footnotesize
\setlength{\tabcolsep}{3.3pt}
\begin{longtable}{@{}lrrrrrrrrr@{}}
\caption{Original-protocol successful-outcome means and coverage for all 27 models. $n$ counts successful cases out of 972 equilibrium, 3,953 elasticity and 1,170 phonon materials, respectively. Formation errors are in meV/atom, bulk and shear errors in GPa, and heat-capacity errors in J mol$^{-1}$ K$^{-1}$. Distance and force-scale deviation are dimensionless. Every mean retains all finite successful outcomes; all failures remain in the source data. Original strains are primary for every model; halved-strain sensitivity results are reported separately.}\label{tab:benchmarks-properties}\label{tab:benchmarks-property-coverage}\\
\toprule
Model & \multicolumn{3}{c}{Equilibrium} & \multicolumn{3}{c}{Elasticity} & \multicolumn{2}{c}{Phonon} & Force scale\\
\cmidrule(lr){2-4}\cmidrule(lr){5-7}\cmidrule(lr){8-9}
 & $n$ & Formation & Distance & $n$ & Bulk & Shear & $n$ & $C_V$ & $|s-1|$\\
\midrule
\endfirsthead
\multicolumn{10}{l}{\textit{Continued}}\\
\toprule
Model & \multicolumn{3}{c}{Equilibrium} & \multicolumn{3}{c}{Elasticity} & \multicolumn{2}{c}{Phonon} & Force scale\\
\cmidrule(lr){2-4}\cmidrule(lr){5-7}\cmidrule(lr){8-9}
 & $n$ & Formation & Distance & $n$ & Bulk & Shear & $n$ & $C_V$ & $|s-1|$\\
\midrule
\endhead
K8-M2/H2x64 & 947 & 77.952 & 0.3628 & 3921 & 26.258 & 18.111 & 1164 & 7.657 & 0.0584\\
K8-M3/H4x64 & 947 & 67.207 & 0.3143 & 3915 & 20.951 & 20.648 & 1162 & 6.037 & 0.0502\\
K16-M2/H2x64 & 946 & 60.659 & 0.3380 & 3924 & 24.624 & 16.967 & 1158 & 7.325 & 0.0559\\
K16-M2/H4x64 & 954 & 58.714 & 0.3296 & 3922 & 23.885 & 17.740 & 1170 & 7.242 & 0.0534\\
K16-M3 & 949 & 60.456 & 0.2962 & 3909 & 19.838 & 15.473 & 1170 & 6.092 & 0.0456\\
K32-M4/H3x128 & 955 & 51.594 & 0.2710 & 3927 & 18.936 & 18.499 & 1170 & 5.924 & 0.0416\\
K32-M4/H3x256 & 960 & 37.739 & 0.2681 & 3931 & 17.943 & 14.766 & 1169 & 5.357 & 0.0439\\
M4-P1/H3x128 & 963 & 40.374 & 0.2686 & 3929 & 16.728 & 15.068 & 1170 & 5.336 & 0.0368\\
M4-P1/H3x256 & 958 & 41.826 & 0.2461 & 3932 & 17.574 & 14.416 & 1169 & 5.808 & 0.0394\\
M6-P1 & 964 & 35.484 & 0.2384 & 3927 & 15.622 & 13.588 & 1166 & 5.708 & 0.0377\\
M8-P1$+$MA & 956 & 30.519 & 0.2265 & 3934 & 14.966 & 15.445 & 1170 & 5.665 & 0.0359\\
M8-P2$+$MA & 963 & 28.282 & 0.2168 & 3936 & 14.044 & 16.896 & 1170 & 5.405 & 0.0354\\
M8-P4$+$MA & 964 & 28.927 & 0.2137 & 3941 & 13.511 & 24.789 & 1170 & 5.240 & 0.0301\\
DPA4.1 Nano & 961 & 36.597 & 0.2591 & 3932 & 12.437 & 19.941 & 1164 & 4.984 & 0.0407\\
DPA4.1 Mini & 966 & 27.370 & 0.2100 & 3937 & 11.399 & 13.805 & 1168 & 4.149 & 0.0302\\
GRACE-1L & 956 & 47.839 & 0.2441 & 3934 & 16.388 & 14.340 & 1167 & 5.329 & 0.0417\\
GRACE-2L & 965 & 30.089 & 0.2206 & 3935 & 14.864 & 14.254 & 1170 & 4.154 & 0.0304\\
Allegro S & 952 & 46.554 & 0.2477 & 3934 & 17.771 & 15.217 & 1167 & 6.329 & 0.0406\\
Allegro M & 953 & 30.529 & 0.2108 & 3931 & 15.483 & 14.818 & 1168 & 5.216 & 0.0304\\
MACE S & 961 & 39.083 & 0.2462 & 3927 & 15.754 & 14.671 & 1168 & 4.927 & 0.0375\\
MACE M & 955 & 33.870 & 0.2410 & 3937 & 14.625 & 14.866 & 1167 & 4.651 & 0.0356\\
NEP89 & 910 & 184.771 & 0.4910 & 3874 & 24.018 & 21.096 & 1127 & 10.727 & 0.1498\\
DPA4C-Nano & 942 & 65.778 & 0.3504 & 3913 & 16.236 & 18.477 & 1157 & 7.988 & 0.0692\\
DPA4C-Mini & 957 & 47.313 & 0.2987 & 3929 & 17.093 & 15.997 & 1167 & 6.298 & 0.0510\\
DPA4C-Neo & 956 & 39.853 & 0.2798 & 3925 & 17.659 & 23.721 & 1166 & 5.933 & 0.0508\\
DPA4C-Air & 959 & 42.022 & 0.2691 & 3931 & 17.033 & 18.121 & 1168 & 5.958 & 0.0457\\
DPA4C-Plus & 959 & 53.786 & 0.2525 & 3932 & 16.714 & 16.253 & 1169 & 5.366 & 0.0428\\
\bottomrule
\end{longtable}
\noindent Force-scale and dynamics coverage is 979/979 and 17/17 for every model. Conductivity scores and failures are in Table~\ref{tab:benchmarks-kappa}. Correlations use fixed material intersections.\par
\endgroup
}

\paragraph{Uniform strain sensitivity.}
For the 10 DTM checkpoints evaluated on both grids, we halved the strain amplitudes to normal strains of $\pm$0.0025 and $\pm$0.005 and tensor-shear strains of $\pm$0.015 and $\pm$0.03.
Across all 3,953 materials we used the same checkpoints, parent relaxations and numerical settings on both grids. 
Main Fig.~4, Supplementary Table~\ref{tab:benchmarks-properties} and Supplementary Fig.~\ref{fig:si-properties} use the original grid for every model. 
Supplementary Table~\ref{tab:half-strain} and Supplementary Fig.~\ref{fig:si-strain} compare the two grids.

Halving the strain amplitudes changed shear MAE from 24.7887 to 14.6529\,GPa for M8-P4+MA and from 16.8963 to 14.7611\,GPa for M8-P2+MA. 
The original P4 mean includes TiPt, whose absolute shear error of 43,219.59\,GPa contributed 10.97\,GPa to that mean. 
Both models' shear medians and bulk means increased on the halved grid. 
Across the 10 models, 4 shear means decreased and 6 increased.
The tensor diagnostics quantify negative Kelvin-stiffness modes and cancellation in the Reuss denominator.

\Needspace{55mm}

\formatOMatTable{{
\scriptsize
\setlength{\tabcolsep}{3.8pt}
\begin{longtable}{@{}lrrrrr@{}}
\caption{Elasticity for the 10 DTM models evaluated at both original and halved strain amplitudes.
Shear and bulk errors are in GPa.
The Negative column counts half-strain fits with a negative Kelvin-stiffness eigenvalue.}\label{tab:half-strain}\\
\toprule
Model & Success original/half & Shear original & Shear half & Bulk original/half & Negative\\
\midrule
\endfirsthead
\multicolumn{6}{l}{\textit{Continued}}\\
\toprule
Model & Success original/half & Shear original & Shear half & Bulk original/half & Negative\\
\midrule
\endhead
\bottomrule
\endfoot
M8-P4+MA & 3941 $\to$ 3941 & 24.7887 & 14.6529 & 13.5113 $\to$ 14.1693 & 127\\
M8-P2+MA & 3936 $\to$ 3937 & 16.8963 & 14.7611 & 14.0441 $\to$ 14.5941 & 137\\
M8-P1+MA & 3934 $\to$ 3932 & 15.4454 & 15.9087 & 14.9659 $\to$ 15.4275 & 149\\
M6-P1 & 3927 $\to$ 3926 & 13.5882 & 14.2156 & 15.6221 $\to$ 15.9344 & 140\\
M4-P1/H3x256 & 3932 $\to$ 3933 & 14.4157 & 17.6430 & 17.5742 $\to$ 17.9125 & 147\\
M4-P1/H3x128 & 3929 $\to$ 3928 & 15.0677 & 15.3938 & 16.7277 $\to$ 16.9071 & 138\\
K32-M4/H3x256 & 3931 $\to$ 3929 & 14.7659 & 14.9397 & 17.9433 $\to$ 18.1057 & 145\\
K32-M4/H3x128 & 3927 $\to$ 3928 & 18.4991 & 14.9820 & 18.9360 $\to$ 19.0659 & 142\\
K16-M3 & 3909 $\to$ 3915 & 15.4726 & 15.8204 & 19.8378 $\to$ 20.0178 & 135\\
K16-M2/H4x64 & 3922 $\to$ 3918 & 17.7403 & 17.7188 & 23.8850 $\to$ 23.9658 & 156\\
\end{longtable}
}
}

\subsection{Finite-temperature dynamics}\label{nav:sub24}\label{sec:si-omat-dynamics}

We scored the trajectories described in Methods with the archived \href{https://github.com/janosh/matbench-discovery/blob/902a309e0f6ec3738f7847fb09785a125ef79aa8/matbench_discovery/metrics/md.py}{Matbench Discovery implementation}. 
Let $g$ denote the radial distribution function, $p$ the normalized angle distribution, $S$ the normalized vibrational spectrum and $W_1$ the one-dimensional Wasserstein distance. The percentage errors are

\begin{equation}\begin{aligned}
e_{\rm RDF}&=100\min\!\left(1,\frac{\int|g_{\rm ref}(r)-g_{\rm pred}(r)|\,dr}{\int|g_{\rm ref}(r)-1|\,dr}\right),\\
e_{\rm ADF}&=100\min\!\left(1,\frac{W_1(p_{\rm ref},p_{\rm pred})}{W_1(p_{\rm ref},p_{\sin\theta})}\right),\\
e_{\rm vDOS}&=100\min\!\left(1,\frac{W_1(S_{\rm ref},S_{\rm pred})}{\sigma_{\rm ref}}\right).
\end{aligned}
\label{eq:si-dynamics-scores}\end{equation}

RDFs use 500 radial bins over the common minimum-image range. ADFs use 180 angular bins and include bonds within 1.25 times the sum of the ASE covalent radii. 
The normalized structureless angular background is $p_{\sin\theta}$. 
To construct the vibrational spectra, we used Hann-windowed, per-degree-of-freedom-normalized power from position-derived velocities. 
We interpolated both spectra onto a common frequency grid, converted them to trapezoidal bin masses and restricted them to the band containing 99.9\% of reference spectral power before normalization. 
Here $\sigma_{\rm ref}$ is the reference frequency standard deviation in that band. 
Dyna-Mat~\cite{dynmat2026} scores vDOS by spectral overlap. 
Our score uses the normalized Wasserstein distance above.

All 27 calculators completed every trajectory. 
Supplementary Table~\ref{tab:benchmarks-dynamics} gives the nonmolecular and molecular-crystal results separately with equal weight per system. 
The five molecular crystals are anthracene, naphthalene, pentacene, picene and tetracene.
Supplementary Fig.~\ref{fig:si-dynamics} gives per-system RDF, ADF and vDOS errors for all 17 systems. 
For the 11 nonmolecular systems with reference pressures, the pressure error is $|\langle p\rangle_{\rm pred}-\langle p\rangle_{\rm ref}|$, with each trajectory averaged separately.

\subsection{Thermal conductivity}\label{nav:sub25}\label{sec:si-omat-conductivity}

For each successful material, $c_{qb}$ is the stored contribution of phonon mode $b$ at wavevector $q$, averaged over the three diagonal conductivity components and weighted by $w_q$.
With $W=\sum_q w_q$, the isotropic conductivity is $\kappa=W^{-1}\sum_{qb}c_{qb}$. The mode-resolved score is

\begin{equation}\kappa_{\rm SRME}=\frac{2\sum_{qb}|c_{qb}^{\rm pred}-c_{qb}^{\rm ref}|}{W(\kappa_{\rm pred}+\kappa_{\rm ref})}.
\label{eq:si-conductivity-score}\end{equation}

Reference and predicted contributions use the same q-point ordering and already include the q-point weights. 
Supplementary Table~\ref{tab:benchmarks-dynamics} reports the mean and median over all attempted materials and the failure count, assigning the prescribed score of 2 to each failed calculation~\cite{pota2026}.
For the conditional comparison, we averaged scores over the materials on which both models succeeded.

DTM-OMat24-L (M8-P2+MA) succeeded on all 103 materials. 
DTM-OMat24-XL (M8-P4+MA) produced imaginary modes for AgBr, AgCl and CuCl. 
The three failure penalties raised the XL mean to 0.501233, above the L mean of 0.469255.
On the 100 jointly successful materials, XL had the lower mean, 0.456270 compared with 0.474869 for L.

\formatOMatTable{\begingroup\footnotesize\setlength{\tabcolsep}{2.5pt}
\begin{longtable}{@{}lrrrrrrrr@{}}
\caption{Finite-temperature dynamics and thermal conductivity for the same 27 models.
RDF and vDOS errors (percent) average 12 nonmolecular or 5 molecular systems, indicated by subscripts.
$|\Delta\bar P|$ is the mean absolute trajectory-mean pressure difference (GPa) on 11 reference-bearing systems.
Conductivity mean and median are $\kappa_{\rm SRME}$ on all 103 materials, including the prescribed failure score of 2.}\label{tab:benchmarks-dynamics}\label{tab:benchmarks-kappa}\\
\toprule
Model & RDF$_{12}$ & RDF$_5$ & vDOS$_{12}$ & vDOS$_5$ & $|\Delta\bar P|$ & \shortstack{Mean\\$\kappa_{\mathrm{SRME}}$} & \shortstack{Median\\$\kappa_{\mathrm{SRME}}$} & Failures\\
\midrule\endfirsthead
\multicolumn{9}{l}{\textit{Supplementary Table \thetable{} (continued)}}\\
\toprule
Model & RDF$_{12}$ & RDF$_5$ & vDOS$_{12}$ & vDOS$_5$ & $|\Delta\bar P|$ & \shortstack{Mean\\$\kappa_{\mathrm{SRME}}$} & \shortstack{Median\\$\kappa_{\mathrm{SRME}}$} & Failures\\
\midrule\endhead\bottomrule\endfoot
K8-M2/H2x64 & 5.877 & 19.512 & 17.361 & 50.423 & 0.6242 & 0.7767 & 0.6591 & 2\\
K8-M3/H4x64 & 6.146 & 8.710 & 17.109 & 43.690 & 0.5270 & 0.5971 & 0.5142 & 2\\
K16-M2/H2x64 & 6.574 & 15.711 & 17.889 & 45.230 & 0.9361 & 0.7019 & 0.5708 & 3\\
K16-M2/H4x64 & 5.631 & 20.154 & 16.388 & 41.553 & 0.8328 & 0.6326 & 0.5545 & 2\\
K16-M3 & 4.840 & 11.687 & 16.113 & 45.627 & 0.6409 & 0.5211 & 0.4848 & 1\\
K32-M4/H3x128 & 4.355 & 7.971 & 15.182 & 44.759 & 0.9273 & 0.5726 & 0.4613 & 2\\
K32-M4/H3x256 & 4.963 & 43.934 & 15.211 & 43.124 & 0.6128 & 0.5620 & 0.4649 & 0\\
M4-P1/H3x128 & 4.421 & 11.491 & 15.684 & 40.163 & 0.5761 & 0.5623 & 0.4396 & 3\\
M4-P1/H3x256 & 4.315 & 20.342 & 15.328 & 43.447 & 0.4908 & 0.5004 & 0.4020 & 2\\
M6-P1 & 3.410 & 17.098 & 14.660 & 41.890 & 0.7451 & 0.5599 & 0.4115 & 2\\
M8-P1$+$MA & 3.681 & 12.203 & 15.301 & 38.456 & 0.5457 & 0.5242 & 0.3525 & 3\\
M8-P2$+$MA & 3.782 & 13.677 & 14.339 & 43.653 & 0.5727 & 0.4693 & 0.3262 & 0\\
M8-P4$+$MA & 3.607 & 6.401 & 14.206 & 38.929 & 0.5174 & 0.5012 & 0.3187 & 3\\
DPA4.1 Nano & 4.779 & 11.096 & 14.774 & 46.992 & 0.7662 & 0.4252 & 0.3342 & 1\\
DPA4.1 Mini & 3.648 & 6.711 & 14.039 & 40.437 & 0.6174 & 0.2520 & 0.1804 & 2\\
GRACE-1L & 3.950 & 8.362 & 15.193 & 40.998 & 0.7076 & 0.3892 & 0.3258 & 1\\
GRACE-2L & 3.578 & 4.224 & 13.907 & 41.810 & 0.6558 & 0.2123 & 0.1700 & 1\\
Allegro S & 3.993 & 9.272 & 14.909 & 40.599 & 0.4178 & 0.6161 & 0.4242 & 3\\
Allegro M & 3.498 & 5.747 & 14.685 & 42.246 & 0.4875 & 0.4211 & 0.3097 & 1\\
MACE S & 3.611 & 9.227 & 14.648 & 44.215 & 0.6249 & 0.2978 & 0.2588 & 0\\
MACE M & 3.649 & 8.378 & 13.893 & 42.269 & 0.3709 & 0.2452 & 0.2359 & 0\\
NEP89 & 9.214 & 7.009 & 22.783 & 42.186 & 3.5793 & 1.2074 & 1.1902 & 13\\
DPA4C-Nano & 6.591 & 24.884 & 18.152 & 49.997 & 0.8966 & 0.6127 & 0.5472 & 1\\
DPA4C-Mini & 4.798 & 19.201 & 16.724 & 43.990 & 0.7945 & 0.5179 & 0.4645 & 0\\
DPA4C-Neo & 4.806 & 15.916 & 16.221 & 43.993 & 0.7803 & 0.5482 & 0.4267 & 2\\
DPA4C-Air & 3.984 & 8.888 & 15.483 & 42.732 & 0.6185 & 0.5388 & 0.3991 & 3\\
DPA4C-Plus & 3.678 & 10.143 & 14.703 & 41.061 & 0.6738 & 0.5000 & 0.3638 & 1\\
\end{longtable}\endgroup
}

\subsection{Validation errors and physical predictions}\label{nav:sub26}
\label{sec:si-mae-benchmarks}
Validation errors summarize sampled energies and derivatives, while physical observables probe the fitted landscape along particular configurations and response directions~\cite{fu2025esen}. 
We assessed this relationship for 26 model identities matched by checkpoint or deployed artifact (13 DTM, 5 DPA4C and 2 each from DPA4.1, GRACE, MACE and Allegro).
All use the official OMat24 validation population. 
Source Data also report the correlations for all 27 models, including NEP89 evaluated against the original OMat24 labels.

\begin{figure}[p]\centering\includegraphics[width=\linewidth,height=0.78\textheight,keepaspectratio]{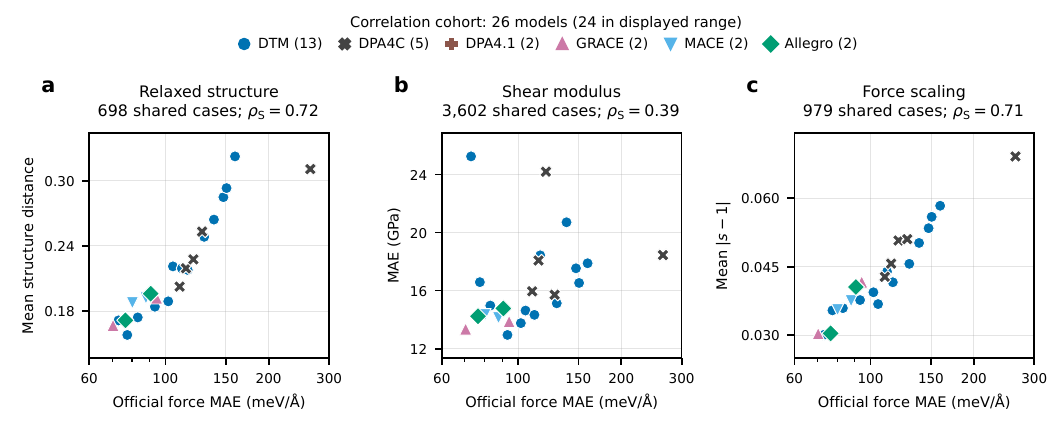}
\caption{\textbf{Validation force MAE and errors in relaxed structure, shear modulus and force scaling.}
\textbf{a}, Mean relaxed-structure distance on 698 materials completed by all models.
\textbf{b}, Shear-modulus MAE on 3,602 materials completed by all models at the original strain amplitudes.
\textbf{c}, Mean force-scale deviation on 979 materials.
The displayed OMat24 validation force-MAE range (60--300\,meV/\AA{}) contains 24 models.
Spearman correlations use all 26 models.
Supplementary Table~\ref{tab:si-mae-benchmarks} gives the correlations for each model group.}\label{fig:si-mae-physical}\end{figure}

\begin{figure}[p]\centering\includegraphics[width=\linewidth,height=0.78\textheight,keepaspectratio]{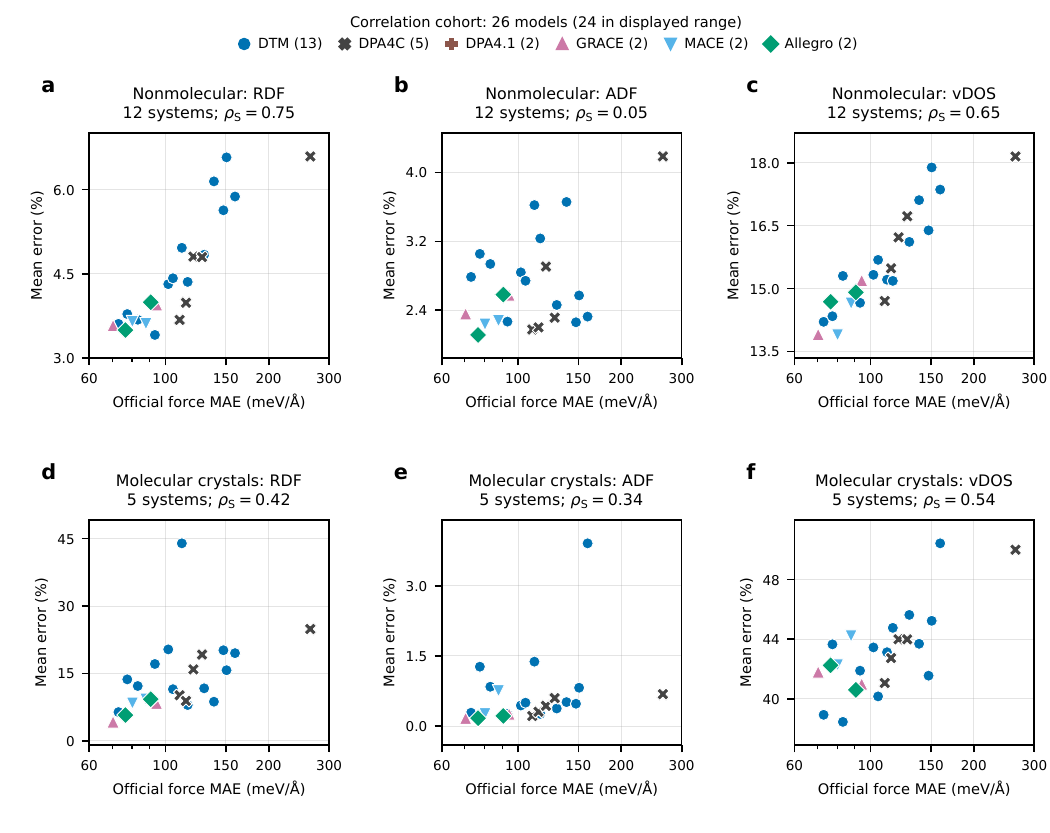}
\caption{\textbf{Validation force MAE and dynamics errors.}
\textbf{a--c}, Mean RDF, ADF and vDOS errors over the 12 nonmolecular systems, with equal weight per system.
\textbf{d--f}, The same errors averaged equally over 5 molecular crystals.
The molecular-crystal reference trajectories include D3, whereas the model calculations apply no D3 correction.
The displayed OMat24 validation force-MAE range (60--300\,meV/\AA{}) contains 24 models.
Spearman correlations use all 26 models.}\label{fig:si-mae-dynamics}\end{figure}

For each benchmark, Spearman $\rho_{\rm S}$ is the Pearson correlation between the average-tie ranks of model validation MAE and benchmark error. 
A positive coefficient associates lower validation error with lower benchmark error.
Each point represents a fixed model.
The coefficients describe this collection of trained models, whose training data, architectures and capacities are related.
Energy MAE, force MAE and MAE pooled over the six stress components give identical rankings for all 13 DTM checkpoints.
Force MAE is used as their representative rank predictor in main Fig.~6 and Supplementary Figs.~\ref{fig:si-mae-physical} and~\ref{fig:si-mae-dynamics}.
Their pairwise rank correlations across all 26 models are 0.960--0.975.
DPA4.1, GRACE, MACE and Allegro each have two measured models and enter the within-family analysis through paired changes.

For each benchmark in Supplementary Table~\ref{tab:si-mae-benchmarks}, we used the same materials for every model group so that differences between correlations reflect the choice of models.
Supplementary Table~\ref{tab:benchmarks-properties} gives the mean errors and the numbers of successful calculations on the full static benchmarks.

\begin{table}[htbp]
\centering
\footnotesize
\caption{Spearman rank correlations between official OMat24 validation force MAE and physical benchmark errors.
Positive values indicate that models with lower force MAE tend to have lower benchmark errors.}
\label{tab:si-mae-benchmarks}
\setlength{\tabcolsep}{4pt}
\begin{tabular}{@{}lrrrrr@{}}
\toprule
 & & \multicolumn{4}{c}{Spearman $\rho_{\mathrm S}$}\\
\cmidrule(lr){3-6}
Benchmark & Cases & All models & External & DTM & DPA4C\\
 & & $n=26$ & $n=13$ & $n=13$ & $n=5$\\
\midrule
Formation energy & 698 & 0.610 & 0.269 & 0.929 & 0.300\\
Structure distance & 698 & 0.720 & 0.555 & 0.973 & 1.000\\
Bulk modulus & 3,602 & 0.434 & -0.005 & 0.995 & -0.300\\
Shear: original (primary) & 3,602 & 0.387 & 0.500 & 0.225 & 0.200\\
Shear: mixed-grid sensitivity & 3,603 & 0.621 & 0.500 & 0.764 & 0.200\\
Force-scale deviation & 979 & 0.707 & 0.489 & 0.978 & 1.000\\
Thermal conductivity & 103 & 0.470 & 0.423 & 0.857 & 0.700\\
Nonmolecular RDF & 12 & 0.748 & 0.692 & 0.907 & 0.900\\
Nonmolecular ADF & 12 & 0.052 & 0.264 & -0.280 & 0.900\\
Nonmolecular vDOS & 12 & 0.646 & 0.544 & 0.885 & 1.000\\
Molecular RDF & 5 & 0.415 & 0.626 & 0.247 & 0.900\\
Molecular ADF & 5 & 0.339 & 0.412 & 0.264 & 1.000\\
Molecular vDOS & 5 & 0.536 & 0.308 & 0.670 & 0.900\\
Nonmolecular pressure & 11 & 0.558 & 0.566 & 0.577 & 0.900\\
\bottomrule
\end{tabular}
\par\smallskip
\begin{minipage}{\linewidth}\footnotesize
For static properties, we used the materials for which all 26 models completed the calculation.
For thermal conductivity, we included every material and assigned a score of 2 to each failed calculation.
The primary shear comparison uses the original strain amplitudes for every model.
The mixed-grid comparison uses halved strain amplitudes for M8-P2+MA and M8-P4+MA and the original amplitudes for all other checkpoints.
We analysed molecular and nonmolecular trajectories separately and used only the nonmolecular systems with reference pressures for the pressure comparison.
Source Data give the heat-capacity correlations separately because the DTM and external models use different acceptance rules (Section~\ref{sec:si-omat-static}).
\end{minipage}
\end{table}

Across all 26 models, the correlations of validation force MAE with formation-energy and bulk-modulus errors are 0.610 and 0.434, respectively.
Across the 13 external models they are 0.269 and $-0.005$.
To test sensitivity to the choice of external models, we omitted DPA4.1 Nano and Mini while retaining the full-split validation MAEs and the same material sets for formation energy and bulk modulus.
Among the remaining 11 models, the corresponding correlations are 0.791 and 0.636.
Within DTM, the formation, bulk and force-scale associations are stronger (Supplementary Table~\ref{tab:si-mae-benchmarks}).

Paired model comparisons also reverse some accuracy rankings. 
MACE Medium has lower force MAE than Small but higher shear error at the original strain amplitudes.
GRACE-2L has lower force MAE than GRACE-1L but higher molecular-crystal vDOS error.
The P2/P4 conductivity comparison in Section~\ref{sec:si-omat-conductivity} additionally separates the full-population outcome from performance conditional on successful calculations.

Shear correlations also depend on the strain grid. 
A separate comparison fixes the 10 DTM models with both grids and the 3,757 materials successful in all 20 evaluations.
The coefficients are 0.018 for the original strain amplitudes and 0.552 for halved amplitudes. 
The three additional fast base models were evaluated only at the original amplitudes.
This comparison isolates the grid change within the 10-model population.

In the nonmolecular benchmarks, one system strongly influences the correlation between force MAE and angular distribution function (ADF) error.
Across 13 DTM checkpoints, the correlation is $-0.280$ for all 12 systems (Supplementary Table~\ref{tab:si-mae-benchmarks}).
Omitting MAPbBr$_3$ at 300\,K changes it to $0.758$.
Omitting any other single system leaves the correlation negative.

\FloatBarrier
\clearpage
\FloatBarrier
\renewcommand{\refname}{References for the Supplementary Information}
\bibliographystyle{ncs-unsrtnat}
\bibliography{references}